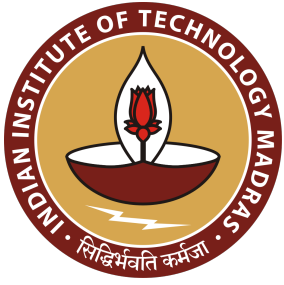

DEPARTMENT OF PHYSICS
INDIAN INSTITUTE OF TECHNOLOGY MADRAS
CHENNAI – 600036

# Information Gain, Operator Spreading, and Sensitivity to Perturbations as Quantifiers of Chaos in Quantum Systems

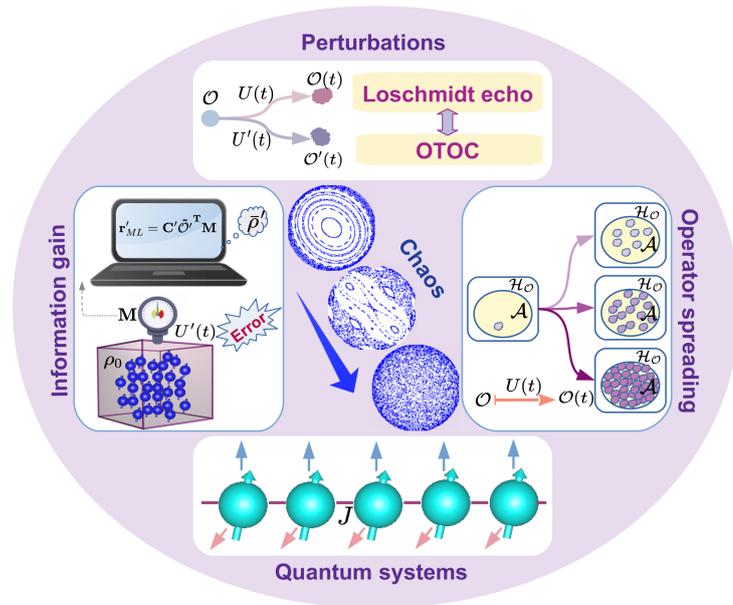

*A Thesis*

*Submitted by*

**ABINASH SAHU**

*For the award of the degree*

*Of*

**DOCTOR OF PHILOSOPHY**

March 2024

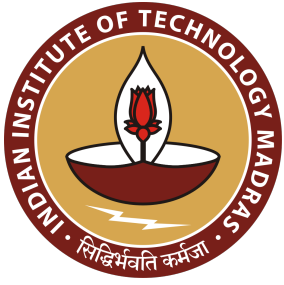

DEPARTMENT OF PHYSICS
INDIAN INSTITUTE OF TECHNOLOGY MADRAS
CHENNAI – 600036

# Information Gain, Operator Spreading, and Sensitivity to Perturbations as Quantifiers of Chaos in Quantum Systems

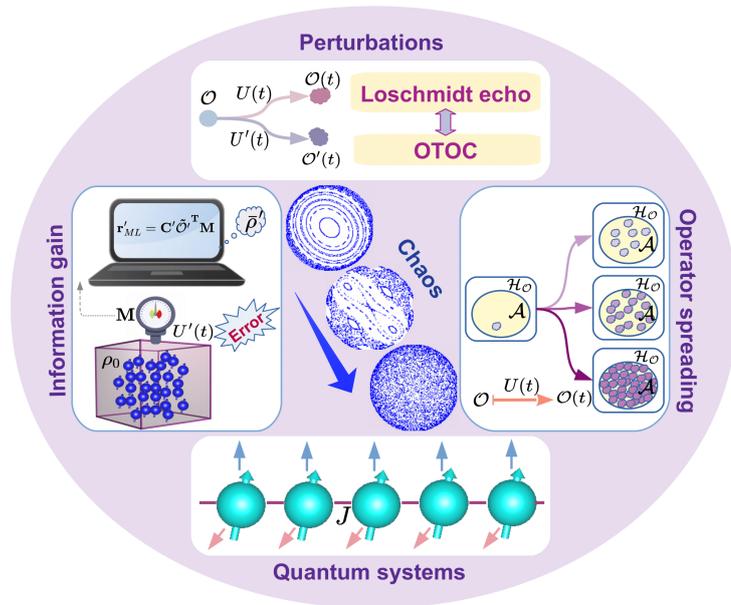

*A Thesis*

*Submitted by*

**ABINASH SAHU**

*For the award of the degree*

*Of*

**DOCTOR OF PHILOSOPHY**

March 2024

# QUOTATION

*pañchaitāni mahā-bāho kāraṇāni nibodha me*

*sānkhye kṛitānte proktāni siddhaye sarva-karmaṇām*

*adhiṣṭhānaṁ tathā kartā karaṇaṁ ca pṛthag-vidham*

*vividhāśh ca pṛthak ceṣṭā daivaṁ caivātra pañcamam*

Translation:

*O mighty-armed Arjuna, learn from Me of the five factors which bring about the accomplishment of all action. These are declared in sāṅkhya philosophy to be the place of action, the performer, the senses, the endeavor, and ultimately the Supersoul.*

**Lord Krishna**

**[Bhagavad Gita, Ch.18, Texts 13-14]**

# DEDICATION

*To the Supreme Lord Jagannath for everything.*

# THESIS CERTIFICATE

This is to undertake that the Thesis titled **INFORMATION GAIN, OPERATOR SPREADING, AND SENSITIVITY TO PERTURBATIONS AS QUANTIFIERS OF CHAOS IN QUANTUM SYSTEMS**, submitted by me to the Indian Institute of Technology Madras, for the award of **Doctor of Philosophy**, is a bona fide record of the research work done by me under the supervision of **Dr. Vaibhav Madhok**. The contents of this Thesis, in full or in parts, have not been submitted to any other Institute or University for the award of any degree or diploma.

**Chennai 600036**  **Abinash Sahu**

**Date: 22 March 2024**

**Dr. Vaibhav Madhok**
Research Supervisor
Associate Professor
Department of Physics
IIT Madras

# LIST OF PUBLICATIONS

## I. REFEREED JOURNALS BASED ON THESIS

1. **Abinash Sahu, Sreeram P G,** and **Vaibhav Madhok.** Effect of chaos on information gain in quantum tomography. *Phys. Rev. E*, **106**, 024209, (2022).

2. **Abinash Sahu, Naga Dileep Varikuti, Bishal Kumar Das,** and **Vaibhav Madhok.** Quantifying operator spreading and chaos in Krylov subspaces with quantum state reconstruction. *Phys. Rev. B*, **108**, 224306 (2023).

3. **Abinash Sahu, Naga Dileep Varikuti,** and **Vaibhav Madhok.** Quantum tomography under perturbed hamiltonian evolution and scrambling of errors– a quantum signature of chaos. Under review with *npj Quantum Information. arXiv:2211.11221*, 2022.

## II. REFEREED JOURNALS (OTHERS)

1. **Naga Dileep Varikuti, Abinash Sahu, Arul Lakshminarayan,** and **Vaibhav Madhok.** Probing dynamical sensitivity of a non-Kolmogorov-Arnold-Moser system through out-of-time-order correlators. *Phys. Rev. E.*, **109**, 014209 (2024).

## III. AWARDS

1. Got the second best poster award for presenting a poster titled "Effect of chaos on information gain in quantum tomography" at the conference CNSD 2022, IISER Pune, India.

## IV. PRESENTATIONS IN INTERNATIONAL CONFERENCES

1. Presented a poster titled "Quantum tomography under perturbed Hamiltonian evolution and scrambling of errors - a quantum signature of chaos" at Quantum Information Processing (QIP 2023), Ghent University, Ghent, Belgium, February 4-10, 2023.

## V. PRESENTATIONS IN NATIONAL CONFERENCES

1. Presented a poster titled "Quantum tomography under perturbed Hamiltonian evolution and scrambling of errors - a quantum signature of chaos" at 7th Physics In-house Symposium 2023, Department of Physics, IIT Madras, October 27-28, 2023.

2. Presented a poster titled "Quantum tomography under perturbed Hamiltonian evolution and scrambling of errors - a quantum signature of chaos" at Complexity and Nonlinear Dynamics in Science, Engineering, Technology, and Mathematics (CNLDS 2023), IIT Hyderabad, Telengana, June 5-7, 2023.

3. Presented a poster titled "Quantum tomography under perturbed Hamiltonian evolution and scrambling of errors - a quantum signature of chaos" at Progress in Quantum Science and Technologies (PiQuST 2023), IIT Madras, Chennai, India, January 23-27, 2023.

4. Presented a poster titled "Effect of chaos on information gain in quantum tomography" at the Conference on Nonlinear Systems and Dynamics (CNSD 2022), IISER Pune, India, December 15-18, 2022.

5. Presented a poster titled "Quantum signature of chaos in tomography of spin coherent states" at Quantum Foundations, Technologies, and Applications (QFTA 2020), IISER Mohali, India, December 04-09, 2020.

# ACKNOWLEDGEMENTS

First and foremost, I am extremely grateful to my advisor Dr. Vaibhav Madhok, who has been very encouraging and supportive right from the very beginning. He has allowed me to work at my own pace and nurtured independent thinking, which helped me to identify my strengths and boundaries. I have enjoyed discussing physics with him sometimes even on the weekends and also late at night. I am forever indebted to him for everything I have learnt from him. Apart from academics, he has also been very cooperative as a good friend and mentor during my difficulties.

I thank my doctoral committee Chair, and our HoD, Prof. Arul Lakshminarayan for all the enlightening discussions at various instances during my research at IIT Madras. I also thank Dr. Prabha Mandayam for insightful discussions during the weekly journal club meetings. I am very thankful to my doctoral committee members- Dr. Sunethra Ramanan, Dr. Pradeep Sarvepalli, and Dr. Sivarama Krishnan, for their support and valuable input.

I am indebted to the Late Prof. SVM Satyanarayana, and Prof. Alok Sharan for their constant guidance and support throughout my MSc days at Pondicherry University. I feel blessed to have them as my teachers in my life. I am thankful to Prof. Manu Jaiswal, and Prof. Birabar Ranjit Kumar Nanda for their keen interest in my progress. I can not thank enough my teachers- Dr. Shukla, Mrs. Subhalaxmi Das, Mr. Akshaya Mishra, Mrs. Anjali Priya Pradhan, Mr. Bibhuti Bhusan Sahu, and Mr. Radhakanta Sahu, who have shaped my early life of education. I acknowledge Dr. Sanat Nalini Roy for being very caring and trying his best to make me fit for my work after my accident through his physiotherapy treatment, counseling and motivating words.

I am thankful to Dr. Chandrashekar Radhakrishnan for his guidance whenever I needed it. Thanks to all my quantum information groupmates- Akshaya, Sreeram, Dileep, Suhail,

Bishal, Dhrubo, Sourav, Vikash, Praveen, Ramdass, and others for being helpful and ready to discuss whenever needed. I thank my other physics friends- Jatin, Bubunu, Bhabani, Nihar, Milan, Mohan, Arpita, Shakshi, Bhawani, Himalaya, Devendar, Santilata, and Mamta for their timely support and motivating words. I also thank my non-physics friends- Abhiseka, Pravash, Harekrushna, Rajasekar, Chhotelal, Ganapathi, Hrishikesh, Ajay, Aman, Dharneesh, Yash, Madhusmita, and Priyadarshini for their time and support.

I am thankful to IIT Madras for providing a vibrant working atmosphere, and a lively greenery campus with innocent deer and naughty monkeys around.

I sincerely thank my Papa and Maa for their unconditional love, support, teachings, and encouragement. I also thank my brother and cousins for always helping me. I express gratitude to all my family members and friends.

# ABSTRACT

**KEYWORDS**    Quantum chaos; kicked top; quantum tomography; Krylov subspace; operator spreading; Loschmidt echo; quantum simulations

Quantum chaos is the study of signatures of classical chaos in corresponding quantum system(s). How does chaos manifest itself in quantum systems? In this quest, we explore footprints of chaos in quantum tomography. We adopt a continuous weak measurement tomography protocol where we generate the measurement record as a series of expectation values of an observable evolving under the desired dynamics, which can show a transition from integrability to complete chaos. Does chaos in the dynamics enable information gain in quantum tomography or impede it? For a given dynamics and an initial Hermitian observable, we observe a completely opposite behavior in tomography of well-localized spin coherent states compared to random states. As the chaos in the dynamics increases, the reconstruction fidelity of spin coherent states decreases. This contrasts with the previous results connecting information gain in tomography of random states with the degree of chaos in the dynamics that drives the system. The rate of information gain and, hence, the fidelity obtained in tomography depends not only on the degree of chaos in the dynamics and to what extent it causes the initial operator to spread in various directions of the operator space but, more importantly, how well these directions are aligned with the density matrix to be estimated.

We further investigate operator spreading, as described above, and its connections to chaos in many-body quantum dynamics by studying its potential to generate an informationally complete measurement record in quantum tomography. We find that the magnitude of operator spreading, which is closely related to and can be quantified by the fidelity of reconstructing random states in quantum tomography and various other metrics of information gain, increases with the degree of chaos in the system. We consider the tilted field Ising model with and without delta kicks and the Heisenberg

XXZ model with a single impurity to explore operator spreading. We find our approach in quantifying operator spreading is a more consistent indicator of quantum chaos than Krylov complexity as the latter may correlate/anti-correlate or show no consistent behavior with the degree of chaos in the dynamics. Our study also gives an operational interpretation for operator spreading in terms of fidelity gain in an actual quantum information processing protocol, which can be realized in experiments with cold atoms and laser fields.

Continuing in our journey of finding footprints of chaos in the quantum domain, we explore the growth of errors in noisy tomography, where the source of noise is in the dynamics employed to drive the system. We find a quantity to capture the "scrambling of errors", an out-of-time-ordered correlator (OTOC) between two operators under perturbed and unperturbed system dynamics, serves as a signature of chaos and quantifies the spread of errors. We find a fundamental link between operator Loschmidt echo and scrambling of errors, as captured by OTOC discussed above, and show that this has a direct bearing on our ability to simulate quantum dynamics, which will inadvertently be accompanied by noise. There is an interesting interplay between information gain due to system dynamics and the growth of errors due to noise. We find that the reconstruction fidelity initially increases regardless of the degree of chaos or strength of noise in the dynamics. Thereafter, the fidelity drops as the information gain due to subsequent evolution cannot offset the accumulation of errors in the measurement record. In particular, for random states, when the measurement record is obtained from a random initial observable, the subsequent drop in the fidelity obtained is inversely correlated to the degree of chaos in the dynamics. More importantly, this also gives us an operational interpretation of Loschmidt echo for operators by connecting it to the performance of quantum tomography.

# CONTENTS

# LIST OF FIGURES

# ABBREVIATIONS

**BGS**    Bohigas-Giannoni-Schmit.

**COE**    Circular orthogonal ensemble.

**CUE**    Circular unitary ensemble.

**GOE**    Gaussian orthogonal ensemble.

**GUE**    Gaussian unitary ensemble.

**IPR**    Inverse participation ratio.

**KAM**    Kolmogorov-Arnold-Moser.

**KS**    Kolmogorov-Sinai.

**NMR**    Nuclear magnetic resonance.

**OTOCs** Out-of-time-ordered correlators.

**PR**    Participation ratio.

**RMT**    Random matrix theory.

# NOTATION

$\mathcal{I}_O$     operator incompatibility

$\bar{\rho}$     reconstructed density operator

**C**     covariance matrix

**M**     measurement record

**r**     Bloch vector

$|\psi_0\rangle$     initial state vector

$|\theta, \phi\rangle$     spin coherent state

$\mathbb{1}$     identity operator

$\mathcal{A}$     subspace spanned by the time evolution of operator $O$

$C_K$     Krylov complexity

$\mathcal{D}_{HS}$     Hilbert-Schmidt distance

$\mathcal{D}_{KL}$     quantum relative entropy

$\mathcal{F}$     reconstruction fidelity

$\mathcal{F}_O$     operator Loschmidt echo

$\mathcal{H}$     Hilbert space

$\mathcal{H}_O$     Hilbert space of operators

$\mathcal{J}$     Fisher information

$O$     initial observable

$O(t), O_n$     operator at time $t$ or at $n_{th}$ time steps

$Q_\rho(\theta, \phi)$  Husimi $Q$ function

$\mathcal{R}$  rank of the covariance matrix

$\mathcal{S}_C$  Shannon entropy

$\mathcal{S}_K$  Krylov entropy

**F**  Fisher information matrix

$\rho_0$  initial density operator for reconstruction

$\sigma^x, \sigma^y, \sigma^z$  Pauli matrices

$\tau$  time period for delta kicks

***J***  angular momentum operator

Tr  trace

$d$  dimension of the Hilbert space

$E_\alpha$  generalized Gellmann matrices, traceless Hermitian operators

$K$  dimension of Krylov subspace

$L$  length of the spin chain

$M(t), M_n$  measurement outcome at time $t$ or at $n_{th}$ time steps

$N_s$  number of particles or identical systems

$S_\rho$  Husimi entropy

$U, U_\tau$  unitary operator

$W(t), W_n$  Gaussian white noise at time $t$ or at $n_{th}$ time steps

$\bar{\mathbf{r}}$  reconstructed Bloch vector

$\mathbf{r}_{ML}$  maximum-likelihood (ML) estimation of the Bloch vector

# CHAPTER 1

# INTRODUCTION

## 1.1 CHAOS IN CLASSICAL AND QUANTUM DYNAMICAL SYSTEMS

Laplace said, standing on Newton's shoulders, "Tell me the force and where we are, and I will predict the future!" That assertion transforms into an important theorem about differential equations- the uniqueness theorem, which guarantees the uniqueness of solutions for a given initial condition. However, the proven existence of a solution does not assure us that we can actually determine the solution. Thus, deterministic time evolution does not promise predictability. At the heart of chaos lies this concept of deterministic unpredictability or deterministic randomness. Formally speaking, chaos is the long-term aperiodic behavior in a deterministic system that shows sensitive dependence on initial conditions [Strogatz (2018)]. Unpredictability is not due to limitations in computational or intellectual capabilities; rather, it is an intrinsic property of certain nonlinear systems. A system having $N_d$ degrees of freedom is called an integrable system if it possesses $N_d$ constants of motions or conserved quantities. For integrable systems, the dynamics is constrained to an $N_d$-dimensional invariant tori from the $2N_d$ dimensional phase space. The nonintegrable system has fewer than $N_d$ number of conserved quantities, and the whole phase space can be filled by complex trajectories showing chaotic behavior.

The motion of planets or the motion of the cricket ball can be perfectly explained by Newton's laws of motion and other laws of classical physics. However, the world is quantum mechanical as everything is made up of quantum particles like atoms and molecules. Quantum theory successfully explains many phenomena, such as the stability of atoms, formation of solids, black-body radiation, photoelectric effect, etc., which cannot be understood by applying classical physics. Thus, the intriguing aspect lies in understanding the quantum-to-classical transition, unraveling how classical mechanics emerges from the foundational quantum framework. Closely related is the question of the origin of classical chaos stemming from nonlinear phenomena, from the underlying quantum world whose evolution is governed by linear and unitary dynamics.

The rapid divergence of neighbouring classical trajectories with time, often described as exponential sensitivity to initial conditions, is the hallmark of deterministic chaos in classical mechanics. However, in the quantum domain, the initial conditions are not points in the phase space because of the Heisenberg uncertainty principle. The quantum state is described by a state vector in the Hilbert space. Thus, the divergence between two possible initial states cannot occur quantum mechanically because of the necessity of preserving the inner product due to the linearity of Schrödinger's equation. Here arises a question: how is classical chaos reflected in quantum dynamical systems whose classical counterpart is chaotic?

One of the primary goals of quantum chaos study is to search for various signatures of chaos in the quantum domain and their consequences in quantum information processing, statistical mechanics, foundational areas like quantum-to-classical transition, and the rate of decoherence under chaotic dynamics. Various signatures of chaos have been discovered. Starting from the behavior of the spectral statistics of the generating Hamiltonian [Haake (1991)] to dynamical signatures of chaos like hypersensitivity of system dynamics to perturbation [Peres (1984); Schack and Caves (1996)] and dynamically generated quantum correlations: quantum discord [Madhok *et al.* (2015, 2018)], and quantum entanglement [Miller and Sarkar (1999); Bandyopadhyay and Lakshminarayan (2002); Wang *et al.* (2004); Trail *et al.* (2008); Furuya *et al.* (1998); Lakshminarayan (2001); Seshadri *et al.* (2018)]. Recently, out-of-time-ordered correlators (OTOCs) have also been used to explore quantum chaos [Maldacena *et al.* (2016); Swingle *et al.* (2016); Hashimoto *et al.* (2017); Kukuljan *et al.* (2017); Swingle (2018); Wang *et al.* (2021); Sreeram *et al.* (2021); Varikuti and Madhok (2022)]. The signatures of chaos are not only explored in the semiclassical limit but also in the deep quantum regime [Neill *et al.*

(2016); Fortes *et al.* (2020); Sreeram *et al.* (2021)].

## 1.2 QUANTUM TOMOGRAPHY AND CHAOS

Tomography of quantum states is essential for quantum information processing tasks like quantum computation, quantum cryptography, quantum simulations, and quantum control. Estimation of quantum states is a highly nontrivial problem because of fundamental restrictions posed by Heisenberg's uncertainty principle and no-cloning theorem [Wootters and Zurek (1982)]. Different protocols have carried out tomography in many systems [Paris and Rehacek (2004); Mauro D'Ariano *et al.* (2003)]. State reconstruction uses the statistics of measurement records on an ensemble of identical systems in order to best estimate the actual state $\rho_0$. An informationally complete set of measurement records is required for high-fidelity tomography. Inverting these records, in principle, should give an estimate of the state. The traditional way has been using projective measurements to extract the information. However, such protocols are resource-intensive since strong measurements destroy the state. To get good fidelity reconstruction, one would then require infinitely many copies of the system. Weak measurement alternatives have been explored in the literature [Lundeen *et al.* (2011); Wu (2013); Hofmann (2010); Shojaee *et al.* (2018); Silberfarb *et al.* (2005)]. Weak measurements, typically act on an ensemble of identically prepared systems that are collectively and coherently evolved and cause minimal disturbance to the system. However, the amount of information gained per measurement is bound to be low in this type of measurement [Busch (2009)]. In this thesis, we are interested in continuous weak measurement tomography [Silberfarb *et al.* (2005); Smith *et al.* (2006); Chaudhury *et al.* (2009); Merkel *et al.* (2010); Smith *et al.* (2004)] where the probe, the laser light, continuously monitors the expectation value of an atomic observable. After its interaction with the atoms, one has a continuous time series as the measurement record. The information of the atomic spin is continuously mapped on to the laser. It then undergoes a homodyne-like detection, and the photocurrent obtained is a continuous time series. It is in this sense that this measurement of the atomic degree of freedom is continuous. We can generate the time series of operators using a random unitary [Merkel *et al.* (2010); PG and Madhok (2021)] or the Floquet operator of a quantum dynamical system [Madhok *et al.* (2014)] in the Heisenberg picture and measurement record is obtained.

One can ask the following question: How is reconstructing quantum states related to the nature of dynamics employed in the tomography process? At first, the connection between chaos and state reconstruction seems distant. Chaos is about the inability to predict the long-term behavior of a dynamical system, while tomography involves information acquisition. However, the flip side of this uncertainty and unpredictability of chaotic dynamics is information. If everything is known about a trajectory, for example, a periodic orbit, we gain no new information. Classically, as one tracks a chaotic trajectory, one gains information at a rate proportional to the magnitude of chaos in the system. This rate is more formally described as the Kolmogorov-Sinai (KS) entropy and is equal to the sum of positive Lyapunov exponents of the system [Kolmogorov (1958, 1959); Sinai (1959); Pesin (1977)]. One might ask what this information is about. The answer is *initial conditions*. One obtains information on increasingly finer scales about the system's initial conditions. In quantum mechanics, this is precisely the goal of tomography. As one follows the archive of the measurement record in a tomography experiment, one gains information about the initial random quantum state. An intriguing question in the quantum case is whether or not the state reconstruction rate is related to the complexity of the dynamics. There seems to be a fascinating and provocative connection between tomography and chaos, as demonstrated in Ref. [Madhok *et al.* (2014)]. While Ref. [Madhok *et al.* (2014)] considered quantum tomography for random states, we find that state reconstruction for localized wave packets remarkably shows the opposite behavior! We show that the rate of state reconstruction is a function of dynamics, the initial state, as well as the time-evolved operators.

## 1.3 OPERATOR SPREADING IN KRYLOV SUBSPACE

Operator spreading characterizes a process in which a local operator evolves under many-body Hamiltonian dynamics in the Heisenberg picture and extends over the entire system [Von Keyserlingk et al. (2018)]. The operator spreading also serves as a probe for quantum information scrambling that is inaccessible to local measurements. Once the information is scrambled, the information is now delocalized over the entire operator space in complex observables. Thus, operator spreading is also connected to the understanding of the questions of chaos, nonintegrability, and thermalization in many-body quantum systems [Deutsch (1991); Srednicki (1994); Tasaki (1998); Rigol et al. (2008); Rigol and Santos (2010); Torres-Herrera and Santos (2013)]. Intense research has been directed towards the study of operator spreading in various fields such as black hole physics [Hayden and Preskill (2007); Sekino and Susskind (2008); Hosur et al. (2016); Shenker and Stanford (2014); McGinley et al. (2022)], holography [Bhattacharyya et al. (2022)], integrable systems [Xu et al. (2020); Rozenbaum et al. (2020); Pilatowsky-Cameo et al. (2020)], random unitary circuits [Nahum et al. (2018a,b); Khemani et al. (2018); Rakovszky et al. (2018)], quantum field theories [Roberts and Stanford (2015); Stanford (2016); Chowdhury and Swingle (2017); Patel et al. (2017)], and chaotic spin-chains [Luitz and Lev (2017); Heyl et al. (2018); Lin and Motrunich (2018); Geller et al. (2022)].

For many quantum systems, operator spreading is a reliable indicator of chaos in the dynamics [Moudgalya et al. (2019); Omanakuttan et al. (2023)]. One can quantify the spreading of operators through out-of-time-ordered correlators (OTOCs) [Maldacena et al. (2016); Swingle (2018); Seshadri et al. (2018); Prakash and Lakshminarayan (2020); Xu and Swingle (2020); Sreeram et al. (2021); Varikuti and Madhok (2022)], operator entanglement [Nie et al. (2019); Wang and Zhou (2019); Alba et al. (2019); Styliaris et al. (2021)], memory matrix formalism [McCulloch and Von Keyserlingk (2022)] or Krylov complexity [Parker et al. (2019); Yates and Mitra (2021); Rabinovici et al. (2021); Noh (2021); Dymarsky and Smolkin (2021); Caputa et al. (2022); Rabinovici et al. (2022b); Avdoshkin et al. (2022); Rabinovici et al. (2022a); Bhattacharya et al.

(2022, 2023); Suchsland *et al.* (2023)]. OTOCs, which measure the incompatibility between a stationary operator and another operator evolving with time in the Heisenberg picture, have been studied extensively to witness operator growth. However, measuring OTOCs in the lab is challenging even with state-of-the-art experimental techniques, which require backward evolution in time, that is, the ability to completely reverse the Hamiltonian [Li *et al.* (2017); Green *et al.* (2022); Gärttner *et al.* (2017); Zhu *et al.* (2016); Swingle *et al.* (2016); Yao *et al.* (2016); Halpern (2017); Bohrdt *et al.* (2017); Tsuji *et al.* (2017); Nie *et al.* (2020); Dressel *et al.* (2018); Joshi *et al.* (2020); Asban *et al.* (2021)]. Recently, certain protocols have been discussed where one can avoid the problem of backward time evolution [Vermersch *et al.* (2019); Blocher *et al.* (2022); Sundar *et al.* (2022)]. Furthermore, the Krylov complexity, another quantifier of operator spreading, is computed when the operator at a given time is expressed in an orthonormal sequence of operators generated from the Lanczos algorithm. The Liouvillian superoperator of a time-independent Hamiltonian is repeatedly applied on the initial operator to construct the Krylov basis. In Sec. 2.4 of Chapter 2, we have detailed the procedure for obtaining Krylov subspace in the Lanczos algorithm and quantifying Krylov complexity. Nevertheless, the saturation value of the Krylov complexity depends on the choice of initial observable. Moreover, the initial growth of Krylov complexity has been observed to be exponential for certain non-chaotic dynamics [Dymarsky and Smolkin (2021); Avdoshkin *et al.* (2022)]. Thus, the Krylov complexity does not serve as an unambiguous indicator of chaos.

We take an alternate route and quantify operator spreading through the performance of a concrete quantum information processing task - quantum tomography. How does the system dynamics drive operator complexity, which affects the information gain in quantum tomography? We answer the above question by quantifying operator spreading in integrable, nonintegrable, and chaotic many-body systems via their ability to generate an optimal measurement record for quantum tomography. Intuitively, an evolution of a fiducial operator with a single random unitary will lead to maximal operator spreading over the entire operator space [Merkel *et al.* (2010); Sreeram and Madhok (2021)]. Krylov subspace for operators is generated by repeated application of a map to an initial operator. Thus, such a random unitary evolution will also saturate the maximum possible dimension of the Krylov subspace. In our study, the operators of Krylov subspace obtained from an initial operator upon time evolution have a simple interpretation. These are elements in the operator Hilbert space that will be measured in tomography. How many of these directions and with what signal-to-noise ratio they are measured as the dynamics becomes nonintegrable and increasingly chaotic will give an operational and physically motivated way to quantify operator spreading.

The Lyapunov exponents quantify the rapid divergence rate of neighbouring trajectories in a classically chaotic system. The quantum counterpart of these diverging trajectories is the growth of incompatibility of operators as quantified by the OTOCs, which give the "quantum Lyapunov exponents" of the dynamics. Therefore, we unify the connections between information gain, scrambling, and chaos through an actual physical process.

A connection between tomography, which is about information gain of an unknown state, and chaos, which represents unpredictability, seems to be at odds with each other. However, a deeper analysis reveals fundamental connections. Classically chaotic systems show exponential sensitivity to perturbations of the initial conditions, as measured by the Kolmogorov-Sinai (KS) entropy [Pesin (1977)], which is equal to the sum of positive Lyapunov exponents of the system. On the flip side, KS entropy also measures the rate at which successive measurements on a classically chaotic system provide information about the initial condition, which is the missing information. A time history of a trajectory at discrete times is an archive of information about the initial conditions given perfect knowledge about the dynamics. Moreover, if the dynamics is chaotic, the rate at which we learn information increases is given by the KS entropy. This is precisely analogous to quantum tomography, where the missing information is the unknown initial quantum state.

This thesis considers the dynamics of the 1D Ising model with a tilted magnetic field [Prosen (2007); Pineda and Prosen (2007); Kukuljan *et al.* (2017); Karthik *et al.* (2007); Prosen and Žnidarič (2007)], and the 1D anisotropic Heisenberg XXZ model with an integrability-breaking field [Santos (2004); Santos *et al.* (2004); Barišić *et al.* (2009); Rigol and Santos (2010); Santos and Mitra (2011); Santos *et al.* (2012); Gubin and F Santos (2012); Brenes *et al.* (2018, 2020); Pandey *et al.* (2020)] to study the growth of operator spreading and its connection to chaos in quantum tomography. They manifest a range of behavior from integrable to fully chaotic. The Hamiltonian of the Ising model with either a time-independent tilted field or a time-dependent delta-kicked tilted field shows integrable to chaos transition, and we explore both models for our tomography process.

## 1.4 NOISY TOMOGRAPHY AND CHAOS

Chaos, classically as well as in quantum mechanics, has an intimate connection with complexity. Classically, chaos implies unpredictability. Time-evolved trajectories twist and wind away from each other at an exponential rate and then fold back to remain confined in a bounded phase space respecting ergodicity. The flip side of these complex trajectories is the potential information that can be obtained if one tracks these trajectories. The perspective, quantified by the KS (Kolmogorv-Sinai) entropy [Pesin (1977)], gives the rate of information gain, at increasingly fine scales, about *the initial conditions* or the missing information of the phase space [Caves and Schack (1997)].

The central goal of quantum chaos is to inform us about the properties of quantum systems having chaotic dynamics in the classical limit. What notions of complexity might be suitable to quantify the manifestation of chaos in the quantum domain? Quantum theory comes with another layer of complexity hitherto unknown in our classical description of reality, the Hilbert space, which is a big space [Caves and Fuchs (1996)]. Quantum theory permits the state of the system to be any vector in this space, even permitting a coherent superposition of possibilities considered mutually exclusive in the classical world. Therefore, while classically chaotic dynamics generate classical information as the classical trajectories, quantum chaotic dynamics produce quantum information as pseudo-random vectors in the Hilbert space, which typically have a high entropy. Here, when one talks about the entropy of pure states, it is always assumed the entropy is calculated according to a fixed fiducial basis [Wootters (1990); Bengtsson and Życzkowski (2017)]. Vigorous thrust in the understanding of quantum many-body dynamical systems through dynamically generated entanglement [Miller and Sarkar (1999); Bandyopadhyay and Lakshminarayan (2002); Wang *et al.* (2004); Trail *et al.* (2008); Furuya *et al.* (1998); Lakshminarayan (2001); Seshadri *et al.* (2018)], and quantum correlations [Madhok *et al.* (2015, 2018)], deeper studies in the ergodic hierarchy of quantum dynamical systems [Gomez and Castagnino (2014); Bertini *et al.* (2019); Aravinda *et al.* (2021); Vikram and Galitski (2022)] have been some recent milestones. Also, the out-of-time-ordered correlators (OTOCs) that attempt to capture operator growth and scrambling of quantum information have been very useful as a probe for chaos in quantum systems [Maldacena *et al.* (2016); Swingle *et al.* (2016); Hashimoto *et al.* (2017); Kukuljan *et al.* (2017); Swingle (2018); Wang *et al.* (2021); Sreeram *et al.* (2021); Varikuti and Madhok (2022)]. These, coupled with the traditional approach to studies of level statistics [Haake (1991)] and Loschmidt echo [Peres (1984, 1997); Goussev *et al.* (2012); Gorin *et al.* (2006)] and complemented by the ability to coherently control and manipulate many-body quantum systems in the laboratory [Gong and Brumer (2001, 2005); Brif *et al.* (2010); Smith *et al.* (2013); Mirkin and Wisniacki (2021)], have brought us to a fork in our path. On the one hand, this is a harbinger of the possibility of building quantum simulators, an important milestone in our quest for the holy grail - a many-body quantum computer. On the other hand, the same properties that make quantum systems generate complexity will make them sensitive to errors that naturally occur in implementing many-body Hamiltonians.

The larger question here is: Can one trust an analog quantum simulator [Hauke *et al.* (2012); Chinni *et al.* (2022)]? For a digital quantum computer, one has a fully developed error correction formalism, and a fault-tolerant quantum computer is, at the very least, a theoretical possibility. Such a guarantee does not exist for a quantum simulator studying, for example, quantum phase transitions in condensed matter systems or structural instabilities in quantum chaotic systems [Chinni *et al.* (2022); Chaudhury *et al.* (2009)]. In these scenarios, the imperfect evolution of the quantum system and trotter errors introduced in simulating the dynamics will cause the actual system dynamics to depart from the simulated dynamics [Lloyd (1996); Heyl *et al.* (2019)]. This is particularly of concern when the trotter errors can be a source of nonintegrability in the dynamics [Sieberer *et al.* (2019)]. This is important as any continuous variable quantum information processing protocol can be viewed as an analog quantum simulator. Thus, the accuracy of any real experiment will decay with time if there are some imperfections.

While chaotic dynamics is a source of information quantified by the positive KS entropy, it is sensitive to errors, as captured by Loschmidt echo. In many body systems, quantum or classical, we must expect the presence of both chaos and errors. In this thesis, we address this scenario; we go on to discover quantum signatures of chaos while shedding light on the larger question of many-body quantum simulations under unavoidable perturbations. While the KS entropy enables a rapid information gain, Loschmidt echo will cause a rapid accumulation of errors, or *error scrambling* as we quantify. This interplay between KS entropy and Loschmidt echo is a generic feature of any many-body system, and we identify and quantify the crossover between these two competing effects.

Quantum tomography gives us a window to study sensitivity to errors in quantum simulations of chaotic Hamiltonians [Lloyd (1996); Johnson *et al.* (2014)]. We consider continuous weak measurement tomography protocol [Silberfarb *et al.* (2005); Smith *et al.* (2006); Riofrío *et al.* (2011); Smith *et al.* (2013)], and the time series of operators can be generated by the Floquet map of a quantum dynamical system to investigate the role of chaos on the information gain in tomography [Madhok *et al.* (2014); Sreeram and Madhok (2021); Sahu *et al.* (2022*a*)]. In this thesis, we show how such errors cause a decline in the reconstruction fidelity, a very well-studied metric obtained in quantum tomography, and how chaotic dynamics plays a role in such a decline. Here, we unify two quantifiers of quantum chaos, namely Loschmidt echo and OTOC, through scrambling of errors in continuous weak measurement tomography.

## 1.5 ORGANIZATION OF THE THESIS

This thesis is organized as follows. The next chapter contains preliminary information for readers. Chapter 3 discusses the effect of chaos on tomography investigated in [Sahu *et al.* (2022*a*)]. Chapter 4 is dedicated to our studies on operator spreading in many-body quantum systems under the dynamics as reported in [Sahu *et al.* (2023)]. Our exploration on error scrambling and the effect of chaos on tomography [Sahu *et al.* (2022*b*)] is explained in Chapter 5. Finally, We conclude with a brief summary of this thesis and outlook in Chapter 6. Appendix A contains the Matlab codes for continuous weak measurement tomography.

# CHAPTER 2

# BACKGROUND

In this chapter, we build a foundation to help understand the content of the remainder of this thesis. This will familiarize the reader with the essential tools, techniques, and concepts required to understand the accomplished work and presented results in the upcoming chapters.

## 2.1 CONTINUOUS WEAK MEASUREMENT TOMOGRAPHY

Quantum state tomography uses the statistics of measurement records on an ensemble of identical systems to best estimate the quantum state $\rho_0$ [Paris and Rehacek (2004); Mauro D'Ariano et al. (2003)]. Strong projective measurements of an informationally complete set of observables have traditionally been used to extract information for state reconstruction. It is a time-consuming and tedious procedure when applied to systems of large dimensions. These protocols require a significant amount of resources, as strong measurements can destroy the state. Also, one needs to reprepare the ensemble and reconfigure the measurement apparatus after each measurement in certain experimental settings [Smith et al. (2006)]. Thus, in projective measurement tomography, one makes measurements on several identically prepared systems and inverts the statistics to estimate the quantum state, which may not always be easy to implement in a laboratory. Also, one would need infinitely many copies of the system to get good reconstruction fidelity. On the other hand, weak measurements help in reducing the number of copies required for the reconstruction process as they cause minimal disturbance to the state. These protocols have been tested in the lab, and very extensive experiments even implementing the quantum kicked top [Chaudhury et al. (2009); Smith et al. (2006)] have employed this. However, the amount of information gained per measurement is bound to be low in this type of measurement. Here, we will review continuous weak measurement tomography protocol [Silberfarb *et al.* (2005); Smith *et al.* (2006); Riofrío *et al.* (2011); Smith *et al.* (2013); Merkel *et al.* (2010); Madhok *et al.* (2014, 2016); Sreeram and Madhok (2021); Sahu *et al.* (2022*a*,*b*)].

We are given an ensemble of $N_s$ identical systems $\rho_0^{\otimes N_s}$, and they undergo separable time evolution by a unitary $U(t)$. A probe is coupled to the ensemble of states that will generate the measurement record by performing weak continuous measurement of the collective observable $O$. Thus, the ensemble is coherently evolved and collectively probed to give the estimate of the state in a single shot. The operator, evolving in Heisenberg fashion, is measured at time *t* as

$$O(t) = U^\dagger(t) O U(t) \tag{2.1}$$

We can exploit this choice of dynamics for time evolution and explore many properties of the quantum systems, as we will see in the subsequent chapters.

The positive operator valued measurement (POVM) elements for measurement outcomes $X(t)$ at time *t* are [Silberfarb *et al.* (2005); Madhok *et al.* (2014)]

$$E_{X(t)} = \frac{1}{\sqrt{2\pi\sigma^2}} \exp\left\{-\frac{1}{2\sigma^2}[X(t) - O(t)]^2\right\}. \tag{2.2}$$

The standard deviation $\sigma$ in $E_{X(t)}$ is because of the probe noise (shot noise). Approximating the entire dynamics as a unitary evolution is an excellent approximation [Smith *et al.* (2013); Deutsch and Jessen (2010)]. The system and probe coupling is sufficiently weak that the entangling effect of measurement backaction is negligible, and the ensemble is well approximated by a product state at all times. The shot noise sets the fundamental resolution of the probe. When the randomness of the measurement outcomes is dominated by the quantum noise in the probe rather than the measurement uncertainty, i.e., the projection noise, quantum backaction is negligible, and the state remains approximately separable. Thus, the measurement records can be approximated

to be

$$M(t) = X(t)/N_s = \text{Tr}\left[O(t)\rho_0\right] + W(t) \qquad (2.3)$$

where $W(t)$ is a Gaussian white noise with spread $\sigma/N_s$.

The density matrix of any arbitrary state having Hilbert space dimension $d$ can be expressed in the orthonormal basis of $d^2 - 1$ traceless and Hermitian operators $\{E_\alpha\}$, and the state lies on the generalized Bloch sphere parametrized by the Bloch vector **r**. Thus, the density matrix can be represented as

$$\rho_0 = \mathbb{1}/d + \Sigma_{\alpha=1}^{d^2-1} r_\alpha E_\alpha, \qquad (2.4)$$

where

$$\Sigma_{\alpha=1}^{d^2-1} r_\alpha^2 = 1 - 1/d.$$

We consider the measurement records at discrete times

$$M_n = M(t_n) = N_s \sum_\alpha r_\alpha \text{Tr}\left[O_n E_\alpha\right] + W_n, \qquad (2.5)$$

where $O_n = U^{\dagger n} O U^n$. This allows one to express the measurement history as

$$\mathbf{M} = \tilde{O}\mathbf{r} + \mathbf{W}, \qquad (2.6)$$

i.e.

$$\begin{pmatrix} M_1 \\ M_2 \\ .. \\ .. \\ M_n \end{pmatrix} = \begin{pmatrix} \tilde{O}_{11} & \tilde{O}_{12} & .. & .. & \tilde{O}_{1d^2-1} \\ \tilde{O}_{21} & \tilde{O}_{22} & .. & .. & \tilde{O}_{2d^2-1} \\ .. & .. & .. & .. & .. \\ .. & .. & .. & .. & .. \\ \tilde{O}_{n1} & \tilde{O}_{n2} & .. & .. & \tilde{O}_{nd^2-1} \end{pmatrix} \begin{pmatrix} r_1 \\ r_2 \\ .. \\ .. \\ r_{d^2-1} \end{pmatrix} + \begin{pmatrix} W_1 \\ W_2 \\ .. \\ .. \\ W_n \end{pmatrix}. \qquad (2.7)$$

Thus, when the backaction is insignificant, the probability distribution corresponding to measurement record **M** for a given Bloch vector **r** is

$$p(\mathbf{M}|\mathbf{r}) \propto \exp\left\{-\frac{N_s^2}{2\sigma^2}\sum_i [M_i - \sum_\alpha \tilde{O}_{i\alpha} r_\alpha]^2\right\}$$
$$\propto \exp\left\{-\frac{N_s^2}{2\sigma^2}\sum_{\alpha,\beta}(\mathbf{r}-\mathbf{r_{ML}})_\alpha\, C^{-1}_{\alpha\beta}\,(\mathbf{r}-\mathbf{r_{ML}})_\beta\right\}, \quad (2.8)$$

where $\mathbf{C} = \left(\tilde{O}^T \tilde{O}\right)^{-1}$ is the covariance matrix with $\tilde{O}_{n\alpha} = \text{Tr}\,[O_n E_\alpha]$ [Silberfarb *et al.* (2005); Smith *et al.* (2006)]. Given the measurement record and the knowledge of the dynamics, one can invert this measurement record to get an estimate of the parameters characterizing the unknown quantum state in Eq. (2.4). The least-square fit of the Gaussian distribution in the parameter space is the maximum-likelihood (ML) estimation of the Bloch vector, $\mathbf{r}_{ML} = \mathbf{C}\tilde{O}^\mathbf{T}\mathbf{M}$. The measurement record is informationally complete if the covariance matrix is of full rank. Suppose the covariance matrix is not of full rank. In that case, the inverse of the covariance matrix is replaced by Moore-Penrose pseudo inverse [Ben-Israel and Greville (2003)], inverting over the subspace where the covariance matrix has support. The eigenvalues of the $\mathbf{C^{-1}}$ determine the relative signal-to-noise ratio with which different observables have been measured. The estimated Bloch vector $\mathbf{r}_{ML}$ may not represent a physical density matrix with non-negative eigenvalues because of the noise present (having a finite signal-to-noise ratio). Therefore, we impose the constraint of positive semidefiniteness [Baldwin *et al.* (2016)] on the reconstructed density matrix and obtain the physical state closest to the maximum-likelihood estimate.

To do this, we employ a convex optimization [Vandenberghe and Boyd (1996)] procedure where the final estimate of the Bloch vector $\bar{\mathbf{r}}$ is obtained by minimizing the argument

$$\|\mathbf{r}_{ML} - \bar{\mathbf{r}}\|^2 = (\mathbf{r}_{ML} - \bar{\mathbf{r}})^T \mathbf{C}^{-1}(\mathbf{r}_{ML} - \bar{\mathbf{r}}) \quad (2.9)$$

subject to the constraint

$$\mathbb{1}/d + \Sigma_{\alpha=1}^{d^2-1}\,\bar{r}_\alpha E_\alpha \geq 0.$$

The positivity constraint also plays a crucial role in compressed sensing tomography.

Any optimization heuristic with positivity constraint is effectively a compressed sensing protocol, provided that the measurements are within the special class associated with compressed sensing [Kalev *et al.* (2015)].

The performance of the quantum state tomography protocol is quantified by the fidelity of the reconstructed state $\bar{\rho}$ relative to the actual state $|\psi_0\rangle$, $\mathcal{F} = \langle\psi_0|\bar{\rho}|\psi_0\rangle$ as a function of time. The reconstruction fidelity $\mathcal{F}$ depends on the informational completeness of the measurement record [Merkel *et al.* (2010); Madhok *et al.* (2014); Sreeram and Madhok (2021)], the choice of observables and quantum states [Sahu *et al.* (2022*a*)] as well as the presence of noise in the measurement outcomes [Sahu *et al.* (2022*b*)].

### 2.1.1 Continuous measurement tomography in real experiments

Continuous weak measurement tomography protocol has been implemented on an ensemble of atomic spin systems of cesium ($^{133}$Cs) atoms [Smith *et al.* (2006); Chaudhury *et al.* (2009); Riofrío *et al.* (2011)]. The Cs ensemble consisting of a few million atoms in a cloud of a small radius was prepared by laser cooling and optical pumping. Efficient quantum control of the ground electronic manifold of cesium atoms helped to perform the state reconstruction properly. A sufficient condition for high fidelity quantum state reconstruction of Hilbert space dimension $d$ is that the dynamics can generate an informationally complete set of observable $\{O_i\}$ from the initial observable $O$ that spans the Lie algebra $\mathfrak{su}(d)$. The desired dynamics is governed by a Hamiltonian, which is a functional of a set of control waveforms of externally applied fields parametrized by frequencies, amplitudes, and phases. The total duration of the experiment is determined by the characteristic time scales for evolution, which take into account the power in the control for the Hamiltonian evolution and the decoherence. The total time is then coarse-grained into slices of duration of $\delta t$ depending on the chosen dynamics.

Two cases were being studied for different control methods: (1) control of quadratic ac Stark shift and quasi-static magnetic fields for quantum tomography on the seven-dimensional $F = 3$ hyperfine manifold of Cs; (2) control of time-dependent radio frequency (RF) and microwave magnetic fields for tomography on 16-dimensional $F = 3 \otimes F = 4$ electronic ground state subspace. Smith et al. [Smith *et al.* (2006)] have carried out the experiments for the former case, and Chaudhuri et al. [Chaudhury *et al.* (2009)] have successfully implemented the kicked top dynamics for the same. However, the latter case is more challenging, and Riofrío et al. have done simulations for the entire 16-dimensional Hilbert space [Riofrío *et al.* (2011)]. Lysne et al. have accomplished the full unitary control over the 16-dimensional subspace of Cs and realized three well-studied models, namely nearest-neighbor transverse Ising, Lipkin-Meshkov-Glick (LMG), and the kicked top models with state-of-the-art experimental setup [Lysne *et al.* (2020)]. Thus, our theoretical work can be executed in these kinds of experiments.

The measurement record is obtained through polarization spectroscopy of the laser probe that is coupled to the ensemble of Cs atoms while it is being controlled. During the experiment, accurate knowledge of the various parameters that characterize the experimental setting is desired. Most importantly, the distribution of probe intensity across the ensemble and the measurement basis used by the polarimeter. Generally, the fidelity of the tomography is limited by noise. Thus, the bandwidth of the Gaussian white noise is carefully chosen to contain relevant information about the state. The signal-to-noise ratio is maximized using appropriate filters to get the optimum information of the state.

In the next section, we document the information-theoretic quantifiers in detail. These information-theoretic quantifiers help us quantify the information gain in quantum tomography and the amount of operator spreading as discussed in this thesis.

## 2.2 INFORMATION-THEORETIC QUANTIFIERS

### 2.2.1 Shannon entropy

In information theory, the uncertainty of a random variable is quantified by entropy known as Shannon entropy in this context. This also quantifies the information one has about the random variable, drawn from a given probability distribution, when it is observed. The Shannon entropy $\mathcal{S}$ is defined as

$$\mathcal{S} = -\sum_i p(i) \log p(i), \tag{2.10}$$

where some random variable sets the probability vector $\{p(1), p(2), ..., p(i)...\}$. While solving some tasks for information processing, Shannon entropy, which measures the amount of information, relates to the physical resources required, for example, the amount of memory needed to store that information. Shannon's noiseless channel coding theorem gives an operational meaning to the Shannon entropy by establishing statistical limits to data compression where the source of data is an *i.i.d* random variable [Shannon (1948)].

For our work in this thesis, we compute the Shannon entropy from the covariance matrix we get in the tomography process. The covariance matrix of the joint probability distribution Eq. (4.3) determines the information gain in the continuous measurement tomography protocol. We have the inverse of the covariance matrix as $\mathbf{C^{-1}} = \tilde{O}^T \tilde{O}$. Thus, in the superoperator picture, we can write

$$\mathbf{C^{-1}} = \sum_{n=1}^{N} |O_n)(O_n|, \tag{2.11}$$

where $|O_n)$ are produced by repeated application of a Floquet map or a single parameter unitary. Each eigenvector of $\mathbf{C^{-1}}$ represents an orthogonal direction in the operator space we have measured until the final time $t = N$. The eigenvalues of $\mathbf{C^{-1}}$ give us the signal-to-noise ratio in that orthogonal direction. This covariance matrix helps us to define and use certain information-theoretic quantifiers for information gain in continuous measurement tomography.

The information obtained about the Bloch vector **r** from the measurement record **M** is the mutual information [Thomas M. Cover (2006)]

$$\mathcal{I}[\mathbf{r}; \mathbf{M}] = \mathcal{S}(\mathbf{M}) - \mathcal{S}(\mathbf{M}|\mathbf{r}), \quad (2.12)$$

where $\mathcal{S}$ is the Shannon entropy for the given probability distribution. The entropy of the measurement record, $\mathcal{S}(\mathbf{M})$, is entirely due to the shot noise of the probe. Thus, it is a constant irrespective of the state, assuming we have perfect knowledge of the dynamics. One can neglect irrelevant constants to get the mutual information between the Bloch vector and a given measurement record to be specified by the entropy of the conditional probability distribution, Eq. (4.3)

$$\mathcal{I}[\mathbf{r}; \mathbf{M}] = -\mathcal{S}(\mathbf{M}|\mathbf{r}) = -\frac{1}{2}\ln(\det(\mathbf{C})) = \ln(\frac{1}{V}). \quad (2.13)$$

Here, V is the volume of the error ellipsoid whose semimajor axes are defined by the covariance matrix. The dynamics that maximizes $1/V = \sqrt{\det(\mathbf{C}^{-1})}$ also maximizes the information gain. An important constraint is that after time $t = n$ [Madhok *et al.* (2014)],

$$\mathrm{Tr}\,(\mathbf{C}^{-1}) = \sum_{i,\alpha} \tilde{O}_{i\alpha}^2 = n\|O\|^2, \quad (2.14)$$

where $\|O\|^2 = \sum_\alpha \mathrm{Tr}\,(OE_\alpha)^2$ is the Euclidean square norm, with initial observable $O$. The Eq. (2.14) is independent of the choice of dynamics, and the quantity $\mathrm{Tr}\,(\mathbf{C}^{-1})$ increases linearly with time. The inequality of arithmetic and geometric means gives us

$$\det(\mathbf{C}^{-1}) \leq \left(\frac{1}{d^2 - 1}\mathrm{Tr}\,(\mathbf{C}^{-1})\right)^{d^2-1} = \left(\frac{n}{d^2 - 1}\|O\|^2\right)^{d^2-1}, \quad (2.15)$$

where the rank of the regularized covariance matrix is $d^2 - 1$. The maximum possible value of mutual information is attained when all eigenvalues are equal and the above inequality is saturated, making the error ellipsoid a hypersphere. At a given time step, the largest mutual information is achieved when the dynamics mixes the eigenvalues most evenly. We quantify this by Shannon entropy $\mathcal{S}_c$ of the normalized eigenvalue spectrum of $\mathbf{C}^{-1}$ as given in Eq. (2.16). One can extract the maximum information about a random state by measuring all components of the Bloch vector with maximum precision. Given finite time, we can obtain the best estimate by dividing equally between operators in all directions of the operator space, which implies the operator dynamics needs to be unbiased. Thus, the information gain in tomography for random states is maximum when all the eigenvalues are uniformly distributed and are equal in magnitude [Madhok *et al.* (2014)].

One can normalize the eigenvalues to get a probability distribution from the eigenvalue spectrum. Shannon entropy quantifies the bias of this distribution as

$$\mathcal{S}_c = -\sum_i \lambda_i \ln \lambda_i, \tag{2.16}$$

where $\{\lambda_i\}$ is the normalized eigenvalue spectrum of $\mathbf{C^{-1}}$ [Madhok *et al.* (2014); Sreeram and Madhok (2021)]. The Shannon entropy increases with time and saturates at a higher value.

### 2.2.2 Fisher information

Fisher information quantifies the amount of information about an unknown parameter $\theta$ present in an observable random variable $X$. Here, the probability distribution of $X$ depends on that unknown parameter $\theta$. For a given data set, Fisher information tells us how well one can estimate $\theta$.

We begin with a few definitions from [Thomas M. Cover (2006)]. Let $\{f(x;\theta)\}, \theta \in \Theta$, denote an indexed family of probability densities, $f(x;\theta) \geq 0$, $\int f(x;\theta)dx = 1$ for all $\theta$. Here $\theta$ is the parameter set.

***Definition:*** An *estimator* for $\theta$ for sample size $n$ is a function $T$, whose input is the sequence of $n$ data values and the output is an estimate, $\theta_{est}$, of the parameter $\theta$. More formally, $T : F^n \rightarrow \Theta$. The estimator gives us the approximate value of the parameter based upon the data $X$. It is desirable to have an idea of how good the estimator is in estimating the parameter. Here, as we shall see, the Fisher information plays a crucial role here.

The Fisher information is defined as

$$\mathcal{J}(\theta) = E_\theta [\frac{\partial}{\partial \theta} \ln f(x;\theta)]^2, \quad (2.17)$$

where the expectation is taken with respect to the density $f(.;\theta)$. The Fisher information gives a lower bound on the error in estimating $\theta$ from the data. Here, we state this result, which shows the significance of the Fisher information, without proof [Thomas M. Cover (2006)].

**Theorem :** The mean squared error of any unbiased estimator $T(X)$ of the parameter $\theta$ is lower bounded by the reciprocal of the Fisher information, i.e.,

$$var(T) \geq \frac{1}{\mathcal{J}(\theta)}. \quad (2.18)$$

This is known as the Cramer-Rao bound [Cramér (1999)], which states that the variance of any unbiased estimator is at least as high as the inverse of the Fisher information. An unbiased estimator that achieves this lower bound is said to be *efficient*. Hence, the Fisher information gives us the idea of the goodness of the estimator.

In this thesis, we try to estimate the Bloch vector components as the parameters and the measurement record is the data set. Thus, we quantify the Fisher information from the covariance matrix as we describe below.

During the reconstruction process, the average Hilbert-Schmidt distance between the true and estimated state in quantum tomography is equal to the total uncertainty in the Bloch vector components [Řeháček and Hradil (2002)]

$$\mathcal{D}_{HS} = \langle \text{Tr} \left[ (\rho_0 - \bar{\rho})^2 \right] \rangle = \sum_\alpha \langle (\Delta r_\alpha)^2 \rangle, \quad (2.19)$$

where $\Delta r_\alpha = r_\alpha - \bar{r}_\alpha$. The Cramer-Rao inequality, $\langle (\Delta r_\alpha)^2 \rangle \geq \left[ \mathbf{F}^{-1} \right]_{\alpha\alpha}$, relates these uncertainties to the the Fisher information matrix, $\mathbf{F}$, associated with the conditional probability distribution, Eq. (4.3), and thus $\mathcal{D}_{HS} \geq \text{Tr}\left[\mathbf{F}^{-1}\right]$. Our probability distribution is a multivariate Gaussian regardless of the state, which helps the bound to saturate in the limit of negligible backaction. In that case, the Fisher information matrix equals the inverse of the covariance matrix, $\mathbf{F} = \mathbf{C}^{-1}$, in units of $N_s^2/\sigma^2$, where $\mathbf{C}^{-1} = \tilde{\mathbf{O}}^T \tilde{\mathbf{O}}$, and $\tilde{O}_{n\alpha} = \text{Tr}\left[O_n E_\alpha\right]$ [Madhok *et al.* (2014)]. Thus, a metric for the total information gained in tomography is the inverse of this uncertainty,

$$\mathcal{J} = \frac{1}{\text{Tr}\left[\mathbf{C}\right]} \qquad (2.20)$$

which measures the total Fisher information. The inverse covariance matrix is never full rank in this protocol. We regularize $\mathbf{C}^{-1}$ by adding to it a small fraction of the identity matrix (see, e.g., [Boyd *et al.* (2004)]). For pure states, the average Hilbert-Schmidt distance $\mathcal{D}_{HS} = 1/\mathcal{J} = 1 - \langle \text{Tr}\,\bar{\rho}^2 \rangle - 2\langle \mathcal{F} \rangle$ [Řeháček and Hradil (2002)]. Fisher information is related to the average reconstruction fidelity; hence, it can be used as a quantifier for the efficiency of the tomography protocol.

## 2.3 MODELS FOR QUANTUM DYNAMICS

In this thesis, we use various time-independent as well as time-dependent Hamiltonians of single-body and many-body systems (see Fig. 2.2) to study signatures of chaos in quantum phenomena.

### 2.3.1 Quantum kicked top

Kicked top is used as a paradigm to explore quantum signatures of chaos in various quantum phenomena both theoretically [Haake *et al.* (1987); Haake (1991); Wang *et al.* (2004); Bandyopadhyay and Lakshminarayan (2004); Ghose *et al.* (2008); Lombardi and Matzkin (2011); Madhok *et al.* (2014, 2016); Ruebeck *et al.* (2017); Bhosale and Santhanam (2017); Kumari and Ghose (2018); Bhosale and Santhanam (2018); Dogra *et al.* (2019); Kumari and Ghose (2019); Sieberer *et al.* (2019); Sreeram *et al.* (2021)] as well as in state-of-the-art experimental settings [Chaudhury *et al.* (2009); Neill *et al.*

(2016); Krithika *et al.* (2019)]. Quantum kicked top is a time-dependent periodic system governed by the Hamiltonian [Haake *et al.* (1987); Haake (1991); Chaudhury *et al.* (2009)]

$$H = \hbar \alpha J_x + \hbar \frac{\lambda}{2j\tau} J_z^2 \sum_n \delta(t - n\tau). \qquad (2.21)$$

Here $J_x$, $J_y$, and $J_z$ are the components of the angular momentum operator $\mathbf{J}$. The first term in the Hamiltonian $H$ describes a linear precession of $\mathbf{J}$ around the $x$-direction by an angle $\alpha$. Each kick is a nonlinear rotation about $z$-direction in a periodic time interval of $\tau$, as given in the second term of the Hamiltonian. The strength of this nonlinear rotation is $\lambda$ and is also the chaoticity parameter. The delta kick allows us to express the Floquet map as a sequence of operations given by

$$U_\tau = \exp\left(-i\frac{\lambda}{2j\tau} J_z^2\right) \exp(-i\alpha J_x). \qquad (2.22)$$

Thus, the time evolution unitary for time $t = n\tau$, $n = 0, 1, 2, ...$ is $U(n\tau) = U_\tau^n$. The Heisenberg evolution of an operator generates a sequence of operators $O_n = U^{\dagger n} O U^n$ for $\tau = 1$.

One can use the angular momentum commutation relations

$$[J_i, J_j] = i\epsilon_{ijk} J_k \qquad (2.23)$$

to find the stroboscopic Heisenberg equation of motion $\mathbf{J}_{t+1} = U^{\dagger n} \mathbf{J}_t U^n$. Thus

$$\begin{aligned}
J_{x,t+1} &= \left\{\tilde{J}_x, \cos\frac{\lambda}{j}\left(\tilde{J}_z - \frac{1}{2}\right)\right\} - \{\tilde{J}_y, \sin ...\} + \frac{i}{2}[\cos ..., \tilde{J}_y] + \frac{i}{2}[\sin ..., \tilde{J}_x], \\
J_{y,t+1} &= \left\{\tilde{J}_x, \sin\frac{\lambda}{j}\left(\tilde{J}_z - \frac{1}{2}\right)\right\} + \{\tilde{J}_y, \cos ...\} + \frac{i}{2}[\sin ..., \tilde{J}_y] - \frac{i}{2}[\cos ..., \tilde{J}_x], \\
J_{z,t+1} &= \tilde{J}_z, \\
\tilde{J}_x &= J_{x,t}, \\
\tilde{J}_y &= J_{y,t} \cos\alpha - J_{z,t} \sin\alpha, \\
\tilde{J}_z &= J_{y,t} \sin\alpha + J_{z,t} \cos\alpha.
\end{aligned} \qquad (2.24)$$

This is clear that $\mathbf{J}_t \to \tilde{\mathbf{J}}$, which is a linear precision about the $x-$ axis by an angle $\alpha$

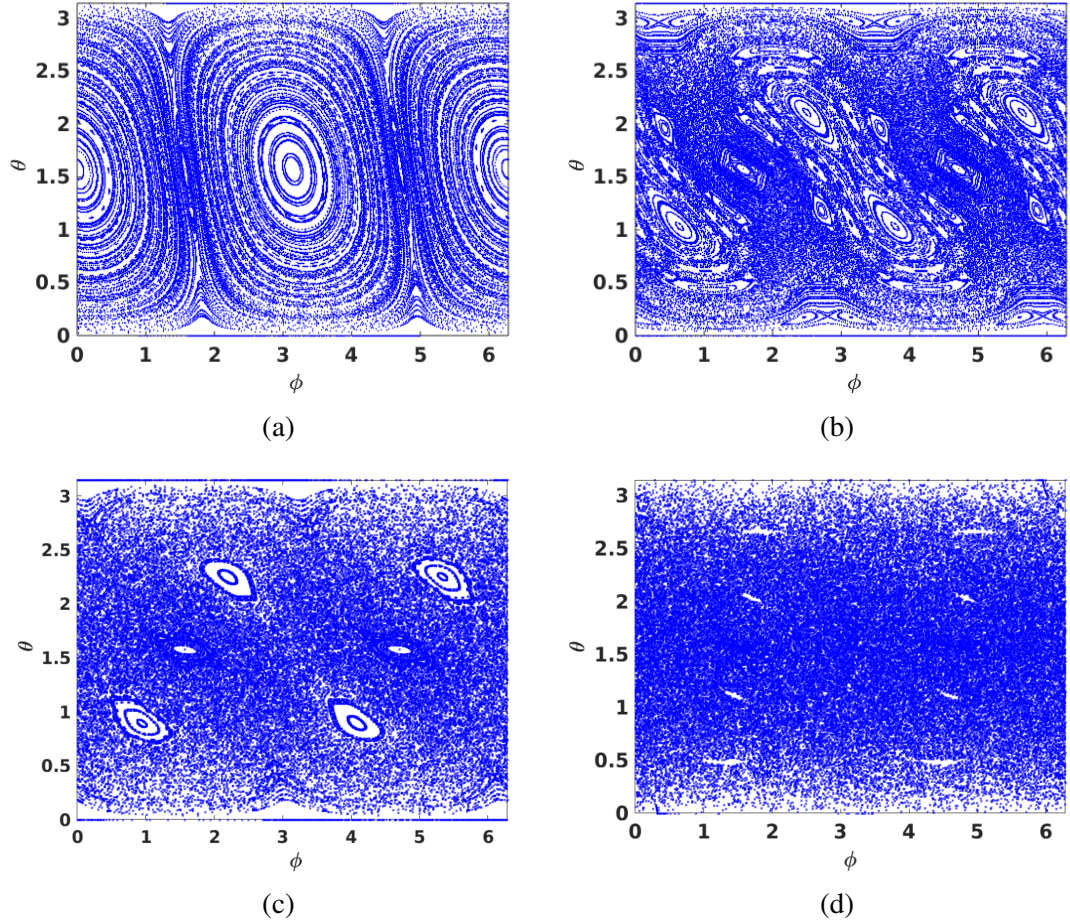

Figure 2.1: Classical phase space for the Kicked top Eq. (2.21) dynamics with an increase in chaoticity parameter for a fixed value of linear precision angle $\alpha = \pi/2$. (a) $\lambda = 0.5$, (b) $\lambda = 2.5$, (c) $\lambda = 3.0$, and (d) $\lambda = 6.5$.

followed by a nonlinear rotation around $z-$ direction as $\tilde{\boldsymbol{J}} \to \boldsymbol{J}_{t+1}$. The classical behavior of this map can be seen by taking the limit $j \to \infty$ for the above equations. The rescaled operator $\boldsymbol{X} = \boldsymbol{J}/j$ tends to a unit vector. Thus, the resulting equations describe the motion of an angular momentum vector on the surface of a sphere, as shown below

$$\begin{aligned}
X_{t+1} &= \tilde{X} \cos \lambda \tilde{Z} - \tilde{Y} \sin \lambda \tilde{Z} \\
Y_{t+1} &= \tilde{X} \sin \lambda \tilde{Z} + \tilde{Y} \cos \lambda \tilde{Z} \\
Z_{t+1} &= \tilde{Z} \\
\tilde{X} &= X_t \\
\tilde{Y} &= Y_t \cos \alpha - Z_t \sin \alpha \\
\tilde{Z} &= Y_t \sin \alpha + Z_t \cos \alpha.
\end{aligned} \quad (2.25)$$

The classical map Eq. (2.25) is effectively 2 - dimensional because of the conservation law $X^2 = 1$, and the dynamics is on the surface of the unit sphere $X^2 = 1$. The motion of the angular momentum on the surface of the sphere is determined by two angles $\theta$, and $\phi$, which defines the orientation of $X$ as

$$X = \sin \theta \, \cos \phi, \ Y = \sin \theta \, \sin \phi, \ Z = \cos \theta. \quad (2.26)$$

The absence of enough constants of motion in the system leads to chaotic dynamics. For our work in this thesis, we fix the value of the parameter $\alpha$ and choose $\lambda$ as the chaoticity parameter. The classical dynamics change from highly regular to fully chaotic as we vary $\lambda$ from 0 to 7. We can see the classical phase space for the kicked top for different values of chaoticity parameter in Fig. 2.1.

### 2.3.2 Ising spin chain with a tilted magnetic field

The Hamiltonian of the tilted field Kicked Ising model consists of the nearest neighbour Ising interaction term, and the system is periodically kicked with a spatially homogenous but arbitrarily oriented magnetic field [Prosen (2007); Pineda and Prosen (2007); Kukuljan *et al.* (2017)]. The Hamiltonian for tilted field kicked Ising model for $L$ spins is given by

$$H_{TKI} = \sum_{j=1}^{L} \left\{ J\sigma_j^z \sigma_{j+1}^z + \left(h_z \sigma_j^z + h_x \sigma_j^x\right) \sum_n \delta(t - n) \right\}, \quad (2.27)$$

where $\sigma_j^\alpha$ are the Pauli spin matrices with $\alpha = x, y, z$. This Hamiltonian has three parameters: the Ising coupling $J$, the transverse magnetic field strength $h_x$, and the longitudinal magnetic field strength $h_z$. The Floquet map for the tilted field kicked Ising model for a time period of $\tau = 1$ is

$$U_{TKI} = e^{-iJ \sum_j \sigma_j^z \sigma_{j+1}^z} e^{-i \sum_j (h_z \sigma_j^z + h_x \sigma_j^x)}. \tag{2.28}$$

We consider the free boundary condition for the model. The model is integrable when either $h_x$ or $h_z$ is zero. The Hamiltonian $H_{TKI}$ is integrable for $h_z = 0$ because of the Jordan-Wigner transformation. There is another non-trivial completely integrable regime found in the tilted field kicked Ising model when the magnitude of the magnetic field is an integer multiple of $\pi/2$, i.e. $h = \sqrt{h_x^2 + h_z^2} = n\pi/2, n \in \mathbb{Z}$ [Prosen (2007)]. Nevertheless, the model is nonintegrable in a general case of a tilted magnetic field when both the components $h_x$ and $h_z$ are non-vanishing, and $2h/\pi$ is non-integer. We fix the Ising coupling strength $J = 1$, and the transverse magnetic field strength $h_x = 1.4$, and vary the longitudinal magnetic field strength $h_z$ to tune the nonintegrability of the dynamics. As we increase the value of $h_z$, the system becomes nonintegrable. Thus, for $h_z = 0.4$, the system is weakly nonintegrable, and it will become strongly nonintegrable for $h_z = 1.4$, i.e., when the strengths of longitudinal and transverse fields become comparable.

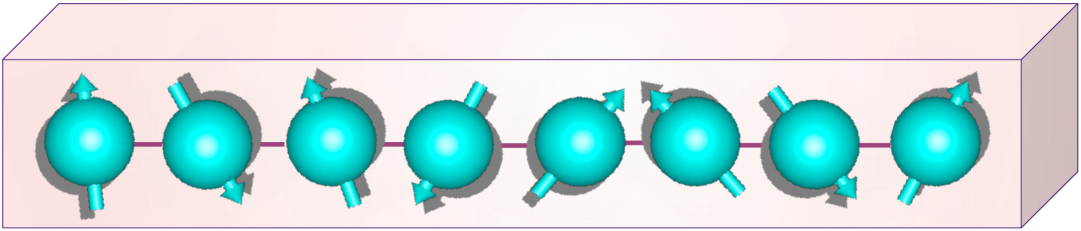

Figure 2.2: Many-body system of spins with nearest-neighbour interaction. One can drive the system from integrable to fully chaotic by tailoring the strength of integrability-breaking fields applied at suitable sites in certain directions.

Interestingly, time dependence is not necessary to make the dynamics nonintegrable. The tilted field Ising model is nonintegrable for non-zero values of $h_z$ even though there are no delta kicks [Prosen and Žnidarič (2007)]. Here, the system is strongly nonintegrable for a small value of $h_z$ even when $h_z$ and $h_x$ are not of comparable strength [Karthik *et al.* (2007)]. Thus, we choose $h_z = 0.1$ for the system to be weakly nonintegrable, and for a higher value of $h_z$, it will obey the random matrix theory predictions. The Hamiltonian for the tilted field Ising model without delta kicks can be expressed as

$$H_{TI} = \sum_{j=1}^{L} \left\{ J\sigma_j^z \sigma_{j+1}^z + h_z \sigma_j^z + h_x \sigma_j^x \right\}. \qquad (2.29)$$

$H_{TI}$ is a time-independent Hamiltonian, so the time evolution unitary operator for this model for time $t$ is given by

$$U_{TI}(t) = e^{-it \sum_j \{J\sigma_j^z \sigma_{j+1}^z + h_z \sigma_j^z + h_x \sigma_j^x\}}. \qquad (2.30)$$

In this thesis, we explore time-dependent and time-independent models to relate the information gain in tomography to the operator spreading and compare them with random matrix theory.

### 2.3.3 Heisenberg XXZ spin chain with integrability breaking field

The 1D anisotropic Heisenberg XXZ spin chain is an integrable model with nearest-neighbour interaction, which can be proved by Bethe ansatz [Shastry and Sutherland (1990); Cazalilla *et al.* (2011)]. The Hamiltonian for the Heisenberg XXZ spin chain is

$$H_{XXZ} = \sum_{j=1}^{L} \frac{J_{xy}}{4} \left\{ \sigma_j^x \sigma_{j+1}^x + \sigma_j^y \sigma_{j+1}^y \right\} + \frac{J_{zz}}{4} \sigma_j^z \sigma_{j+1}^z, \qquad (2.31)$$

where $s_j^\alpha = \frac{1}{2}\sigma_j^\alpha$. There are various ways in which we can make the XXZ model nonintegrable. One can introduce a single magnetic impurity at one of the sites [Santos (2004); Santos and Mitra (2011); Barišić *et al.* (2009); Brenes *et al.* (2020); Pandey *et al.* (2020); Rabinovici *et al.* (2022a)], a global staggered transverse field [Brenes *et al.* (2018)] or next-to-nearest-neighbour interaction [Santos *et al.* (2012); Gubin and F Santos (2012); Rabinovici *et al.* (2022a)] to make the dynamics nonintegrable. In this thesis, we consider the single magnetic impurity at one of the sites and explore the operator spreading with an increase in the integrability-breaking field strength. The Hamiltonian for this nonintegrable Heisenberg model is

$$H_{HNI} = H_{XXZ} + \frac{g}{2}H_{si}, \qquad (2.32)$$

where the integrability-breaking field with strength $g$ is $H_{si} = \sigma_l^z$, for site $j = l$. For our analysis, we consider $J_{xy} = 1$, $J_{zz} = 1.1$, and vary the strength of the integrability-breaking field $g$ as the chaoticity parameter. While changing the value of $g$ from 0 to 1, the fully integrable XXZ model becomes chaotic, which is clear from the level statistics and other properties [Santos (2004); Rabinovici *et al.* (2022*a*)]. The time evolution unitary for time $t$ for this nonintegrable time-independent Hamiltonian is

$$U_{HNI} = e^{-it(H_{XXZ} + \frac{g}{2}H_{si})}. \qquad (2.33)$$

The XXZ model with periodic boundary condition respects many symmetries, including translation symmetry in the space because of the conservation of linear momentum [Joel *et al.* (2013)], and we can find many degenerate states [Santos (2004)]. Thus, we choose the free boundary condition for the XXZ model. For a spin chain with a very large number of spins, the boundary conditions have no effects, but for numerical calculations, we have to take a finite number of spins. However, even in the deep quantum regime, we can still witness integrability to chaos transition. The Hamiltonian has a reflection symmetry about the center of the chain if a single impurity is placed at the center of the chain. The Hamiltonian $H_{HNI}$ commutes with the total spin along $z$ direction $S_z = \frac{1}{2}\sum_{j=1}^{L}\sigma_j^z$ which makes the system invariant under rotation around the $z-$ axis [Joel *et al.* (2013)].

## 2.4 KRYLOV SUBSPACE AND KRYLOV COMPLEXITY

This section aims to briefly review Krylov subspace and Krylov complexity based on the known results in the literature. Krylov subspace is generated by repeated application of a map $\mathcal{M}_K$ on an initial operator $O$ as $\mathcal{A} = \text{span}\{O, \mathcal{M}_K(O), \mathcal{M}_K^2(O), \mathcal{M}_K^3(O), ...\}$.

Here, we are interested in studying the Krylov subspace of linear operators for the Hilbert space $\mathcal{H}$. Here, we have outlined the two methods for generating the Krylov subspace: the Lanczos algorithm and the Arnoldi iteration method. In the Arnoldi iteration method [Arnoldi (1951)], a unitary operator $U$ generates the Krylov subspace.

### 2.4.1 Lanczos algorithm

In the Lanczos scheme, the generator of the Krylov basis is a Hermitian operator (a Hamiltonian $H$) [Viswanath and Müller (2008)]. The time evolution of an operator $\mathcal{O}$ is given by

$$\mathcal{O}(t) = e^{iHt}\mathcal{O}e^{-iHt} = e^{i\mathcal{L}t}\mathcal{O}, \qquad (2.34)$$

where the Liouvillian operator is defined as $\mathcal{L} = [H, .]$. In the superoperator notation the operator $\mathcal{O}$ is written as $|\mathcal{O})$, which allows us to write the Eq. (2.34) as

$$|\mathcal{O}(t))) = \sum_{k=0}^{\infty} \frac{(it)^k}{k!} \mathcal{L}^k |\mathcal{O}). \qquad (2.35)$$

Now, one can orthonormalize the operators $\{\mathcal{L}^k |\mathcal{O})\}_{k=0}^{\infty}$ using Gram-Schmidt orthogonalization procedure to get the Krylov basis $\mathcal{A}_L = \{|Q_k)\}_{k=0}^{K-1}$, where $K$ is the dimension of Krylov subspace and $K \leq d^2 - d + 1$ [Rabinovici *et al.* (2021, 2022*b*)]. This is an iterative process to get the orthonormal basis known as Lanczos algorithm for a well-defined Hilbert-Schmidt inner product $(Q_k|Q_l) = \text{Tr}\,(Q_k^\dagger Q_l) = \delta_{kl}$. In this process the first two basis operators are $|Q_0) = |\mathcal{O})/(\mathcal{O}|\mathcal{O})^{1/2}$ and $|Q_1) = b_1^{-1}\mathcal{L}|Q_0)$, where $b_1 = \sqrt{|\mathcal{L}|Q_0)|^2}$. Following the Lanczos algorithm, we can get the other basis operators $|Q_k)$ in a recursive method as

$$|Q_k) = b_k^{-1}\left(\mathcal{L}|Q_{k-1}) - b_{k-1}|Q_{k-2})\right), \qquad (2.36)$$

where $b_k = \sqrt{|\mathcal{L}|Q_{k-1}) - b_{k-1}|Q_{k-2})|^2}$ are called the Lanczos coefficients.

The Liouvillian operator is a tridiagonal matrix in the Krylov basis, as it is apparent from

Eq. (2.36). Thus, we can express the time-evolved operator in the Krylov basis as

$$|O(t)) = \sum_{k}^{K-1} i^k \varphi_k(t) |Q_k),  \qquad (2.37)$$

where $\varphi_k = i^{-k} (Q_k|O(t))$ are the time-dependent real probability amplitudes that describe the distribution of the time-evolved operator over the Krylov basis. Recently, specific features of this probability distribution have been explored to quantify nonintegrability and chaos in the dynamics [Parker *et al.* (2019); Rabinovici *et al.* (2021); Noh (2021); Caputa *et al.* (2022); Rabinovici *et al.* (2022b,a); Bhattacharya *et al.* (2022, 2023); Español and Wisniacki (2023)]. Krylov complexity is a measure of the average position of the operator distribution on the ordered Krylov basis, which is defined as

$$C_K(t) = \sum_{k=0}^{K-1} k|\varphi_k(t)|^2. \qquad (2.38)$$

In the thermodynamic limit, the Krylov complexity $C_K$ grows exponentially with time initially, and the Lanczos coefficients grow linearly as $b_k \propto k$ when the dynamics is chaotic[Parker *et al.* (2019); Caputa *et al.* (2022)]. For finite-dimensional systems, the long-time saturation value of the Krylov complexity is higher when the dynamics is chaotic, which is identified as a signature of chaos [Rabinovici *et al.* (2022b)]. However, these signatures depend highly on the choice of initial observable [Español and Wisniacki (2023)] as shown in Fig. 2.3. Another measure known as Krylov entropy quantifies how evenly the operator is distributed over the Krylov subspace

$$S_K(t) = -\sum_{k=0}^{K-1} |\varphi_k(t)|^2 \ln |\varphi_k(t)|^2. \tag{2.39}$$

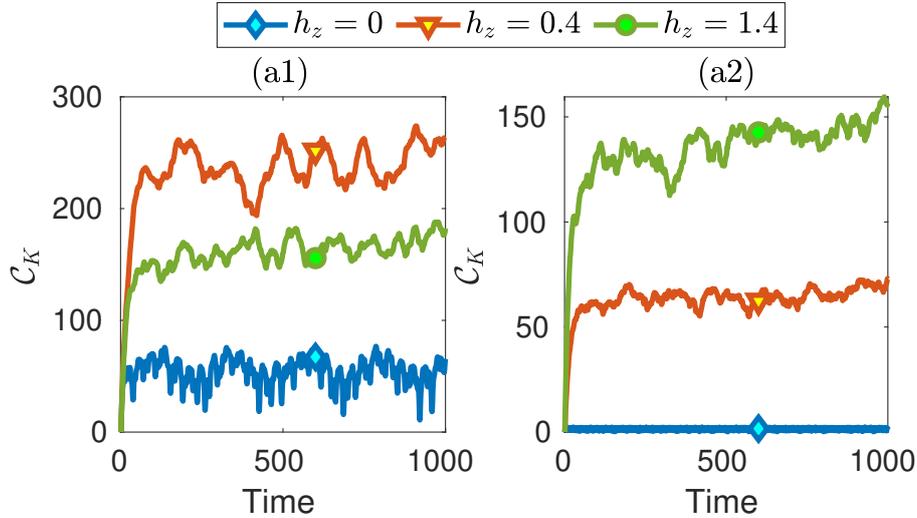

Figure 2.3: Krylov complexity $C_K$ as a function of time for the dynamics of the 1D Ising model with a tilted magnetic field for an increase in the field strength $h_z$. We have considered $L = 5$ spins with $J = 1$ and $h_x = 1.4$ for the numerical simulations. Two different initial observables (a1) $O = S_z$ and (a2) $O = S_x$. For generating this plot, we have not desymmetrized the Hamiltonian.

Lanczos algorithm is suitable for analytical calculations of the orthogonal operators $|Q_k)$ and the Lanczos coefficients $b_k$. Unfortunately, it is numerically not feasible to generate the Krylov basis and the coefficients because of the unavoidable errors accumulated from floating-point rounding in the Hilbert-Schmidt inner products. Thus, alternative methods are required to address this issue. Full orthogonalization method [Parlett (1998)] is such a method that performs a brute-force re-orthogonalization of the newly constructed Krylov element concerning the previous ones at every iteration of the Lanczos algorithm. In the next section, we present the full orthogonalization method to compute the Lanczos coefficients numerically.

### 2.4.2 Full orthogonalization algorithm for Lanczos scheme

Lanczos algorithm, which makes use of the two previous operators while constructing each Krylov element, encounters numerical instabilities because of the accumulation of errors from the finite precision arithmetic, and the orthogonality of the Krylov basis is lost in a few steps. Residual overlaps between the Krylov elements grow rapidly with the number of iterations $k$, giving rise to unreliable Lanczos coefficients $b_k$. The full orthogonalization method [Parlett (1998)] performs a brute-force re-orthogonalization of the newly constructed Krylov element concerning the previous ones at every iteration of the Lanczos algorithm that ensures the orthonormality of the Krylov basis up to machine precision $\epsilon$. Full orthogonalization performs Gram-Schmidt orthogonalization at every iteration in the Lanczos algorithm to ensure orthogonality up to the machine precision $\epsilon$.

For optimality, it is better to adopt Gram-Schmidt twice every time. The algorithm reads as follows:

1. $|Q_0) = |O) / (O|O)^{1/2}$.

2. For $k \geq 1$: compute $|B_k) = \mathcal{L} |Q_{k-1})$.

3. Re-orthogonalize $|B_k)$ explicitly with respect to all previous Krylov elements:

$$|B_k) \longmapsto |B_k) - \sum_{m=0}^{k-1} |Q_m) (Q_m|B_k).$$

4. Repeat step 3.

5. Set $b_k = \sqrt{(B_k|B_k)}$.

6. If $b_k = 0$ stop; otherwise set $|Q_k) = b_k^{-1} |B_k)$, and go to step 2.

We can find that the values of Lanczos coefficients $b_k$ initially grow as a function of $k$, then drop slowly and become zero at $k = K$, after which the Lanczos algorithm can not generate any more orthogonal operators. Figure 2.4 illustrates the Lanczos sequence for initial observable $\sigma_1^y$ different number of spins of the tilted field Ising model Hamiltonian Eq.(2.30).

### 2.4.3 Krylov subspace generated from repeated application of a unitary

In the Arnoldi method of construction for Krylov subspace is generated from the repeated application of a unitary $U$ on the initial operator $O$ [Arnoldi (1951); Yates and Mitra (2021)]. For a unitary map $U$, the time evolved operator after $n$ time steps is

$$O_n = U^{\dagger n} O U^n. \tag{2.40}$$

In the superoperator picture, we can write the operator $O_n$ as

$$|O_n) = \mathcal{U}_K^n |O), \tag{2.41}$$

where $\mathcal{U}_K = U^\dagger \otimes U^T$. The Krylov basis can be obtained by orthonormalizing the operators $\{|O_n)\}_{n=0}^{K-1}$ using Gram-Schmidt orthogonalization procedure. In this method

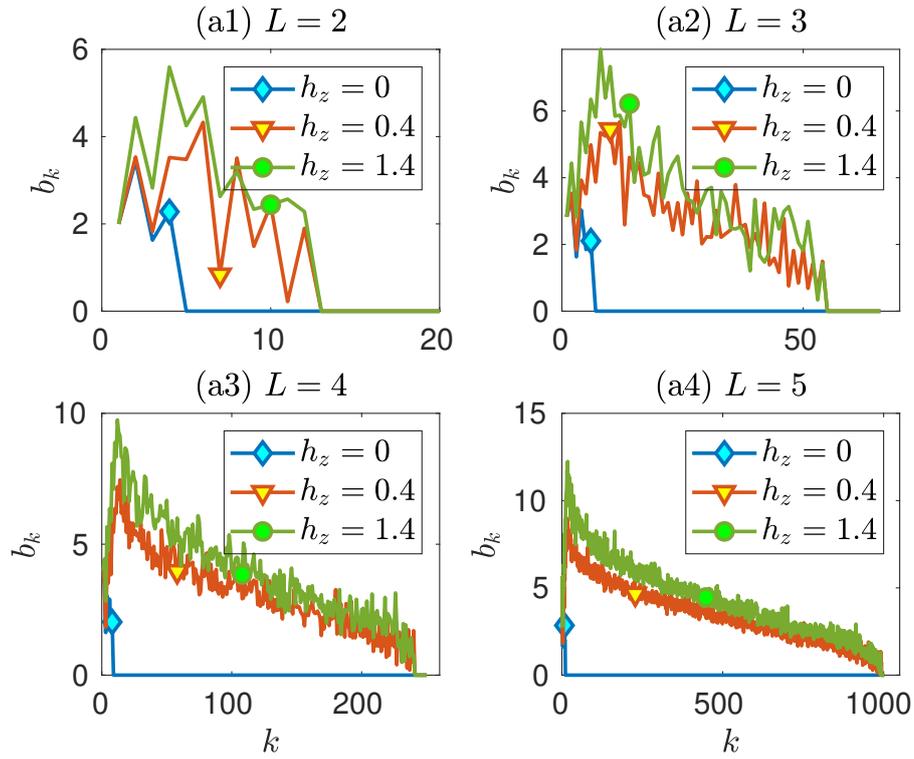

Figure 2.4: Lanczos coefficients sequence as a function of $k$. The Krylov subspace is obtained for initial observable $s_1^y$ using tilted field Ising model Hamiltonian Eq. (2.30) with $J = 1$, and $h_z = 1.4$. We vary $h_x$ as the nonintegrability parameter. The Lanczos sequence is plotted for different numbers of spins $L$. $K$ is the dimension of the Krylov subspace when the dynamics is fully nonintegrable. (a1) $L = 2$, $K = 13$. (a2) $L = 3$, $K = 57$. (a3) $L = 4$, $K = 241$. (a4) $L = 5$, $K = 993$.

also, one can have maximum $K \leq d^2 - d + 1$ linearly independent basis operators, as we see in the Lanczos algorithm. The operator $\mathcal{U}_K$ is a unitary, unlike the Liouvillian $\mathcal{L}$, which is a Hermitian tridiagonal matrix.

Here, we delineate the proof of how the repeated application of a single unitary can generate $K \leq d^2 - d + 1$ number of orthogonal operators as demonstrated by Merkel et al. in Ref. [Merkel *et al.* (2010)]. The Krylov subspace of operators is obtained as a time series of operators $O_n = U^{\dagger n} O U^n$, where $O$ is a Hermitian operator and $U$ is a fixed unitary. We can determine the dimension of the Krylov subspace of orthogonal operators $\mathcal{A}_A \equiv \text{span}\{O_n\}$. Let us consider the subspace of operators that are preserved under unitary conjugation by $U$, $\mathcal{G} \equiv \{g \in \mathfrak{su}(d) | UgU^\dagger = g\}$. Let us define $C \equiv \{g \in \mathcal{G} | \text{Tr}(gO) = 0\}$. The subspace whose elements are orthogonal to the elements of the Krylov subspace $\mathcal{A}_A$ is $\mathcal{A}_\perp$, and operators in this set are not included in the time series of operators. Thus it is clear that $C \subseteq \mathcal{A}_\perp$ since $\forall g \in C$

$$\text{Tr}(O_n g) = \text{Tr}[U^{\dagger n} O U^n g] = \text{Tr}(Og) = 0. \tag{2.42}$$

Now $\dim(\mathcal{A}_A) + \dim(C) \leq \dim[\mathfrak{su}(d)] = d^2 - 1$ as the two spaces are orthogonal. Thus, for $U$ having non-degenerate eigenvalues $\mathcal{G}$ will be isomorphic to the Cartan subalgebra of $\mathfrak{su}(d)$ which is the largest commuting subalgebra. However, if the eigenspectrum of $U$ has degeneracy, $\mathcal{G}$ will have some additional elements. The dimension of Cartan subalgebra is $d-1$ and hence $\dim(\mathcal{G}) \geq d-1$. By definition, $C$ is obtained from $\mathcal{G}$ by projecting out one direction in operator space and thus $\dim(C) = \dim(\mathcal{G}) - 1 \geq d-2$. Therefore,

$$\dim(\mathcal{A}_A) \leq \dim[\mathfrak{su}(d)] - \dim(C) \leq d^2 - d + 1. \tag{2.43}$$

We can see the bound on the dimension of the Krylov subspace in another way. There is certain condition on the eigenphases of the unitary chosen. Here, we show the proof for the same as explored in Ref. [Merkel *et al.* (2010); Sreeram and Madhok (2021)]. The unitary $U$ is diagonal in some basis as $U = \sum_{j=1}^d e^{-i\phi_j} |j\rangle \langle j|$, that allows us to represent the time evolved operator $O_n$ to be

$$O_n = \sum_{j,k=1}^{d} e^{-in(\phi_j - \phi_k)} \langle k| O |j\rangle |k\rangle \langle j|. \quad (2.44)$$

The diagonal component of $O_n$ does not depend on $n$ which helps us to write it as

$$O_n = \sum_{j=1}^{d} \langle j| O |j\rangle |j\rangle \langle j| + \sum_{j \neq k}^{d} e^{-in(\phi_j - \phi_k)} \langle k| O |j\rangle |k\rangle \langle j|. \quad (2.45)$$

The set of operators $\{O_n\}$ will span $\mathcal{A}_A$ with the condition

$$\sum_{n=0}^{d^2-d} a_n O_n = 0 \text{ iff } a_n = 0 \ \forall \ n, \quad (2.46)$$

which confirms the linear independence of the operators. The above condition Eq. (2.46) can be written explicitly as

$$O_n = \left(\sum_{n=0}^{d^2-d} a_n\right) \sum_{j=1}^{d} \langle j| O |j\rangle |j\rangle \langle j| + \sum_{j \neq k}^{d} \left[\sum_{n=0}^{d^2-d} a_n e^{-in(\phi_j - \phi_k)}\right] \langle k| O |j\rangle |k\rangle \langle j| \quad (2.47)$$

It has been shown that if the following conditions are satisfied by the eigenphases $\phi_j$, and the eigenvectors $|j\rangle$ of $U$, the set $\{O_n\}$ is linearly independent [Merkel *et al.* (2010)].

1. $\exists \ j$ such that $\langle j| O_0 |j\rangle \neq 0$,

2. $\forall \ j \neq k$, $\langle k| O_0 |j\rangle \neq 0$,

3. $\forall \ j \neq j'$, $\phi_j - \phi_k \neq \phi_{j'} - \phi_{k'} (\text{mod}(2\pi))$.

Now, the dimension of the Krylov subspace becomes $d^2 - d + 1$. The single parameter random unitary evolution can generate $d^2 - d + 1$ linearly independent operators. Our numerical simulations show that the Floquet map of a dynamical system with fully chaotic dynamics also saturates this bound. In this thesis, we explore this using strongly nonintegrable quantum many-body systems.

## 2.5 RANDOM MATRIX THEORY (RMT)

The primary goal of random matrix theory is to help us understand various properties (like eigenvalue and eigenvector statistics) of the matrices with entries drawn from different probability distributions known as random matrix ensembles [Mehta (2004)]. Eugene Wigner, in the 1950's used RMT with the motivation to understand the higher excitation spectra of heavy nuclei [Wigner (1993*a*,*b*)]. Since the Hamiltonian of heavy nuclei with many strongly interacting quantum particles is very complex, it is merely impossible to understand those nuclear spectra in detail. Later, in 1984, Bohigas, Giannoni, and Schmit (BGS) conjectured that when the classical dynamics is chaotic and ergodic, its quantum counterpart has statistical properties of a random matrix taken from the appropriate ensemble [Bohigas *et al.* (1984*b*,*a*); Bohigas and Giannoni (1984)]. The appropriate ensemble depends on the symmetries of the system, for example, whether it is time reversal invariant or not [Haake (1991)].

To explore quantum chaos, we generally consider random Hermitian matrices and random unitary matrices. Random Hermitian matrices are sampled when we are interested in understanding the energy levels of a complex chaotic system. The properties of time reversal invariant systems show excellent agreement with the properties of random Hermitian matrices sampled from the Gaussian orthogonal ensemble (GOE). Similarly, the ensemble of random Hermitian matrices used to describe the Hamiltonanis without having any time reversal symmetry is called the Gaussian unitary ensemble (GUE). However, random unitary matrices represent the Floquet operators of time-periodic systems showing chaotic dynamics. The ensemble of random unitaries is called the circular ensemble. The time reversal invariant Floquet operators are described by matrices drawn from the circular orthogonal ensemble (COE), and the Floquet operators without having time reversal symmetry are studied using random unitary matrices from the circular unitary ensemble (CUE). These random matrix ensembles were first introduced by Dyson [Dyson (1962*a*,*b*)]. Time reversal symmetric Hamiltonians or unitaries are real matrices, whereas the matrices without the time reversal symmetry are generally complex.

The construction of Gaussian orthogonal matrices is very simple. One would need to create a square matrix $A$ with all the entries chosen from the standard Gaussian distribution. To get a Hermitian matrix, we have to make it symmetric as $H_O = (A + A^T)/2$. For generating random matrices from GUE, all elements of $A$ are complex numbers with real and imaginary parts of the complex number chosen from standard Gaussian distribution. Later the Hermitian matrix is constructed as $H_H = (A + A^\dagger)/2$. CUE is the ensemble of matrices picked from U(d) according to the Haar measure. The eigenvalues of CUE matrices lie on the unit sphere in the complex plane.

### 2.5.1 An example of Gaussian Ensemble of Hermitian Matrices

Here, we illustrate the above description of the Gaussian ensemble by providing an example. We construct real symmetric $2 \times 2$ matrices with the orthogonal group as their group of canonical transformations as discussed in [Haake (1991)]. Consider the matrix,

$$\begin{bmatrix} H_{11} & H_{12} \\ H_{21} & H_{22} \end{bmatrix}$$

We want to determine the probability density, $P(H)$, for the three independent matrix elements $H_{11}$, $H_{22}$, and $H_{12}$ normalised as

$$\int_{-\infty}^{+\infty} P(H) dH_{11} dH_{22} dH_{12} = 1 \tag{2.48}$$

To determine $P(H)$ uniquely, we need

(1) $P(H)$ must be invariant under any canonical transformation, i.e.,

$$P(H) = P(H'), \ H' = O^T H O, \ O^T = O^{-1}. \tag{2.49}$$

(2) The three independent elements must be uncorrelated. This implies that $P(H)$ be of

the form

$$P(H) = P_{11}(H_{11})P_{22}(H_{22})P_{12}(H_{12}). \tag{2.50}$$

Considering an infinitesimal orthogonal change of basis,

$$O = \begin{bmatrix} 1 & \theta \\ -\theta & 1 \end{bmatrix}$$

for which $H' = O^T H O$ gives

$$H'_{11} = H_{11} - 2\theta H_{12} \tag{2.51a}$$

$$H'_{22} = H_{22} + 2\theta H_{12} \tag{2.51b}$$

$$H'_{12} = H_{12} + \theta(H_{12} - H_{22}). \tag{2.51c}$$

Using the above equations, for infinitesimal $\theta$, we get $P_{11}(H'_{11}) = P_{11}(H_{11}) - 2\theta H_{12}\frac{dP(H_{11})}{dH_{11}}$, with similar expressions for $P_{22}(H'_{22})$ and $P_{12}(H'_{12})$. Since $P(H)$ is invariant under the orthogonal transformation we get,

$$P(H) = P(H) \left\{ 1 - \theta \left[ 2H_{12}\frac{d\ln P_{11}}{dH_{11}} - 2H_{12}\frac{d\ln P_{22}}{dH_{22}} - (H_{11} - H_{22})\frac{d\ln P_{12}}{dH_{12}} \right] \right\}. \tag{2.52}$$

Since the angle $\theta$ is arbitrary, its coefficient in the above equation must vanish,

$$\frac{1}{H_{12}}\frac{d\ln P_{12}}{dH_{12}} - \frac{2}{H_{11} - H_{22}}\left[ \frac{d\ln P_{11}}{dH_{11}} - \frac{d\ln P_{22}}{dH_{22}} \right] = 0. \tag{2.53}$$

This gives three differential equations for three independent functions $P_{ij}(H_{ij})$. The solutions are Gaussian and have the product

$$P(H) = C \exp\left[ -A(H_{11}^2 + H_{22}^2 + 2H_{12}^2) - B(H_{11} + H_{22}) \right]. \tag{2.54}$$

Eliminating the constant $B$ is possible by choosing an appropriate zero of energy. Then $P(H)$ becomes

$$P(H) = C \exp\left( -A\text{Tr}\{H^2\} \right). \tag{2.55}$$

Similarly, the construction of Gaussian Hermitian matrices is done where the probability

$P(H)$ is invariant under the unitary transformation. Thus, different ensembles of random matrices follow the requirements of (1) invariance of $P(H)$ under the different possible groups of canonical transformations and (2) complete statistical non-correlations between all matrix elements. We now discuss briefly the circular ensembles for unitary matrices.

### 2.5.2 Circular Ensembles

We consider an example of a $2 \times 2$ random unitary matrix from the orthogonal ensemble. One can express a general symmetric unitary matrix as [Nazarov and Blanter (2009)]

$$\begin{bmatrix} \sqrt{R}e^{i\gamma} & \sqrt{T}e^{i\eta} \\ \sqrt{T}e^{i\eta} & -\sqrt{R}e^{i(2\eta-\gamma)} \end{bmatrix},$$

where $T + R = 1$. The phases $\gamma$ and $\eta$ are assumed to be independent and uniformly distributed between 0 and $2\pi$. The eigenvalues of the above matrix lie on the unit circle in the complex plane and are given by

$$\lambda_1 = e^{i(\eta+\phi)} = e^{i\alpha_1}, \tag{2.56a}$$

$$\lambda_2 = -e^{i(\eta-\phi)} = e^{i\alpha_2}, \tag{2.56b}$$

where $\sin\phi = \sqrt{R}\sin(\gamma-\eta)$. The difference between the phases is, $\alpha = \alpha_1 - \alpha_2 = \pi - 2\phi$ which depends only on the difference $\gamma - \eta$. The probability distribution of $\gamma - \eta$ is a constant given by $P(\gamma - \eta) = \frac{1}{2\pi}$. Now, the distribution function of the phase difference becomes

$$P(\alpha) = \frac{1}{2\pi}\left|\frac{\partial(\gamma-\eta)}{\partial\alpha}\right| = \frac{\sin(\alpha/2)}{4\pi\sqrt{R-\cos^2(\alpha/2)}}. \tag{2.57}$$

It is clear that the probability distribution is suppressed as the phase differences become more negligible. This property is known as the repulsion of eigenvalues for random unitary matrices. This level repulsion occurs in higher dimensions and also for random unitary matrices invariant under the unitary transformation [Nazarov and Blanter (2009)]. The eigenvalues of chaotic Hamiltonians and the eigenphases of chaotic Floquet maps also show this level repulsion, which agrees amazingly well with the predictions of random matrices.

In this thesis, we explore the information gained in tomography for random states chosen from the Haar measure using random unitary dynamics chosen from an appropriate ensemble. The seminal work on how complex chaotic dynamics generates pseudo-random pure states [Wootters (1990)], the random nature of eigenvectors [Życzkowski *et al.* (1992)] of the Hamiltonian and a volume of work on quantum signatures of chaos now a standard text in the literature [Haake and Życzkowski (1990); Lenz and Zyczkowski (1992); Lakshminarayan *et al.* (2008); Miller and Sarkar (1999); Trail *et al.* (2008)]. The pseudo-random vectors in the Hilbert space typically have a high entropy. Here, when one talks about the entropy of pure states, it is always assumed the entropy is calculated according to a fixed fiducial basis [Wootters (1990); Bengtsson and Życzkowski (2017)]. We show that the information-theoretic quantifiers for fully chaotic dynamics show excellent agreement with the random matrix prediction as reported in [Sahu *et al.* (2023)].

## 2.6 QUANTIFIERS OF QUANTUM CHAOS

In this section, we mention some of the quantifiers of quantum chaos that have been studied in the literature.

### 2.6.1 Loschmidt echo

Classically chaotic systems show exponential sensitivity to initial conditions. Two nearby trajectories go away from each other after some time as $\|\delta(t)\| \sim \|\delta_0\|e^{\lambda t}$. However, this is not the case in quantum mechanics. Linearity and unitarity of the quantum equation motion (Schrdinger's equation) makes the quantum dynamics always stable against small perturbation of the initial state described by the wave function. Thus, the inner product is preserved as

$$\langle \psi(0)|\psi'(0)\rangle = \langle \psi(t)|\psi'(t)\rangle \tag{2.58}$$

where $|\psi(t)\rangle = U |\psi(0)\rangle$, and $|\psi'(t)\rangle = U |\psi'(0)\rangle$.

A slight variation in the Hamiltonian can yield exciting and non-trivial effects on quantum time evolution, as suggested by Asher Peres [Peres (1984)]. As time progresses, the overlap of the perturbed and unperturbed states decays, and the decay rate gives us information about the stability of the quantum dynamics. This is the notion of sensitivity to perturbation in quantum dynamical systems. The overlap is defined as Loschmidt echo

$$\mathcal{F}(t) = |\langle\psi_0|e^{iH't/\hbar}e^{-iHt/\hbar}|\psi_0\rangle|^2 \qquad (2.59)$$

where $H$ and $H'$ are unperturbed and perturbed Hamiltonian respectively, and $|\psi_0\rangle$ is the initial state. The Hamiltonian $H$ governs the forward evolution of the state. The Hamiltonian $H'$ governs the backward evolution, and the overlap of the initial and final state is determined as fidelity $\mathcal{F}(t)$ as a function of time.

Peres introduced this Loschmidt echo and illustrated the stability of quantum evolutions for the kicked top and the coupled rotator model in [Peres (1984, 1997)]. Peres inspects the time evolution of the same initial quantum state for two slightly different values of the kicking strength in the kicked top Hamiltonian [Peres (1997)]. He shows that, after certain time steps, the overlap between the evolved states drops to a lower value if the initial spin coherent state is located in a chaotic region of the classical phase space. Still, the two evolved states remain close to each other if the initial spin coherent state is centered near a stable fixed point. Later, Loschmidt echo has been studied rigorously, theoretically, and experimentally for many complex quantum systems [Gorin *et al.* (2006); Goussev *et al.* (2012)]. It is verified that for typical quantum states, the fidelity decay is much faster for regular dynamics than chaotic dynamics, which is entirely opposite to Peres's conclusion as he had considered a very special initial coherent state. Thus, for random quantum states, the quantum chaotic dynamics is less prone to error than for regular dynamics. In this thesis, we give an operational interpretation for Loschmidt echo by connecting it to continuous weak measurement tomography [Sahu *et al.* (2022*b*)].

### 2.6.2 Out-of-time-ordered correlators (OTOC)

Out-of-time-ordered correlator (OTOC) is another quantifier of chaos that has been explored as a probe for dynamical signatures of chaos [García-Mata *et al.* (2022); Maldacena *et al.* (2016); Swingle *et al.* (2016); Hashimoto *et al.* (2017); Kukuljan *et al.* (2017); Swingle (2018); Wang *et al.* (2021); Sreeram *et al.* (2021); Varikuti and Madhok (2022)]. OTOC is defined as the modulus square of the commutator of two operators. So

$$\text{OTOC} = \langle |[A(t), B(0)]|^2 \rangle, \tag{2.60}$$

where $A(t)$ is an operator evolved with time and $B(0)$ is a stationary operator. The operators $A$ and $B$ can be either Hermitian observables or unitary operators. Usually, the expectation value is taken with respect to a thermal state at an inverse temperature $\beta$. The above expression for OTOC becomes simple and depends only on the four-point correlator when the operators are unitaries as

$$\text{OTOC} = 2\text{Re}\left(1 - \langle A^\dagger(t) B^\dagger(0) A(t) B(0) \rangle\right). \tag{2.61}$$

Because of the unusual time ordering of this four-point correlator, it is termed as out-of-time-ordered correlator.

This OTOC is intimately connected to the chaos and Lyapunov exponent as one can see when the operators are position $x$ and momentum $p$ observables.

$$\text{OTOC} = \langle |[x(t), p(0)]|^2 \rangle. \tag{2.62}$$

In classical mechanics, the commutator is replaced by a Poisson bracket. For a classically chaotic system, we know that

$$\{x(t), p(0)\} = \left(\frac{\delta x(t)}{\delta x(0)}\right) \sim e^{\lambda_c t}, \tag{2.63}$$

where $\lambda_c$ is classical Lyapunov exponent. Thus, for quantum systems, the growth of OTOC as shown in Eq. (2.62) is exponential with the exponent $\lambda_q \simeq 2\lambda_c$, where $\lambda_q$ is the quantum Lyapunov exponent.

Operator spreading and information scrambling can be quantified by the growth of out-of-time-ordered correlators (OTOCs) [Maldacena *et al.* (2016); Swingle (2018); Seshadri *et al.* (2018); Prakash and Lakshminarayan (2020); Xu and Swingle (2020); Sreeram *et al.* (2021); Varikuti and Madhok (2022)]. This growth of OTOC persists till Ehrenfest time, and after that, it saturates. It is challenging to measure OTOCs in the lab even with state-of-the-art experimental techniques [Li *et al.* (2017); Green *et al.* (2022); Gärttner *et al.* (2017); Zhu *et al.* (2016); Swingle *et al.* (2016); Yao *et al.* (2016); Halpern (2017); Bohrdt *et al.* (2017); Tsuji *et al.* (2017); Nie *et al.* (2020); Dressel *et al.* (2018); Joshi *et al.* (2020); Asban *et al.* (2021)].

In recent literature, people have tried to connect different quantifiers of chaos. In [Styliaris *et al.* (2021)], Styliaris et al. showed that the operator entanglement of the evolution is equal to the bipartite OTOC for a pair of local random operators. Yan et al. demonstrated a direct link between the OTOC and Loschmidt echo in [Yan *et al.* (2020)]. This thesis illustrates a connection between Loschmidt echo and OTOC by connecting the former to the scrambling of errors in continuous weak measurement tomography [Sahu *et al.* (2022*b*)].

# CHAPTER 3

# EFFECT OF CHAOS ON INFORMATION GAIN IN QUANTUM TOMOGRAPHY

## 3.1 INTRODUCTION

Tomography of quantum states is crucial for quantum information processing tasks like quantum computation, quantum cryptography, quantum simulations, and quantum control. In this chapter, we are interested in continuous weak measurement tomography [Silberfarb *et al.* (2005); Smith *et al.* (2006); Chaudhury *et al.* (2009); Merkel *et al.* (2010); Smith *et al.* (2004)]. The central focus of our work in this chapter is the following question: How reconstructing quantum states is related to the nature of dynamics employed in the tomography process? At first, the connection between chaos and state reconstruction seems odd. Chaos is about the inability to predict the long-term behavior of a dynamical system, while tomography involves information acquisition. However, the flip side of this uncertainty and unpredictability of chaotic dynamics is information [Caves and Schack (1997)]. Classically, as one follows a chaotic trajectory, one gains information at a rate proportional to the extent of chaos in the system. This rate is more formally described as the Kolmogorov-Sinai (KS) entropy and is equal to the sum of positive Lyapunov exponents of the system [Kolmogorov (1958, 1959); Sinai (1959); Pesin (1977)]. One might ask what this information is about? The answer is *initial conditions*. One obtains information on increasingly finer scales about the system's initial conditions. In quantum mechanics, this is precisely the goal of tomography. As one follows the archive of the measurement record in a tomography experiment, one gains information about the initial random quantum state. An intriguing question in the quantum case is whether or not the rate of information gain is related to the degree of chaos in the dynamics. There seems to be a fascinating and provocative connection between tomography and chaos, as demonstrated in Ref. [Madhok *et al.* (2014)]. While

[Madhok *et al.* (2014)] considered quantum tomography for random states, we find that state reconstruction for localized wave packets remarkably shows the opposite behavior! We show that the rate of state reconstruction is a function of dynamics and the initial state, as well as the relationship between time evolved operators and the initial state.

The remainder of this chapter is organized as follows. In Sec. 3.2, we provide background information on the concepts and tools we use for the results presented in this chapter. Section 3.3 gives a brief overview of continuous weak measurement tomography. In Sec. 3.4, the heart of the chapter, we explore the relationship between tomography and dynamics for localized spin coherent states and contrast them with that for random states. Finally, in Sec. 3.5 we conclude by discussing our findings followed by outlook.

## 3.2 BACKGROUND

### 3.2.1 Quantum kicked top

Quantum kicked top is a time-dependent periodic system governed by the Hamiltonian [Haake *et al.* (1987); Haake (1991); Chaudhury *et al.* (2009)]

$$H = \hbar\alpha J_x + \hbar\frac{\lambda}{2j\tau}J_z^2 \sum_n \delta(t - n\tau). \tag{3.1}$$

Here $J_x$, $J_y$, and $J_z$ are the components of the angular momentum operator **J**. The delta kick allows us to express the Floquet map as a sequence of operations given by

$$U_\tau = \exp\left(-i\frac{\lambda}{2j\tau}J_z^2\right)\exp(-i\alpha J_x). \tag{3.2}$$

Thus, the time evolution unitary for time $t = n\tau$, $n = 0, 1, 2, ...$ is $U(n\tau) = U_\tau^n$. The Heisenberg evolution of an operator generates a sequence of operators $O_n = U^{\dagger n}OU^n$.

The classical behavior of this map can be seen by expressing the Heisenberg equations of motion for the angular momentum operators $J_x$, $J_y$, and $J_z$ and then taking the limit $j \to \infty$. The resulting equations describe the motion of an angular momentum vector on the surface of a sphere. The dynamics can be realized as a linear precession by an angle $\alpha$ about the $x$-axis, followed by a nonlinear precession about the $z$-axis. The absence of enough constants of motion in the system leads to chaotic dynamics. For this work, we fix $\alpha = \pi/2$ and choose $\lambda$ as the chaoticity parameter. The classical dynamics change from highly regular to fully chaotic as we vary $\lambda$ from 0 to 7. In Sec 2.3.1 we have explained this model in more detail.

### 3.2.2 Spin coherent states

To address the problem, we will explore the reconstruction of random states and coherent states which are on two extremes as far as localization in phase space is concerned. Being the closest analog of classical minimum uncertainty wave packets, coherent states [Bohr (1920)] are particular quantum states of a quantum harmonic oscillator that, despite the quantum mechanical uncertainty in position and momentum, follow classical-like dynamics. Similarly, spin coherent states are minimum uncertainty wave packets that satisfy the Heisenberg uncertainty principle for angular momentum operators.

Spin coherent states are highly localized and serve as the closest analog for points in a classical phase space. The spin coherent states point in a particular direction to the extent allowed by the angular momentum commutation relation. For a given point $(\theta, \phi)$ in the classical phase space, the spin coherent state is defined as [Arecchi *et al.* (1972); Radcliffe (1971)]

$$|\theta, \phi\rangle = (1 + |\mu|^2)^{-j} e^{\mu J_-} |j, j\rangle \equiv |\mu\rangle, \qquad (3.3)$$

where $\mu = e^{i\phi} \tan \frac{\theta}{2}$; $0 \leq \theta \leq \pi$, $0 \leq \phi \leq 2\pi$, and the spin lowering operator $J_- = J_x - iJ_y$. The spin coherent state $|j, j\rangle$ is one of the eigenbasis of $J^2$ and $J_z$ from the set $\{|j, m\rangle\}$, $m = -j, -j+1, ..., j$. Other directed angular momentum states can be generated by rotating the state $|j, j\rangle$ as

$$|\theta, \phi\rangle = \exp\left(i\theta \left(J_x \sin \phi - J_y \cos \phi\right)\right) |j, j\rangle. \qquad (3.4)$$

The uncertainty in $J$ for the state $|\theta, \phi\rangle$ is

$$\left(\langle J^2\rangle - \langle J\rangle^2\right)/j^2 = 1/j. \qquad (3.5)$$

Thus, the uncertainty goes to zero as the $j$ value becomes very large, and the spin coherent states are highly localized at the point $(\theta, \phi)$ in the phase space.

### 3.2.3 Husimi entropy

The Husimi $\mathcal{Q}$-function of a density matrix $\rho$ is a quasiprobability distribution in phase space. Husimi $\mathcal{Q}$-function is defined as [Husimi (1940)]

$$\mathcal{Q}_\rho(\theta, \phi) = \langle\theta, \phi|\rho|\theta, \phi\rangle. \qquad (3.6)$$

A notion of entropy can be associated with any density matrix through Husimi $\mathcal{Q}$-function, called the Husimi entropy (also known as Wehrl entropy) [Wehrl (1979)],

$$S_\rho = -\frac{2j+1}{4\pi}\int_\Omega d\Omega\, \mathcal{Q}_\rho \ln[\mathcal{Q}_\rho]. \qquad (3.7)$$

where $\Omega = \{\theta, \phi\}$, $0 \leq \theta \leq \pi$, $0 \leq \phi \leq 2\pi$. To treat both observables and density operators on equal footing, we determine the Husimi entropy for an operator after doing some regularization as follows. We construct a positive operator from an observable by retaining its eigenvectors and taking the modulus of its eigenvalues. To normalize this operator, we divide by its trace. Now, we can calculate the Husimi entropy and analyze the localization of the operator in the phase space. For regularized Hermitian observables, the Husimi function is the expectation value with respect to spin coherent state $|\theta, \phi\rangle$. Thus, for a regularized operator $\mathcal{O}$ the Husimi function is

$$\mathcal{Q}_\mathcal{O}(\theta, \phi) = \langle\theta, \phi|\mathcal{O}|\theta, \phi\rangle, \qquad (3.8)$$

and the Husimi entropy is given by

$$S_\mathcal{O} = -\frac{2j+1}{4\pi}\int_\Omega d\Omega\, \mathcal{Q}_\mathcal{O} \ln[\mathcal{Q}_\mathcal{O}]. \qquad (3.9)$$

## 3.3 CONTINUOUS MEASUREMENT TOMOGRAPHY

In this protocol, a collective operator $O$ of an ensemble of quantum systems is measured through a weakly coupled probe to generate the measurement record. The reader may refer to Sec. 2.1 for more details. In the Heisenberg picture, and the operator that is measured at time $t$ is

$$O(t) = U^\dagger(t) O U(t). \tag{3.10}$$

Thus, in the negligible backaction limit, the measurement records can be approximated to be

$$M(t) = X(t)/N_s = \text{Tr}\,[O(t)\rho_0] + W(t), \tag{3.11}$$

where $W(t)$ is a Gaussian white noise with spread $\sigma/N_s$.

The density matrix can be represented as

$$\rho_0 = \mathbb{1}/d + \Sigma_{\alpha=1}^{d^2-1} r_\alpha E_\alpha, \tag{3.12}$$

where $\Sigma_{\alpha=1}^{d^2-1} r_\alpha^2 = 1 - 1/d$ as $\mathbf{r}$ is the generalized Bloch vector and $\{E_\alpha\}$ are the traceless Hermitian operators. We consider the measurement records at discrete times

$$M_n = M(t_n) = N_s \sum_\alpha r_\alpha \text{Tr}\,[O_n E_\alpha] + W_n, \tag{3.13}$$

where $O_n = U^{\dagger n} O U^n$. Thus, in the negligible backaction limit, the probability distribution associated with measurement history $\mathbf{M}$ for a given state vector $\mathbf{r}$ is [Silberfarb *et al.* (2005); Smith *et al.* (2006)]

$$\begin{aligned}
p(\mathbf{M}|\mathbf{r}) &\propto \exp\left\{-\frac{N_s^2}{2\sigma^2} \sum_i [M_i - \sum_\alpha \tilde{O}_{i\alpha} r_\alpha]^2\right\} \\
&\propto \exp\left\{-\frac{N_s^2}{2\sigma^2} \sum_{\alpha,\beta} (\mathbf{r} - \mathbf{r_{ML}})_\alpha\, C^{-1}_{\alpha\beta}\, (\mathbf{r} - \mathbf{r_{ML}})_\beta\right\},
\end{aligned} \tag{3.14}$$

where $\mathbf{C} = \left(\tilde{O}^T \tilde{O}\right)^{-1}$ is the covariance matrix with $\tilde{O}_{n\alpha} = \text{Tr}\,[O_n E_\alpha]$. The least-square fit of the Gaussian distribution in the parameter space is the maximum-likelihood (ML) estimation of the Bloch vector, $\mathbf{r}_{ML} = \mathbf{C}\tilde{O}^\mathbf{T}\mathbf{M}$. If the covariance matrix is not full rank, the inverse of the covariance matrix is replaced by Moore-Penrose pseudo inverse [Ben-Israel and Greville (2003)], inverting over the subspace where the covariance matrix has support. The estimated Bloch vector $\mathbf{r}_{ML}$ may not represent a physical density matrix with non-negative eigenvalues because of the noise present (having a finite signal-to-noise ratio). Therefore, we impose the constraint of positive semidefiniteness [Baldwin *et al.* (2016)] on the reconstructed density matrix and obtain the physical state closest to the maximum-likelihood estimate.

To do this, we employ a convex optimization [Vandenberghe and Boyd (1996)] procedure where the final estimate of the Bloch vector $\bar{\mathbf{r}}$ is obtained by minimizing the argument

$$\|\mathbf{r}_{ML} - \bar{\mathbf{r}}\|^2 = (\mathbf{r}_{ML} - \bar{\mathbf{r}})^T \mathbf{C}^{-1} (\mathbf{r}_{ML} - \bar{\mathbf{r}}) \quad (3.15)$$

subject to the constraint

$$\mathbb{1}/d + \Sigma_{\alpha=1}^{d^2-1} \bar{r}_\alpha E_\alpha \geq 0.$$

In this chapter, we use the periodic application of a Floquet map $U_\tau$ for simplicity, and the unitary at $n^{\text{th}}$ time step is $U(n\tau) = U_\tau^n$. The measurement record generated by such periodic evolution is not informationally complete, and it leaves out a subspace of dimension $\geq d - 2$, out of $d^2 - 1$ dimensional operator space. We employ a well-studied kicked top model [Haake *et al.* (1987); Haake (1991); Chaudhury *et al.* (2009)] described by the Floquet map $U_\tau = e^{-i\lambda J_z^2/2J} e^{-i\alpha J_x}$ as the unitary.

## 3.4 QUANTUM CHAOS AND TOMOGRAPHY: SPIN COHERENT STATES VS. RANDOM STATES

In this section, we come to the central question we ask. What is the effect of the degree of chaos on the tomography of states? For our analysis, we study the dynamics of quantum kicked top for angular momentum $j = 20$ for spin coherent states and $j = 10$ for random states. For the tomography, we consider the initial observable as $O = J_y$, and the subsequent observables whose expectation values we measure are acquired by evolving under the quantum kicked top Floquet unitary operator. The fidelity of the reconstructed state $\bar{\rho}$ is determined relative to the actual state $|\psi_0\rangle$, $\mathcal{F} = \langle\psi_0|\bar{\rho}|\psi_0\rangle$ as a function of time.

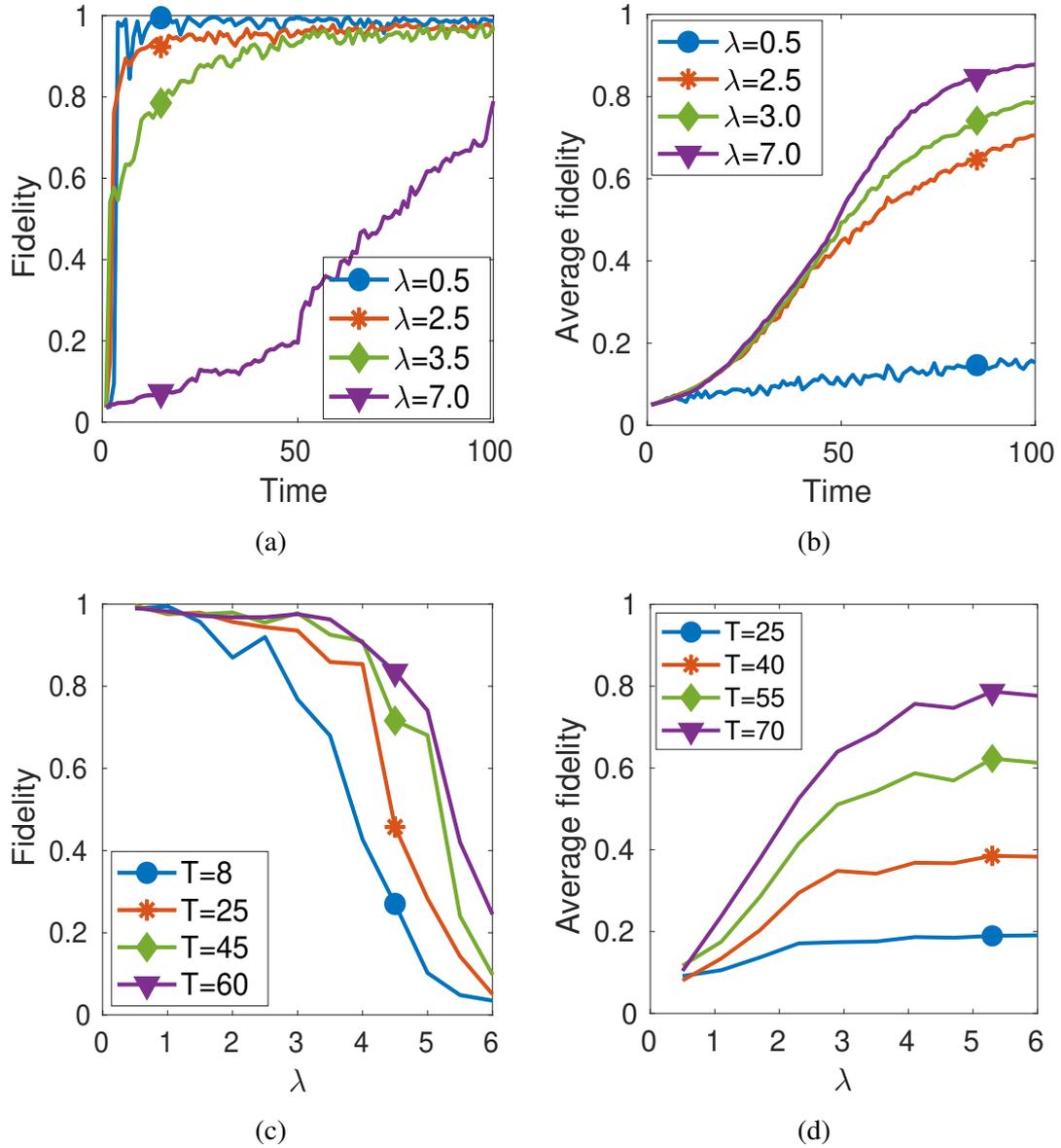

Figure 3.1: Column-wise comparison showing the contrasting behaviour of reconstruction fidelity for spin coherent state and random states. (a and b) Fidelity as a function of time for different chaoticity parameters. (c and d) Fidelity as a function of chaoticity at different time steps of the tomography process. The fidelity of spin coherent state (a random choice from the phase space) decreases with the increase in the values of the chaoticity parameter $\lambda$, whereas the average fidelity of the random states increases with an increase in $\lambda$. Here, we consider a spin coherent state with $\theta = 2.04$ and $\phi = 2.42$ for $j = 20$ and 50 random states for $j = 10$ chosen from the Haar measure for reconstruction.

We discover interesting, contrasting, and counterintuitive effects of chaos in the tomography fidelities depending on whether the states involved are random states spread across the phase space or localized coherent states. Figure 3.1a and Fig. 3.1b show the reconstruction fidelities as a function of time for coherent states and random states, respectively, with different degrees of chaos. A common observation in both cases is that as time increases, the fidelity rises. This is because of more measurements and information gain with time. However, with chaoticity, the coherent and random states show opposite behavior. It is evident from Fig. 3.1a that for spin coherent states, the fidelity decreases with the increase in the level of chaos, which is in contrast to the nature of random states [Madhok *et al.* (2014)], as shown in Fig. 3.1b. This is made more explicit in Fig. 3.1c and 3.1d, where we plot fidelity against chaoticity at different instances of time. We set out to investigate this distinctive behavior with respect to chaoticity.

First, we ask the following question: what constitutes information gain in tomography? More precisely, the rate of state reconstruction? It is important to make this crucial distinction between information gain and its acquisition rate for the following reason. In the limit of vanishing shot noise in Eq. (3.11) and assuming informationally completeness of the measurement record, we can reconstruct the quantum state with unit fidelity irrespective of the dynamics involved. This is because we are able to determine the components of the $d^2 - 1$ dimensional generalized Bloch vector completely from such a noiseless and informationally complete measurement record. This can be seen, for example, in the case of quantum tomography for a single qubit on the usual Bloch vector on the 2-sphere. Here, we need three expectation values in the direction of the Pauli matrices to determine the components of this vector and completely specify the state.

However, even in the case of vanishing shot noise (that gives us the maximal signal-to-noise ratio), the order in which we measure various operators matters as far as the rate of information gain is concerned. For example, let us consider the density matrix as a vector and express it as Eq. (3.12). We ask the following question: What is the order in which one should measure various $E_\alpha$'s to get the most rapid information gain about the unknown state? It is easy to see that the order of $\{E_\alpha\}$, which corresponds to the Bloch vector components $\{r_\alpha\}$ in the descending order of magnitude gives the maximum rate of information gain. Figure 3.3 shows the effect of the ordering of $\{E_\alpha\}$ on the fidelity and information gain in tomography.

The above discussion helps us qualitatively understand why spin coherent states do worse in reconstruction as one increases the chaos in the dynamics. Chaos scrambles and delocalizes the operators such that the subsequent operators generated in the Heisenberg picture have less support over the density matrix, which hinders rapid information gain. The fidelities obtained are a function of the dynamics and the degree to which the operators generated yield information about the Bloch vector components. However, we need to elucidate this intuition with a more concrete analysis.

The probability distribution of observing a measurement record $\mathbf{M}$ given an initial state $\rho_0$, the dynamics $\mathcal{L}$ (that involves application of unitaries), and the measurement process $\mathcal{M}$ (the choice of operators $\mathcal{O}$ to be measured) is $p(\mathbf{M}|\rho_\mathbf{0}, \mathcal{L}, \mathcal{M})$. Thus, the probability of reconstructing the state $\rho_0$ is

$$p(\rho_\mathbf{0}|\mathbf{M}, \mathcal{L}, \mathcal{M}) = A\ p(\mathbf{M}|\rho_\mathbf{0}, \mathcal{L}, \mathcal{M})\ p(\rho_0|\mathcal{L}, \mathcal{M})\ p(\mathcal{L}, \mathcal{M}). \qquad (3.16)$$

Here, $A$ is a normalization constant, $p(\rho_0|\mathcal{L}, \mathcal{M})$ is the posterior probability distribution conditioned upon the knowledge of the dynamics and the measurement operators. In the limit of zero noise, and given measurement observables $\{E_\alpha\}$, this conditional probability is constantly updated. Eventually, it becomes a product of Dirac-delta functions, each of them specifying a particular Bloch vector component, once we obtain an informationally complete measurement record. The term, $p(\mathcal{L}, \mathcal{M})$, in the above expression, is the prior information about the choice of dynamics and measurement operators and can be absorbed in the constant. Equation (3.16) is illuminating as it separates the probability of estimation into a product of two terms (up to a constant). The first term $p(\mathbf{M}|\rho_0, \mathcal{L}, \mathcal{M})$, which is identical to Eq. (3.14), contains the errors due to shot noise and quantifies the signal-to-noise ratio in various directions in the operator space *independent* of the state to be estimated. Therefore, this term estimates the information gained, given a density matrix, in different directions in the operator space. The second term quantifies how likely this particular density matrix is to be the actual unknown initial state. This gives a constant factor for random states, as there is no correlation between the measurement observables and the initial state chosen randomly.

However, for spin coherent state tomography, the term $p(\rho_0|\mathcal{L}, \mathcal{M})$ becomes crucial, as we see in the discussion below. Let us look at a measure of information gain that is oblivious to the choice of initial state and reordering of measurement operators. One can quantify how much the system dynamics is correlated with information gain in quantum tomography by calculating the Fisher information corresponding to the measurement process. For the case of random states, this measure perfectly characterizes the effect of chaos on tomography [PG and Madhok (2021); Madhok *et al.* (2014)]. Quantum tomography is equivalent to "parameter estimation", i.e., estimation of the Bloch vector components of $\rho_0$. The Fisher information quantifies how well our estimator can predict these parameters from the data, regardless of the state.

The Hilbert-Schmidt distance between the true and estimated state in quantum tomography, averaged over many runs of the estimator, $\mathcal{D}_{HS} = \langle \text{Tr}\left[(\rho_0 - \bar{\rho})^2\right] \rangle$ [Řeháček and Hradil (2002)], can be shown equal to the total uncertainty in the Bloch vector components, $\mathcal{D}_{HS} = \sum_\alpha \langle (\Delta r_\alpha)^2 \rangle$. The Cramer-Rao inequlaity, $\langle (\Delta r_\alpha)^2 \rangle \geq \left[\mathbf{F}^{-1}\right]_{\alpha\alpha}$, relates these uncertainties to the the Fisher information matrix, **F**, Eq. (3.14), and thus $\mathcal{D}_{HS} \geq \text{Tr}\,[\mathbf{F}^{-1}]$. Since we have a multivariate Gaussian distribution regardless of the state, in the negligible backaction limit, this bound is saturated. In that case, the Fisher information matrix equals the inverse of the covariance matrix, $\mathbf{F} = \mathbf{C}^{-1}$, in units of $N_s^2/\sigma^2$, where $\mathbf{C}^{-1} = \tilde{O}^T \tilde{O}$, and $\tilde{O}_{n\alpha} = \text{Tr}\,[O_n E_\alpha]$ [Madhok *et al.* (2014)]. Thus, a metric for the total information gained in tomography is the inverse of this uncertainty,

$$\mathcal{J} = \frac{1}{\text{Tr}\,[\mathbf{C}]} \tag{3.17}$$

which measures the total Fisher information.

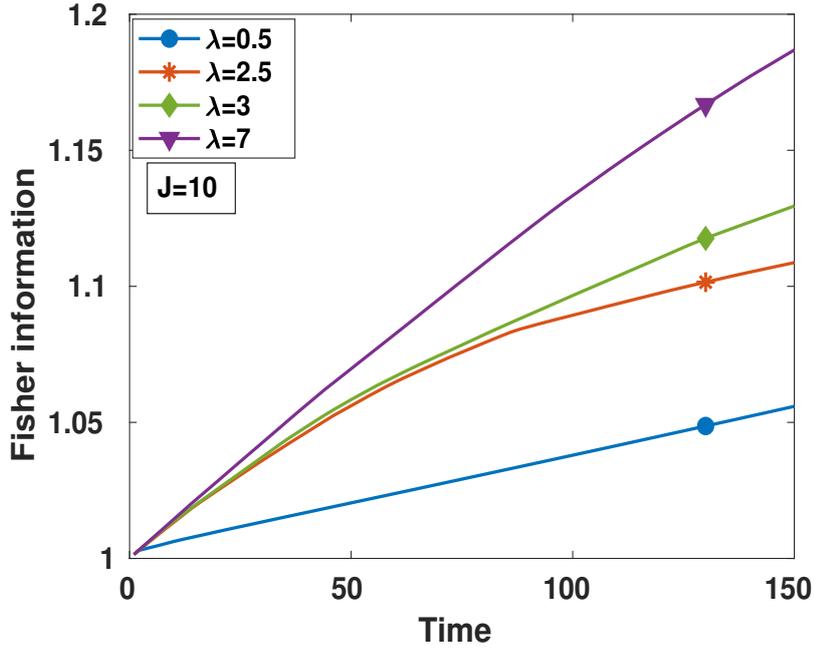

Figure 3.2: The Fisher information of the parameter estimation in tomography as a function of time for different degrees of chaos.

In Fig. 3.2 we show $\mathcal{J}$ as a function of time, generated by repeatedly applying kicked top Floquet unitary. We see the level of chaos and the information gain in tomography for random states are closely related. Since the inverse covariance matrix is never full rank in this protocol, we regularize $\mathbf{C}^{-1}$ by adding to it a small fraction of the identity matrix (see e.g., [Boyd *et al.* (2004)]). For pure states, the average Hilbert-Schmidt distance $\mathcal{D}_{HS} = 1/\mathcal{J} = 1 - \langle \text{Tr } \bar{\rho}^2 \rangle - 2\langle \mathcal{F} \rangle$ [Řeháček and Hradil (2002)]. A correlation between chaos in the dynamics and the information gain, as seen in the average fidelity (Fig. 3.1b), implies that the Fisher information shows the behavior.

Based on how much the dynamics generate Fisher information, the above analysis explains the reconstruction fidelity and its correlation with chaos for random states. However, the fact that Fisher information cannot capture all aspects of the problem can be easily seen by calculating it for the case discussed in Fig. 3.3. Since the Fisher information is independent of the order in which $E_\alpha$'s are measured, it gives no information about the reconstruction procedure as shown in Fig. 3.4. Therefore, we need to re-look at Eq. (3.14) and the prior information captured by the second term of Eq. (3.16), $p(\rho_0|\mathcal{L},\mathcal{M})$.

As we have discussed, $p(\rho_0|\mathcal{L},\mathcal{M})$ is the Bayesian estimate of the density matrix parameters at a particular time in the estimation process based on the information obtained. This is independent of the shot noise and depends on the nature of the observables measured and the dynamics employed to generate these operators (choice of unitary). Thus, combining Eq. (3.14) and Eq. (3.16), we get

$$p(\rho_0|\mathbf{M},\mathcal{L},\mathcal{M}) \propto \exp\left\{-\frac{N_s^2}{2\sigma^2} \sum_i [M_i - \sum_\alpha O_{i\alpha} r_\alpha]^2\right\} p(\rho_0|\mathcal{L},\mathcal{M})$$

$$\propto \exp\left\{-\frac{N_s^2}{2\sigma^2} \sum_{\alpha,\beta} (\mathbf{r} - \mathbf{r_{ML}})_\alpha \, C_{\alpha\beta}^{-1} \, (\mathbf{r} - \mathbf{r_{ML}})_\beta\right\} p(\rho_0|\mathcal{L},\mathcal{M}).$$

(3.18)

In the limit of zero shot noise, the errors due to the first term are zero, and we may purely focus on the conditional probability distribution, $p(\rho_0|\mathcal{L},\mathcal{M})$. In terms of the observables in continuous measurement tomography, one can express $p(\rho_0|\mathcal{L},\mathcal{M}) = p(\mathbf{r}|O_1,O_2,...,O_n)$, giving the conditional probability of the density matrix parameters $\mathbf{r}$ till the time step $n$. For example, consider the measurement operator at the first $k$ time steps are the ordered set $\{E_1, E_2,...,E_k\}$, giving precise information about Bloch vector components $\{r_1, r_2, ..., r_k\}$. The conditional probability distribution at time $k$ is,

$$p(\mathbf{r}|E_1, E_2, ..., E_k) = \delta(r_1 - \text{Tr}\,[E_1\rho_0])\,\delta(r_2 - \text{Tr}\,[E_2\rho_0])\,...\,\delta(r_k - \text{Tr}\,[E_k\rho_0])$$
$$\delta\left(\sum_{\alpha \neq 1,2,...k}^{d^2-1} r_\alpha^2 = 1 - 1/d - r_1^2 - r_2^2... - r_k^2\right).$$
(3.19)

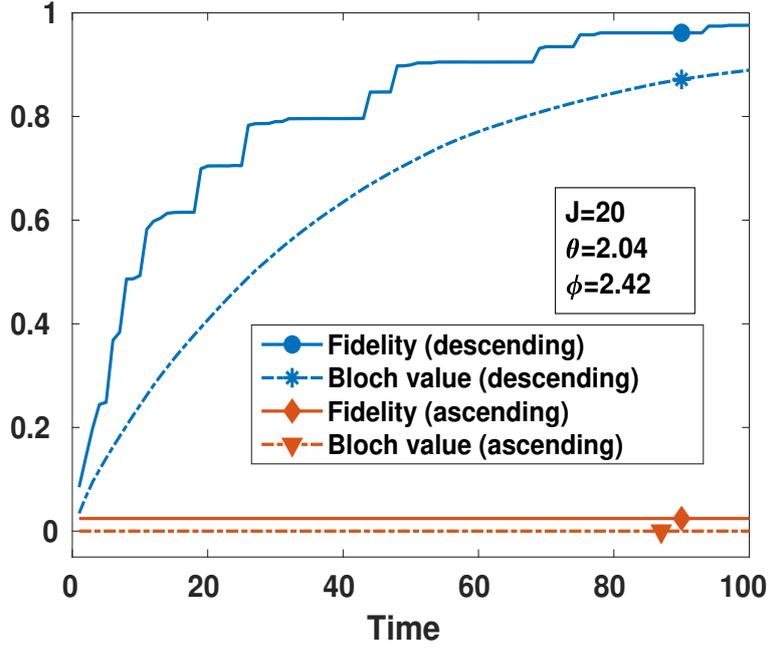

Figure 3.3: Information gain (Bloch values) with ordered Bloch vector components (i.e., that corresponds to the Bloch vector components, $r_\alpha = \text{Tr}\,[\rho_0 E_\alpha]$ in the descending and ascending order of magnitude), and fidelity in the limit of vanishing shot noise. The Bloch value at time $k$ refers to the quantity $r_i^2 + r_j^2 + ... + r_k^2$, where $\{r_i, r_j, ..., r_k\}$ is the ordered set of Bloch vector components as described above.

Each noiseless measurement above gives us complete information in one of the orthogonal directions. For example, at the first time step,

$$p(\mathbf{r}|E_1) = \delta(r_1 - \text{Tr}\,[E_1\rho_0])\,\delta\left(\sum_{\alpha \neq 1}^{d^2-1} r_\alpha^2 = 1 - 1/d - r_1^2\right). \qquad (3.20)$$

Hence, once $r_1$ is determined, the rest of the $d^2 - 2$ Bloch vector components are constrained to lie on a surface given by the equation $\sum_{\alpha \neq 1}^{d^2-1} r_\alpha^2 = 1 - 1/d - r_1^2$. The state estimation procedure under incomplete information shall pick a state consistent with $r_1$ as determined by the first measurement and the remaining Bloch vector components from a point on this surface. Therefore, qualitatively speaking, the average fidelity of the estimated state is correlated with the area of this surface. This area depends on the magnitude of $r_1$ that appears in the scaling factor mentioned above. Hence, the order of measuring operators $\{E_\alpha\}$ that corresponds to the Bloch vector components $\{r_\alpha\}$ in the descending order of magnitude gives the maximum rate of information gain as shown in Fig. 3.3. After $k$ time steps, the error is proportional to the area of the surface consistent with the equation $1 - 1/d - r_i^2 - r_j^2 - ... - r_k^2$. This area, quantifying the average error, shrinks with each measurement. The shrinkage rate of this error area for spin coherent states is more when the dynamics is regular. On the other hand, for random states, chaotic dynamics reveals more information about the initial condition as discussed above [Madhok *et al.* (2014)].

To see it in another way, consider the fidelity between the actual and reconstructed state. The fidelity $\mathcal{F} = \langle \psi_0 | \bar{\rho} | \psi_0 \rangle$, combined with Eq. (3.12) for expressing both $\rho_0$ and $\bar{\rho}$, is

$$\mathcal{F} = 1/d + \Sigma_{\alpha=1}^{d^2-1} \bar{r}_\alpha r_\alpha. \tag{3.21}$$

As one makes measurements, $E_1, E_2, ..., E_k$ and gets information about the corresponding Bloch vector components (with absolute certainty in the case of zero noise for example), one can express the fidelity as

$$\mathcal{F} = 1/d + \Sigma_{i=1}^{k} r_i^2 + \Sigma_{\alpha \neq 1,2,...k}^{d^2-1} \bar{r}_\alpha r_\alpha. \tag{3.22}$$

The term $1/d + \Sigma_{i=1}^{k} r_i^2$ puts a lower bound on the fidelity obtained after $k$ measurements and, therefore, the rate of information gain in tomography is intimately tied with the extent of alignment between the measurement operators and the density matrix.

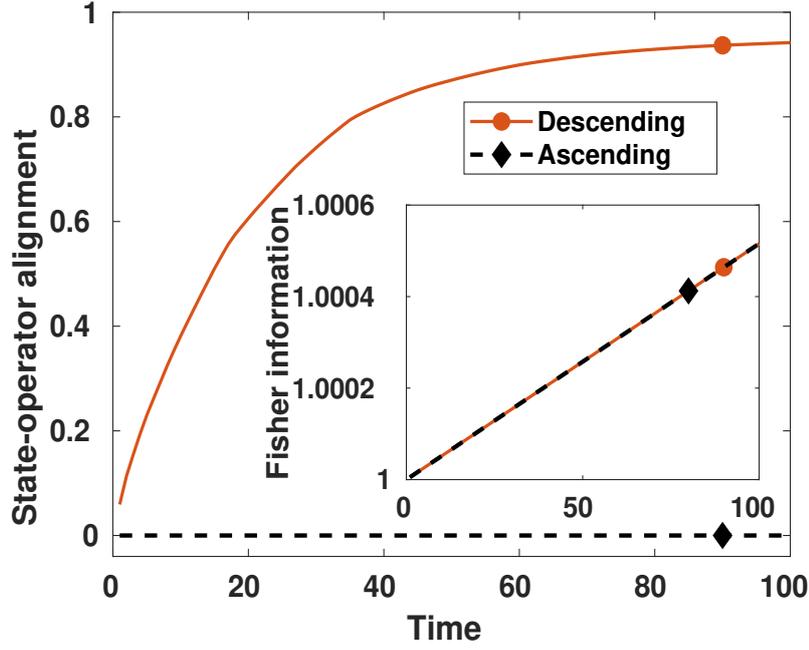

Figure 3.4: State-operator alignment and Fisher information (the inset figure) for ordered $\{E_\alpha\}$ as a function of time. The solid line indicates the behaviour for the operators in descending order, and the dotted line is for ascending order (i.e., that corresponds to the Bloch vector components, $r_\alpha = \text{Tr}\,[\rho_0 E_\alpha]$ in the descending and ascending order of magnitude).

The foregoing discussion helps us understand how ordering operators facilitates the fidelity gain. Specifically, the overlap of the operators with the density matrix can be captured with the help of an "alignment matrix"

$$\tilde{S} = \begin{pmatrix} r_1 \tilde{O}_{11} & r_2 \tilde{O}_{12} & .. & .. & r_{d^2-1} \tilde{O}_{1d^2-1} \\ r_1 \tilde{O}_{21} & r_2 \tilde{O}_{22} & .. & .. & r_{d^2-1} \tilde{O}_{2d^2-1} \\ .. & .. & .. & .. & .. \\ .. & .. & .. & .. & .. \\ r_1 \tilde{O}_{n1} & r_2 \tilde{O}_{n2} & .. & .. & r_{d^2-1} \tilde{O}_{nd^2-1} \end{pmatrix}, \quad (3.23)$$

where $\tilde{S}_{n\alpha} = r_\alpha \tilde{O}_{n\alpha} = r_\alpha \text{Tr}\,[O_n E_\alpha]$, and $O_n = U^{\dagger n} O U^n$. We quantify the extent of alignment of the time evolved operators with the state at a given time as $\text{Tr}\,[\mathcal{T}]$, where $\mathcal{T} = \tilde{S}^T \tilde{S}$. State-operator alignment as shown in Fig. 3.4, explains the correlation between the information gain and the ordering of operators $\{E_\alpha\}$, while Fisher information is oblivious to that. Figure 3.5 illustrates how the alignment of the operators with respect to the density matrix decreases with an increase in the degree of chaos, in agreement with the reconstruction rate of coherent states (Fig. 3.1a).

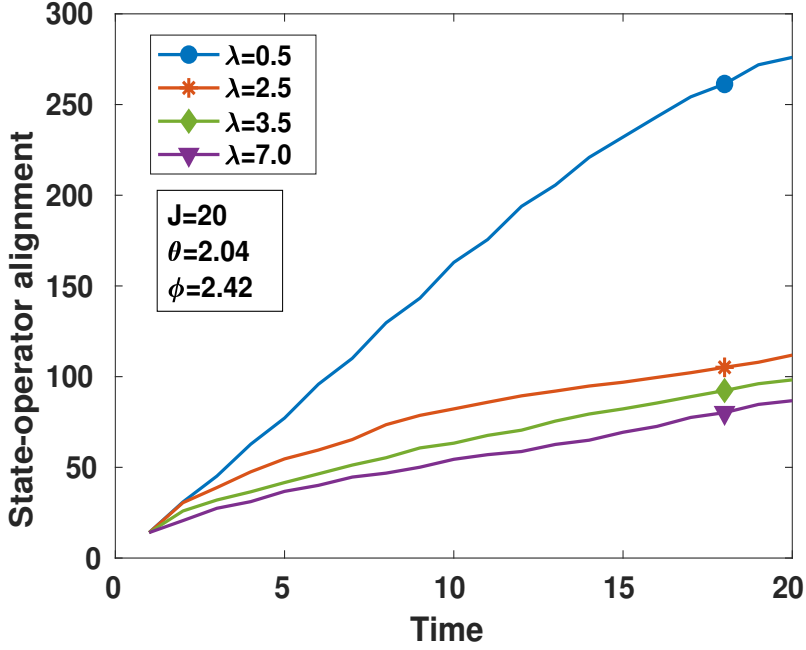

Figure 3.5: State-operator alignment as a function of time for different degrees of chaos.

To understand the connection between state-operator alignment and the nature of the dynamics, one can look at the localization of operators in the over-complete basis of spin coherent states. We notice that at a given time the operator becomes more delocalized as the chaos increases. This delocalization is captured by the Husimi entropy defined in Eq. (3.9). The operator spreads more in the phase space as the chaoticity increases, which is apparent from Fig. 3.6. The Husimi entropy increases and saturates at a higher value for a high value of chaoticity. A spin coherent state is localized in phase space and with the increase in chaos, the overlap of the state and the time evolved operator gets distributed in the phase space. As the operator dynamics become more chaotic, more spin coherent states make up the operator, and the amount of information one gains about a particular state of interest is low. Thus, the reconstruction of localized spin coherent states becomes difficult as the chaos in the dynamics shoots up. This behavior is also true for phase space averaged reconstruction fidelity of spin coherent states.

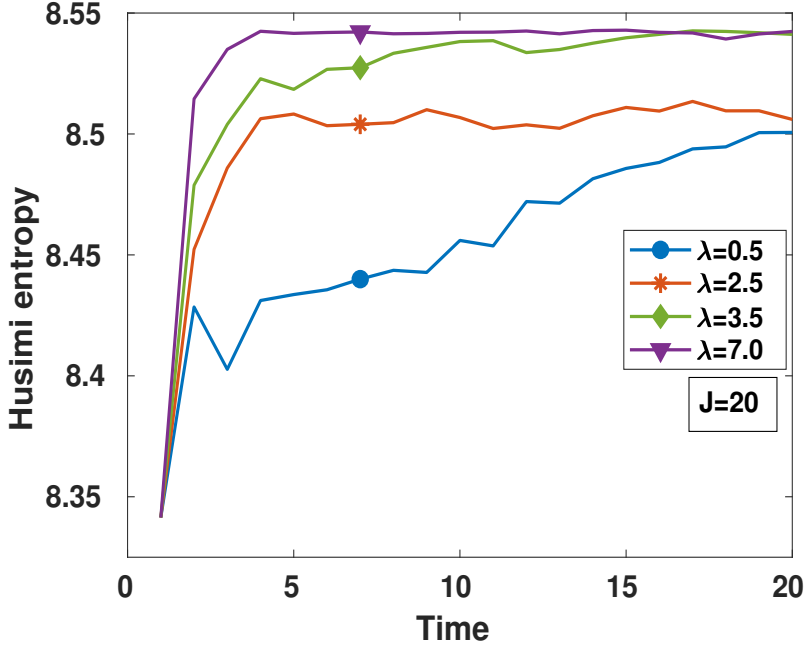

Figure 3.6: Husimi entropy of the operators evolved from the initial observable $O = J_y$ as a function of time.

In contrast, the operator delocalization in phase space is positively correlated with the fidelity gain for random states. In this case, the most optimal measurement is the one that evenly measures all possible directions in the operator space and hence explains the positive correlation of information gain with the degree of operator spread in phase space. Interpreting this way, tomography and information gain give us an operational interpretation of operator spreading and scrambling of information which is being vigorously pursued through the study of out-of-time-ordered correlators and tripartite mutual information [Maldacena *et al.* (2016); Swingle *et al.* (2016); Hashimoto *et al.* (2017); Kukuljan *et al.* (2017); Swingle (2018); Wang *et al.* (2021); Sreeram *et al.* (2021); Varikuti and Madhok (2022); Seshadri *et al.* (2018)]. This is in close resemblance with the classical Kolmogorov-Sinai (KS) entropy which relates the increasingly fine-grained knowledge about the initial conditions as one monitors a chaotic trajectory [Sinai (1959); Caves and Schack (1997)].

## 3.5 DISCUSSION

In this chapter, we have given a complete picture of the role of chaos in information gain in order to perform tomography via weak continuous measurements. Remarkably, the reconstruction rate of spin coherent states decreases with the increase in chaos, in contrast to the behavior of random quantum states. The fact is that spin coherent states are localized in the phase space as a Gaussian wave packet with a minimal spread, unlike the random states, which are spread all over the phase space. The Fisher information serves as a suitable quantifier of information gain for random states where we need to consider only the dynamics assuming a uniform prior information for all random states. However, Fisher information does not account for the decrease in the reconstruction fidelity of spin coherent states with an increase in chaos. We include the prior knowledge for these highly localized states and define a measure called state-operator alignment, which captures the decline in the fidelity rate as the dynamics become chaotic. Furthermore, we show that the ordering of operators also plays a role in the reconstruction rate. The angular momentum operators get delocalized in the phase space as we evolve them with chaotic dynamics. We see that the degree of delocalization of operators increases with chaos. Hence, the information gain in the measurement decreases, making reconstruction of spin coherent states more difficult.

Though quantum systems show no sensitivity to initial conditions, due to unitarity of evolution, they do show sensitivity to parameters in the Hamiltonian [Peres (1984)]. This leads to an interesting question for quantum tomography and, more generally, quantum simulations. Under what conditions are the system dynamics sensitive to perturbations, and how does this affect our ability to perform quantum tomography? Can quantum tomography say something about the notion of sensitivity to perturbations in system dynamics in quantum systems? In particular, one may ask, how do the effects of perturbations manifest in the reconstruction algorithm, and how are they affected by the chaoticity of the system?

The connection between information gain, quantum chaos, and the spreading of operators is an exciting avenue providing an operational interpretation to operator scrambling, which is more popularly captured by out-of-time-ordered correlators (OTOCs). The information gain in tomography quantifies the amount of new information added as one follows the trajectory of operators generated by the dynamics in the Heisenberg picture. However, the OTOC is a quantum analog of divergence of two trajectories which is captured by Lyapunov exponents in the classical picture and operator incompatibility in the quantum counterpart [Larkin and Ovchinnikov (1969); Maldacena *et al.* (2016); Swingle (2018)]. Therefore, a natural direction is to connect the information gain in tomography to the Lyapunov exponents, thereby unifying the connections between information gain, scrambling, and chaos and connecting it to an actual physical process. We will address these questions in the next two chapters.

# CHAPTER 4

# OPERATOR SPREADING AND CHAOS IN KRYLOV SUBSPACE

## 4.1 INTRODUCTION

Operator spreading characterizes a process in which the Heisenberg evolution of a local operator under the dynamics of a many-body Hamiltonian extends over the entire system [Von Keyserlingk *et al.* (2018)]. The operator spreading is a probe for scrambling of quantum information that is inaccessible to local measurements. Once the information is scrambled, the information is now delocalized over the entire operator space in complex observables. Thus, operator spreading is also connected to the understanding of the questions of chaos, nonintegrability, and thermalization in many-body quantum systems [Deutsch (1991); Srednicki (1994); Tasaki (1998); Rigol *et al.* (2008); Rigol and Santos (2010); Torres-Herrera and Santos (2013)].

For many quantum systems, operator spreading is a reliable indicator of chaos in the dynamics [Moudgalya *et al.* (2019); Omanakuttan *et al.* (2023)]. One can quantify the spreading of the operator through out-of-time-ordered correlators (OTOCs) [Maldacena *et al.* (2016); Swingle (2018); Seshadri *et al.* (2018); Prakash and Lakshminarayan (2020); Xu and Swingle (2020); Sreeram *et al.* (2021); Varikuti and Madhok (2022)], operator entanglement [Nie *et al.* (2019); Wang and Zhou (2019); Alba *et al.* (2019); Styliaris *et al.* (2021)], memory matrix formalism [McCulloch and Von Keyserlingk (2022)] or Krylov complexity [Parker *et al.* (2019); Yates and Mitra (2021); Rabinovici *et al.* (2021); Noh (2021); Dymarsky and Smolkin (2021); Caputa *et al.* (2022); Rabinovici *et al.* (2022*b*); Avdoshkin *et al.* (2022); Rabinovici *et al.* (2022*a*); Bhattacharya *et al.* (2022, 2023); Suchsland *et al.* (2023)]. The Krylov complexity is computed when the operator at a given time is expressed in an orthonormal sequence of operators generated from the

Lanczos algorithm. The Liouvillian superoperator of a time-independent Hamiltonian is repeatedly applied on the initial operator to construct the Krylov basis. In the Sec. 2.4 of Chapter 2, we have detailed the procedure for obtaining Krylov subspace in the Lanczos algorithm and quantifying Krylov complexity. Nevertheless, the saturation value of the Krylov complexity depends on the choice of initial observable. Also, the initial growth of Krylov complexity has been observed to be exponential for certain non-chaotic dynamics [Dymarsky and Smolkin (2021); Avdoshkin *et al.* (2022)]. Thus, the Krylov complexity does not serve as an unambiguous indicator of chaos.

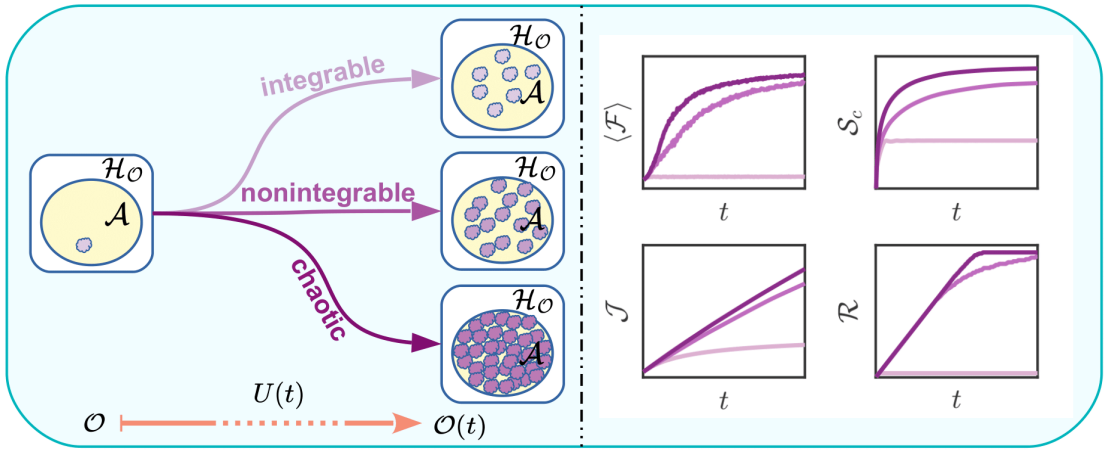

Figure 4.1: An illustration of operator spreading in the Hilbert space of operators $\mathcal{H}_O$. The initial operator $\mathcal{O}$ evolves under the system dynamics and generates a set of operators that spans the subspace $\mathcal{A}$. The dimension of the subspace is more as the dynamics becomes nonintegrable and finally chaotic. However, fully chaotic dynamics helps to span the largest subspace possible with $\dim(\mathcal{A}) = d^2 - d + 1$. In this chapter, we quantify the operator spreading through the rate of reconstruction with certain information-theoretic quantifiers like average reconstruction fidelity $\langle \mathcal{F} \rangle$, Shannon entropy $\mathcal{S}_c$, Fisher information $\mathcal{J}$, and rank of covariance matrix $\mathcal{R}$.

In this chapter, we take an alternate route and quantify operator spreading through the performance of a concrete quantum information processing task quantum tomography. How does the system dynamics drive operator complexity, which affects the information gain in quantum tomography? We answer the above question by quantifying operator spreading in integrable, nonintegrable, and chaotic many-body systems via their ability to generate an optimal measurement record for quantum tomography. Intuitively, an evolution of a fiducial operator with a single random unitary will lead to maximal operator spreading over the entire operator space [Merkel *et al.* (2010); Sreeram and Madhok (2021)]. Krylov subspace for operators is generated by repeated application of a map to an initial operator. Thus, such a random unitary evolution will also saturate the maximum possible dimension of the Krylov subspace. In our study, the operators of Krylov subspace obtained from an initial operator upon time evolution have a simple interpretation. These are elements in the operator Hilbert space that will be measured in tomography. How many of these directions and with what signal-to-noise ratio they are measured as the dynamics becomes nonintegrable and increasingly chaotic will give an operational and physically motivated way to quantify operator spreading as illustrated in Fig. 4.1.

In this chapter, we consider the dynamics of the 1D Ising model with a tilted magnetic field [Prosen (2007); Pineda and Prosen (2007); Kukuljan *et al.* (2017); Karthik *et al.* (2007); Prosen and Žnidarič (2007)], and the 1D anisotropic Heisenberg XXZ model with an integrability-breaking field [Santos (2004); Santos *et al.* (2004); Barišić *et al.* (2009); Rigol and Santos (2010); Santos and Mitra (2011); Santos *et al.* (2012); Gubin and F Santos (2012); Brenes *et al.* (2018, 2020); Pandey *et al.* (2020)] to study the growth of operator spreading and its connection to chaos in quantum tomography. They manifest a range of behavior from integrable to fully chaotic. The Hamiltonian of the Ising model with either a time-independent tilted field or a time-dependent delta-kicked tilted field shows integrable to chaos transition, and we explore both models for our tomography process.

This chapter is organized as follows. In Sec. 4.2, we briefly review the continuous measurement tomography protocol. We then describe both the models we have considered in this chapter: the Ising model with a tilted magnetic field and the Heisenberg XXZ model with an integrability-breaking field. In Sec. 4.3, the heart of the chapter, we quantify the operator spreading using various information-theoretic quantifiers by connecting it to the rate of information gain in tomography for both models. We also show that our quantifiers for operator spreading are independent of the choice of the initial operator and act as an unambiguous way of measuring chaos, unlike Krylov complexity. We conclude in Sec. 4.4 with a summary and some final remarks.

## 4.2 BACKGROUND

### 4.2.1 Continuous weak measurement tomography

Quantum state tomography uses the statistics of measurement records to estimate an unknown quantum state $\rho_0$ [Paris and Rehacek (2004); Mauro D'Ariano *et al.* (2003)]. Here, we are interested in continuous weak measurement tomography [Silberfarb *et al.* (2005); Smith *et al.* (2006); Riofrío *et al.* (2011); Smith *et al.* (2013); Merkel *et al.* (2010); Madhok *et al.* (2014, 2016); Sreeram and Madhok (2021); Sahu *et al.* (2022a,b)]. Over a period of time, one can generate an informationally complete set of measurement records by continuously probing an ensemble of identically prepared, collectively and coherently evolved systems.

A time series of operators is generated by evolving a physical observable under the dynamics of a many-body system in the Heisenberg picture. We exploit this choice of dynamics for time-evolution and explore operator spreading across the entire system or in the Hilbert space of operators. An ensemble of $N_s$ identical systems undergo separable time evolution by a unitary $U(t)$ where all the systems of the ensemble evolve independently under the chosen dynamics. In this chapter, we consider an ensemble of systems consisting of $N_s$ number of identical tilted field Ising spin chains or Heisenberg XXZ models. The collective observable $\sum_{j}^{N_s} O^{(j)}$ is a sum of single system operators, which is being evolved under the collective dynamics and probed. We use $O$, a single system operator, for all our calculations since all the systems evolve independently under the chosen dynamics. We generate the measurement record from weak continuous measurement of an observable $O$ through a probe coupled to the ensemble of identical systems. For sufficiently weak coupling, the randomness of the measurement outcomes is dominated by the quantum noise (shot noise) in the probe rather than the quantum fluctuations of measurement outcomes intrinsic to the state (known as projection noise). In such a case, quantum backaction on the state is insignificant throughout the measurement, and the state of the ensemble remains approximately separable. We get the measurement record

$$M(t) = X(t)/N_s = \text{Tr}\,[O(t)\rho_0] + W(t), \tag{4.1}$$

where $W(t)$ is a Gaussian white noise with spread $\sigma/N_s$, and $O(t) = U^\dagger(t)OU(t)$ is the time evolved operator in Heisenberg picture.

A generalized Bloch vector $\mathbf{r}$ describes any arbitrary density matrix of Hilbert-space dimension $\dim(\mathcal{H}) = d$, when expressed in an orthonormal basis of $d^2 - 1$ traceless Hermitian operators $\{E_\alpha\}$ as $\rho_0 = I/d + \Sigma_{\alpha=1}^{d^2-1} r_\alpha E_\alpha$. We consider the measurement record at discrete times as $M_n = M(t_n) = \text{Tr}\,(O_n\rho_0) + W_n$, that allows one to express the measurement history as

$$\mathbf{M} = \tilde{O}\mathbf{r} + \mathbf{W}, \tag{4.2}$$

where $\tilde{O}_{n\alpha} = \text{Tr}\,(O_n E_\alpha)$. In the negligible backaction limit, the probability distribution associated with measurement history $\mathbf{M}$ for a given state vector $\mathbf{r}$ is Gaussian with spread $\sigma$ [Silberfarb *et al.* (2005); Smith *et al.* (2006)]

$$\begin{aligned}
p(\mathbf{M}|\mathbf{r}) &\propto \exp\left\{-\frac{N_s^2}{2\sigma^2}\sum_i [M_i - \sum_\alpha \tilde{O}_{i\alpha}r_\alpha]^2\right\} \\
&\propto \exp\left\{-\frac{N_s^2}{2\sigma^2}\sum_{\alpha,\beta}(\mathbf{r}-\mathbf{r}_{\text{ML}})_\alpha\, C_{\alpha\beta}^{-1}\, (\mathbf{r}-\mathbf{r}_{\text{ML}})_\beta\right\},
\end{aligned} \tag{4.3}$$

where $\mathbf{C} = \left(\tilde{O}^T \tilde{O}\right)^{-1}$ is the covariance matrix and the inverse is Moore-Penrose pseudo inverse [Ben-Israel and Greville (2003)] in general. The peak of the distribution is the maximum likelihood estimate $\mathbf{r}_{\text{ML}}$ of the unknown Bloch vector parameters $\{r_\alpha\}$ which is equal to the least-square fit as given by

$$\mathbf{r}_{\mathrm{ML}} = \mathbf{C}\tilde{\mathcal{O}}^{\mathbf{T}}\mathbf{M}. \qquad (4.4)$$

In the presence of measurement noise, or when the measurement record is incomplete, the estimated Bloch vector $\mathbf{r}_{\mathrm{ML}}$ may represent an unphysical density matrix $\rho_{ML}$ with negative eigenvalues. Therefore, we impose the positivity constraint on the reconstructed density matrix and obtain the physical state closest to the maximum likelihood estimate, which is the most consistent with our measured data. We employ a convex optimization procedure [Vandenberghe and Boyd (1996)], a semidefinite program to obtain the final estimate of the Bloch vector $\bar{\mathbf{r}}$ by minimizing the argument

$$\|\mathbf{r}_{\mathrm{ML}} - \bar{\mathbf{r}}\|^2 = (\mathbf{r}_{\mathrm{ML}} - \bar{\mathbf{r}})^T \mathbf{C}^{-1}(\mathbf{r}_{\mathrm{ML}} - \bar{\mathbf{r}}) \qquad (4.5)$$

subject to the constraint

$$I/d + \Sigma_{\alpha=1}^{d^2-1} \bar{r}_\alpha E_\alpha \geq 0.$$

The performance of the quantum state tomography protocol is quantified by the fidelity of the reconstructed state $\bar{\rho}$ relative to the actual state $|\psi_0\rangle$, $\mathcal{F} = \langle\psi_0|\bar{\rho}|\psi_0\rangle$ as a function of time.

### 4.2.2 Models

We consider two different many-body quantum systems as spin chains, as shown in Fig. 2.2. The models are the Ising spin chain with a tiled magnetic field and the Heisenberg XXZ spin chain with an integrability-breaking field, as we describe below.

**Ising spin chain with a tilted magnetic field**

The Hamiltonian of the tilted field Kicked Ising model consists of the nearest neighbour Ising interaction term, and the system is periodically kicked with a spatially homogenous but arbitrarily oriented magnetic field [Prosen (2007); Pineda and Prosen (2007); Kukuljan *et al.* (2017)]. The Hamiltonian for tilted field kicked Ising model for $L$ spins is given by

$$H_{TKI} = \sum_{j=1}^{L} \left\{ J\sigma_j^z \sigma_{j+1}^z + \left(h_z \sigma_j^z + h_x \sigma_j^x\right) \sum_n \delta(t-n) \right\}, \tag{4.6}$$

where $\sigma_j^\alpha$ are the Pauli spin matrices with $\alpha = x, y, z$. This Hamiltonian has three parameters: the Ising coupling $J$, the transverse magnetic field strength $h_x$, and the longitudinal magnetic field strength $h_z$. The Floquet map for the tilted field kicked Ising model for a time period of $\tau = 1$ is

$$U_{TKI} = e^{-iJ \sum_j \sigma_j^z \sigma_{j+1}^z} \, e^{-i \sum_j (h_z \sigma_j^z + h_x \sigma_j^x)}. \tag{4.7}$$

We consider the free boundary condition for the model. The model is integrable when either $h_x$ or $h_z$ is zero. The Hamiltonian $H_{TKI}$ is integrable for $h_z = 0$ because of the Jordan-Wigner transformation. There is another non-trivial completely integrable regime found in the tilted field kicked Ising model when the magnitude of the magnetic field is an integer multiple of $\pi/2$, i.e. $h = \sqrt{h_x^2 + h_z^2} = n\pi/2, n \in \mathbb{Z}$ [Prosen (2007)]. Nevertheless, the model is nonintegrable in a general case of a tilted magnetic field when both the components $h_x$ and $h_z$ are non-vanishing, and $2h/\pi$ is non-integer. We fix the Ising coupling strength $J = 1$, the transverse magnetic field strength $h_x = 1.4$, and vary the longitudinal magnetic field strength $h_z$ to tune the nonintegrability of the dynamics. As we increase the value of $h_z$, the system becomes nonintegrable. Thus, for $h_z = 0.4$, the system is weakly nonintegrable, and it will become strongly nonintegrable for $h_z = 1.4$, i.e., when the strengths of longitudinal and transverse fields become comparable.

Interestingly, time dependence is not necessary for making the dynamics nonintegrable. The tilted field Ising model is nonintegrable for non-zero values of $h_z$ even though there are no delta kicks [Prosen and Žnidarič (2007)]. Here, the system is strongly nonintegrable for a small value of $h_z$ even when $h_z$ and $h_x$ are not of comparable strength [Karthik *et al.* (2007)]. Thus, we choose $h_z = 0.1$ for the system to be weakly nonintegrable, and for a higher value of $h_z$, it will obey the random matrix theory predictions. The Hamiltonian for the tilted field Ising model without delta kicks can be expressed as

$$H_{TI} = \sum_{j=1}^{L} \left\{ J\sigma_j^z \sigma_{j+1}^z + h_z \sigma_j^z + h_x \sigma_j^x \right\}. \tag{4.8}$$

$H_{TI}$ is a time-independent Hamiltonian, so the time evolution unitary operator for this model for time $t$ is given by

$$U_{TI}(t) = e^{-it \sum_j \{J\sigma_j^z \sigma_{j+1}^z + h_z \sigma_j^z + h_x \sigma_j^x\}}. \tag{4.9}$$

For our current work of this chapter, we explore time-dependent and time-independent models to relate the information gain in tomography to the operator spreading and compare them with random matrix theory.

**Heisenberg XXZ spin chain with an integrability-breaking field**

The 1D anisotropic Heisenberg XXZ spin chain is an integrable model with nearest-neighbour interaction, which can be proved by Bethe ansatz [Shastry and Sutherland (1990); Cazalilla *et al.* (2011)]. The Hamiltonian for the Heisenberg XXZ spin chain is

$$H_{XXZ} = \sum_{j=1}^{L} \frac{J_{xy}}{4} \left\{ \sigma_j^x \sigma_{j+1}^x + \sigma_j^y \sigma_{j+1}^y \right\} + \frac{J_{zz}}{4} \sigma_j^z \sigma_{j+1}^z, \tag{4.10}$$

where $s_j^\alpha = \frac{1}{2}\sigma_j^\alpha$. There are various ways in which we can make the XXZ model nonintegrable. One can introduce a single magnetic impurity at one of the sites [Santos (2004); Santos and Mitra (2011); Barišić *et al.* (2009); Brenes *et al.* (2020); Pandey *et al.* (2020); Rabinovici *et al.* (2022a)], a global staggered transverse field [Brenes *et al.* (2018)] or next-to-nearest-neighbour interaction [Santos *et al.* (2012); Gubin and F Santos (2012); Rabinovici *et al.* (2022a)] to make the dynamics nonintegrable. We consider the single magnetic impurity at one of the sites and explore the operator spreading with an increase in the integrability-breaking field strength. The Hamiltonian for this nonintegrable Heisenberg model is

$$H_{HNI} = H_{XXZ} + \frac{g}{2} H_{si}, \tag{4.11}$$

where the integrability-breaking field with strength g is $H_{si} = \sigma_l^z$, for site $j = l$. For our analysis, we consider $J_{xy} = 1$, $J_{zz} = 1.1$, and vary the strength of the integrability-breaking field g as the chaoticity parameter. While changing the value of g from 0 to 1, the fully integrable XXZ model becomes chaotic, which is clear from the level statistics and other properties [Santos (2004); Rabinovici *et al.* (2022a)]. The time evolution unitary for time *t* for this nonintegrable time-independent Hamiltonian is

$$U_{HNI} = e^{-it(H_{XXZ} + \frac{g}{2} H_{si})}. \tag{4.12}$$

The XXZ model with periodic boundary condition respects many symmetries, including translation symmetry in the space because of the conservation of linear momentum [Joel *et al.* (2013)], and we can find many degenerate states [Santos (2004)]. Thus, we choose the free boundary condition for the XXZ model. For a spin chain with a very large number of spins, the boundary conditions have no effects, but for numerical calculations, we have to take a finite number of spins. However, even in the deep quantum regime, we can still witness integrability to chaos transition. The Hamiltonian has a reflection symmetry about the center of the chain if a single impurity is placed at the center of the chain. The Hamiltonian $H_{HNI}$ commutes with the total spin along z direction $S_z = \frac{1}{2} \sum_{j=1}^{L} \sigma_j^z$ which makes the system invariant under rotation around the $z-$ axis [Joel *et al.* (2013)].

## 4.3 INFORMATION GAIN AS A PARADIGM: OPERATOR COMPLEXITY, NONINTEGRABILITY, AND CHAOS

In this section, we come to the central question we ask. What are the consequences of operator spreading in quantum information theory? We use continuous weak measurement tomography as a paradigm to study the operator spreading. The measurement record is acquired as expectation values of operators generated by the Heisenberg evolution of a chosen dynamics. We exploit the freedom of choosing the dynamics to explore and explain the effect of chaos in the operator spreading. For our analysis, we consider both time-dependent delta kicked and time-independent 1D tilted field Ising model dynamics and the dynamics of 1D anisotropic Heisenberg XXZ spin chain with an integrability-breaking field to investigate the operator complexity through various information-theoretic metrics. We also relate our information-theoretic way of quantifying operator complexity to the Krylov complexity.

Krylov subspace is generated by repeated application of a map $\mathcal{M}_K$ on an initial operator $O$ as $\mathcal{A} = \text{span}\{O, \mathcal{M}_K(O), \mathcal{M}_K^2(O), \mathcal{M}_K^3(O), ...\}$. Here, we are interested in studying the Krylov subspace of linear operators for the Hilbert space $\mathcal{H}$. We have outlined the Lanczos algorithm for constructing Krylov subspace in Sec. 2.4 of Chapter 2. The dimension of the operator Hilbert space is $\dim(\mathcal{H}_O) = d^2 - 1$. However, the maximum dimension of Krylov subspace is $d^2 - d + 1$, which leaves out a subspace of dimension at least $d - 2$ from $\mathcal{H}_O$ (see Sec. 2.4.3 of Chapter 2 for proof as given in Ref. [Merkel *et al.* (2010)] also Ref. [Rabinovici *et al.* (2021, 2022*b*)] for an alternate proof). In this chapter, we use a unitary map for certain dynamical systems to generate the Krylov subspace that helps to quantify operator spreading due to the desired dynamics. We apply a single parameter unitary map $U$ repeatedly to get the time evolved operator after $n$ time steps

$$O_n = U^{\dagger n} O U^n. \tag{4.13}$$

In the superoperator picture, one can write the operator $O_n$ as

$$|O_n) = \mathcal{U}_K^n |O), \tag{4.14}$$

where $\mathcal{U}_K = U^\dagger \otimes U^T$. Thus, we get the Krylov subspace, which is $\text{span}\{|O_n)\}$, and quantify the operator spreading through various metrics.

We choose $L$ qubit random states with Hilbert space dimension $d = 2^L$ for state reconstruction. We evolve an initial local operator $O$ and get the archive of operators that help in state reconstruction. In the beginning, the observable $O$ can be a local observable with access to the spin at site $j$; hence, it does not gain any information about other sites.

However, we need an informationally complete set of global observables to reconstruct any arbitrary random pure states. We notice that the reconstruction fidelity increases with time, which implies the growth or spread of the initial local operator across the spin chain as a complex operator. Thus, the average reconstruction fidelity serves as a quantifier for the operator complexity. We take the average over 80 Haar random pure states on SU($d$), where the Hilbert space dimension $d = 2^L$.

To further quantify the operator complexity, we study certain information-theoretic metrics. The covariance matrix of the joint probability distribution Eq. (4.3) determines the information gain in the continuous measurement tomography protocol. We have the inverse of the covariance matrix as $\mathbf{C^{-1}} = \tilde{O}^T \tilde{O}$. Thus, in the superoperator picture, we can write

$$\mathbf{C^{-1}} = \sum_{n=1}^{N} |O_n)(O_n|, \quad (4.15)$$

where $|O_n)$ are produced by applying the Floquet unitary repeatedly as given in Eq. (4.14) or by applying the time evolution unitary of a time-independent Hamiltonian for time $t = n$. Each eigenvector of $\mathbf{C^{-1}}$ represents an orthogonal direction in the operator space we have measured until the final time $t = N$. The eigenvalues of $\mathbf{C^{-1}}$ give us the signal-to-noise ratio in that orthogonal direction. Given a finite time, the operator dynamics needs to be unbiased to get equal information in all the directions of the operator space, which requires the eigenvalues of $\mathbf{C^{-1}}$ to be equal. Thus, the information gain in tomography for random states is maximum when all the eigenvalues are equal in magnitude [Madhok *et al.* (2014)]. We have given a detailed explanation on Shannon entropy in Sec. 2.2 of Chapter 2. One can normalize the eigenvalues to get a probability distribution from the eigenvalue spectrum. Shannon entropy quantifies the bias of this distribution as

$$\mathcal{S}_c = -\sum_i \lambda_i \ln \lambda_i, \quad (4.16)$$

where $\{\lambda_i\}$ is the normalized eigenvalue spectrum of $\mathbf{C^{-1}}$ [Madhok *et al.* (2014); Sreeram and Madhok (2021)]. The Shannon entropy is maximum when the eigenvalues are uniformly distributed. The Shannon entropy increases with time and saturates at a higher value. The saturation value of Shannon entropy is higher when the dynamics is fully nonintegrable, and the saturation value increases with an increase in the degree of chaos. As the dynamics becomes chaotic, the operators spread uniformly in the operator space because of the ergodicity. Thus, Shannon entropy $\mathcal{S}_c$ of the spectrum $\{\lambda_i\}$ can be used as a quantifier for operator complexity. Krylov entropy $\mathcal{S}_K$ as quantified in Eq. (2.39) (see Sec. 2.4 of Chapter 2 for the details), measures the complexity of the time-evolved operator in the Krylov basis generated from the Lanczos algorithm. In contrast, Shannon entropy $\mathcal{S}_c$ determines the spreading of the operators along the orthogonal directions in the operator space measured till time $t = N$. Shannon entropy $\mathcal{S}_c$ is a very good quantifier of operator complexity that comes naturally while doing a quantum information processing task, the quantum tomography.

During the reconstruction process, the average Hilbert-Schmidt distance between the true and estimated state in quantum tomography is equal to the total uncertainty in the Bloch vector components [Řeháček and Hradil (2002)]

$$\mathcal{D}_{HS} = \langle \text{Tr}\left[(\rho_0 - \bar{\rho})^2\right]\rangle = \sum_\alpha \langle (\Delta r_\alpha)^2 \rangle, \tag{4.17}$$

where $\Delta r_\alpha = r_\alpha - \bar{r}_\alpha$. The Cramer-Rao inequality, $\langle (\Delta r_\alpha)^2 \rangle \geq \left[\mathbf{F}^{-1}\right]_{\alpha\alpha}$, relates these uncertainties to the Fisher information matrix, $\mathbf{F}$, and thus $\mathcal{D}_{HS} \geq \text{Tr}\left[\mathbf{F}^{-1}\right]$. We have a multivariate Gaussian distribution regardless of the state, which helps the bound to saturate in the negligible backaction limit. Thus, we get $\mathbf{F} = \mathbf{C}^{-1}$, in units of $N_s^2/\sigma^2$, where $\mathbf{C}^{-1} = \tilde{O}^T \tilde{O}$, and $\tilde{O}_{n\alpha} = \text{Tr}\left[O_n E_\alpha\right]$ [Madhok *et al.* (2014)]. Thus, a metric for the total information gained in tomography is the inverse of this uncertainty,

$$\mathcal{J} = \frac{1}{\text{Tr}\left[\mathbf{C}\right]} \tag{4.18}$$

which measures the total Fisher information. The inverse covariance matrix is never full rank in this protocol. We regularize $\mathbf{C}^{-1}$ by adding to it a small fraction of the identity matrix (see, e.g., [Boyd *et al.* (2004)]). For pure states, the average Hilbert-Schmidt distance $\mathcal{D}_{HS} = 1/\mathcal{J} = 1 - \langle \text{Tr}\, \bar{\rho}^2 \rangle - 2\langle \mathcal{F} \rangle$ [Řeháček and Hradil (2002)]. Fisher information and the average reconstruction fidelity are related as shown above; hence, $\mathcal{J}$ can be used as a quantifier for the efficiency of the tomography protocol.

To further elucidate the operator spreading, we calculate the rank of the covariance matrix $\mathcal{R}$. The rank of the covariance matrix determines the dimension of the operator space spanned under the evolution of the system dynamics. Repeatedly applying a single parameter unitary can generate $K \leq d^2 - d + 1$ number of linearly independent operators [Merkel *et al.* (2010)]. Therefore, $\mathcal{R} \leq d^2 - d + 1$, and the maximum rank of the covariance matrix increases with an increase in the extent of chaos, and we can adopt it as a measure of operator spreading. Our quantifiers work very well for various models irrespective of the choice of initial observables. Here, we have shown the results for certain generic observables. However, these results are valid for other observables as well. In Fig. 4.6 of Sec. 4.3.2, we have illustrated our findings even for a local random initial observable. We have examined various local and global observables; however, they are not presented here.

### 4.3.1 Results for Ising spin chain with a tilted magnetic field

We have considered an ensemble of systems consisting of $N_s$ number of identical tilted field Ising spin chains with $L$ number of spins. Thus, $\mathcal{O}$ acts on the site $j$ of each of the systems for collective evolution and measurement. We evolve an initial local operator $\mathcal{O} = s_1^y$ to generate the measurement record. In the beginning, the observable $\mathcal{O} = s_1^y$ has access to the spin at site $j = 1$; hence, it does not gain any information about other sites. We notice that the reconstruction fidelity increases with time, as it is apparent in Fig. 4.2 (a1) for the time-dependent tilted field kicked Ising model and Fig. 4.3 (a1) for time-independent tilted field Ising model. Thus, the average reconstruction fidelity serves as a quantifier for the operator spreading. For the kicked Ising model we use the parameters $h_z = \{0.0, 0.4, 1.4\}$, and for the time-independent Ising model we consider the parameter set $h_z = \{0.0, 0.1, 1.4\}$ since the later becomes strongly nonintegrable for a small value of $h_z$ even when $h_z$ and $h_x$ are not of comparable strength [Karthik *et al.* (2007)]. It is evident from Fig. 4.2 (a2) and Fig. 4.3 (a2) that the saturation value of Shannon entropy is more when the dynamics is fully nonintegrable, and the saturation value increases with an increase in the degree of chaos. Figure 4.2 (a3) and Fig. 4.3 (a3) display the Fisher information for random states as a function of time with an increase in the level of chaos. We can observe how the rise in Fisher information is correlated with the chaos in the dynamics, making it fit as a quantifier for operator complexity. It is pretty clear from Fig. 4.2 (a4), and Fig. 4.3 (a4) that the rank is more when the dynamics is chaotic as opposed to when it is integrable. The dimension of the Krylov subspace $K$ also matches with the maximal $\mathcal{R}$ when the dynamics is fully chaotic (see Fig. 2.4 of Sec. 2.4.2, and Fig. 4.7 of Sec. 4.3.3 for this). Thus, rank $R$ represents a natural measure for operator spreading. In Fig. 4.6 of Sec. 4.3.2, we show the values of our information-theoretic quantifiers in the strongly nonintegrable regime are consistent with random matrix predictions.

Operator spreading measured from the saturation value of Krylov complexity $C_K$ as quantified in Eq. (2.38) depends on the choice of initial observable (see Ref. [Español and Wisniacki (2023)]). Previously it is demonstrated that the late-time saturation value of Krylov complexity correlates with the level of chaos for some operators like the collective spin operator $S_x = \frac{1}{2} \sum_{j=1}^{L} \sigma_j^x$, and anti-correlates with some other operators like $S_z = \frac{1}{2} \sum_{j=1}^{L} \sigma_j^z$ (see Fig. 2.3 in Sec. 2.4 of Chapter 2). Also, there are operators for which the Krylov complexity does not exhibit any systematic behavior with the level of chaos. Remarkably, the information-theoretic measures we use here give us unambiguous signatures of chaos, as is evident in Fig. 4.4, and Fig. 4.5. We have shown the behavior of the information-theoretic quantifiers for the collective spin operators $S_x$ and $S_z$. The rate of information gain in tomography for random states increases with an increase in the level of chaos in the dynamics, which is valid for any initial physical observable. We have not desymmetrized the Hamiltonian in this work or considered any symmetric subspace for our calculations. Here, the collective spin operators respect the reflection symmetry of the tilted field Ising model. The set of operators from the time evolution of $S_x$ or $S_z$ will not generate the informationally complete set of operators as the operators will be restricted to respective symmetric subspaces. Thus, one can not achieve informationally completeness of the measurement record, which leads to a lower value of reconstruction fidelity, and the saturation value of $\mathcal{R}$ is less than the maximum value. One has to study the effects of symmetry on the reconstruction closely to have a better understanding. Nevertheless, the Shannon entropy $\mathcal{S}_c$, the Fisher information $\mathcal{J}$, and the rank $\mathcal{R}$ serve as natural measures for operator complexity through a concrete physical task.

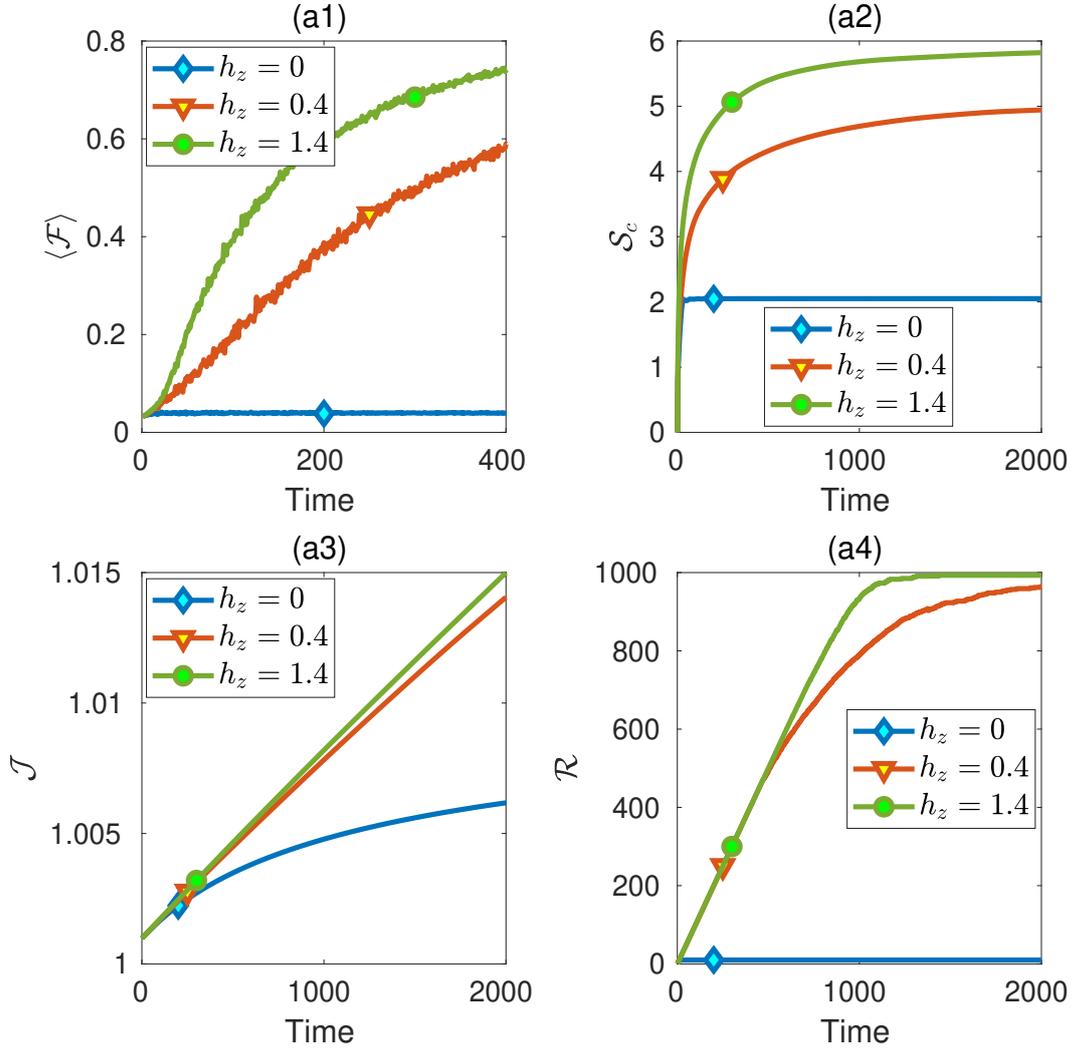

Figure 4.2: Quantifying operator spreading through various information-theoretic metrics as a function of time with an increase in the extent of chaos. The time series of operators are generated by repeatedly applying the Floquet operator of the time-dependent tilted field kicked Ising model $U_{TKI}$ as shown in Eq. (4.7) for plots (a1)-(a4). All numerical simulations are carried out for the Ising model of $L = 5$ spins with $J = 1$, $h_x = 1.4$, and for the initial observable $s_1^y$. (a1) Average reconstruction fidelity $\langle \mathcal{F} \rangle$ as a function of time. (a2) The Shannon entropy $\mathcal{S}_C$. (a3) The Fisher information $\mathcal{J}$ (a4) Rank $\mathcal{R}$ of the covariance matrix. In all cases, the values of the quantifiers are higher for a higher nonintegrability parameter $h_z$.

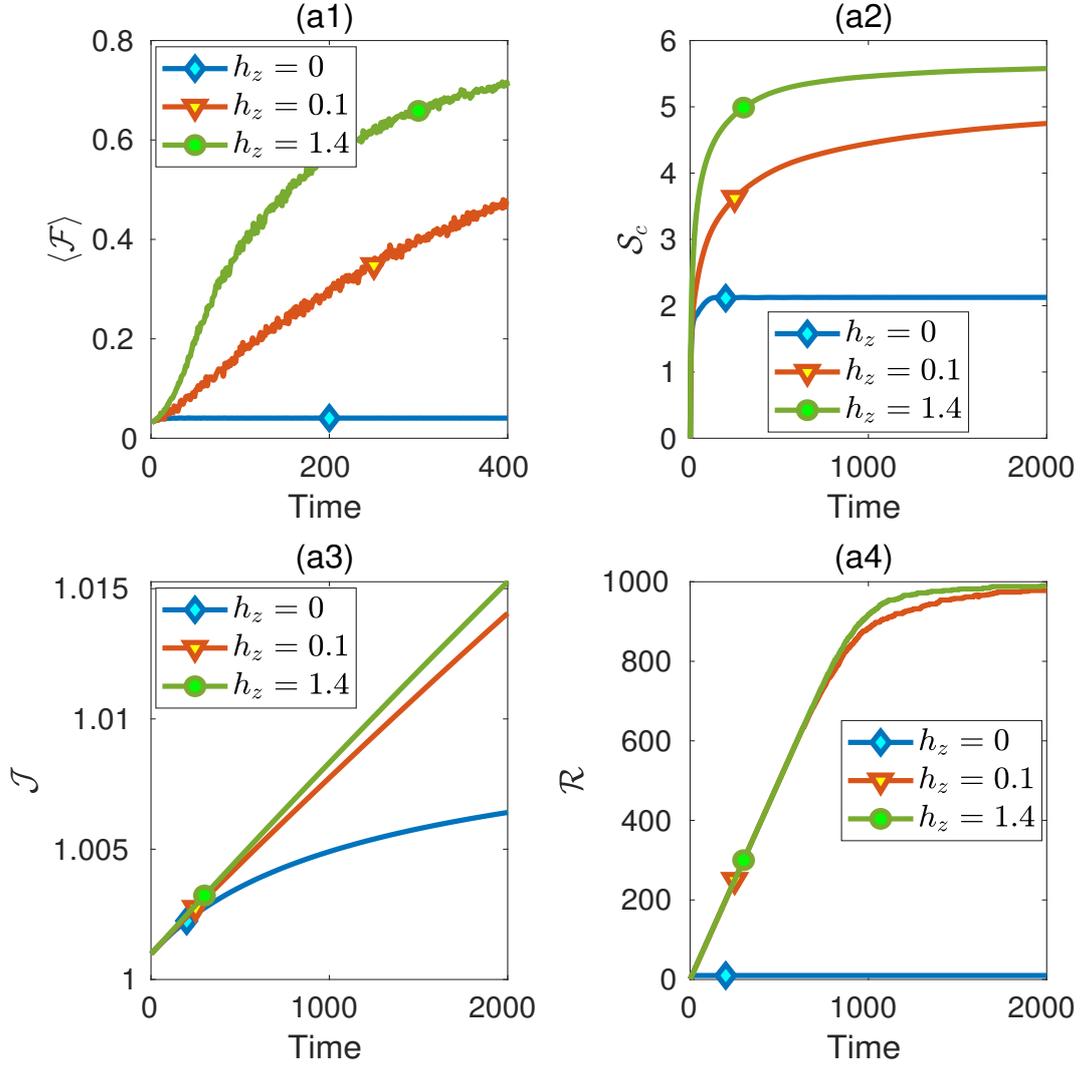

Figure 4.3: Quantifying operator spreading through various information-theoretic metrics as a function of time with an increase in the extent of chaos. The unitary for time-independent tilted field Ising model Eq. (4.9) generates the time evolved operators for plots (a1)-(a4). All numerical simulations are carried out for the Ising model of $L = 5$ spins with $J = 1$, $h_x = 1.4$, and for the initial observable $s_1^y$. (a1) Average reconstruction fidelity $\langle \mathcal{F} \rangle$ as a function of time. (a2) The Shannon entropy $\mathcal{S}_C$. (a3) The Fisher information $\mathcal{J}$. (a4) Rank $\mathcal{R}$ of the covariance matrix. In all cases, the values of the quantifiers are higher for a higher nonintegrability parameter $h_z$.

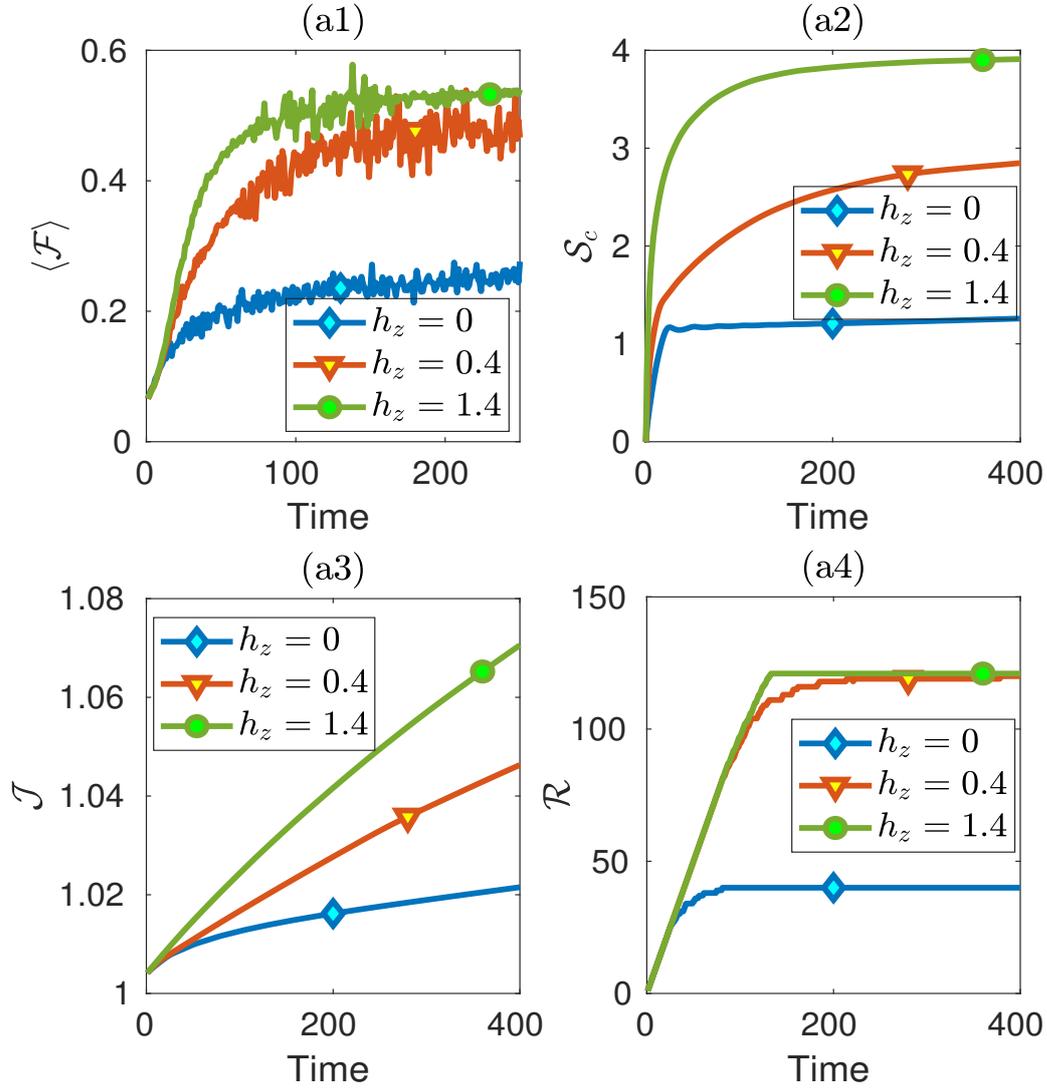

Figure 4.4: Quantifying operator spreading through various information-theoretic metrics as a function of time with an increase in the strength of the nonintegrability field. The time series of operators are generated by repeatedly applying the Floquet operator of the time-dependent tilted field kicked Ising model $U_{TKI}$ as visible in Eq. (4.7) for all plots. All numerical simulations are carried out for the Ising model of $L = 4$ spins with $J = 1$, $h_x = 1.4$. All plots are for the initial observable $S_z$. (a1) Average reconstruction fidelity $\langle \mathcal{F} \rangle$ as a function of time. (a2) The Shannon entropy $\mathcal{S}_C$. (a3) The Fisher information $\mathcal{J}$. (a4) Rank $\mathcal{R}$ of the covariance matrix. In all cases, the values of the quantifiers are higher for a higher nonintegrability parameter $h_z$.

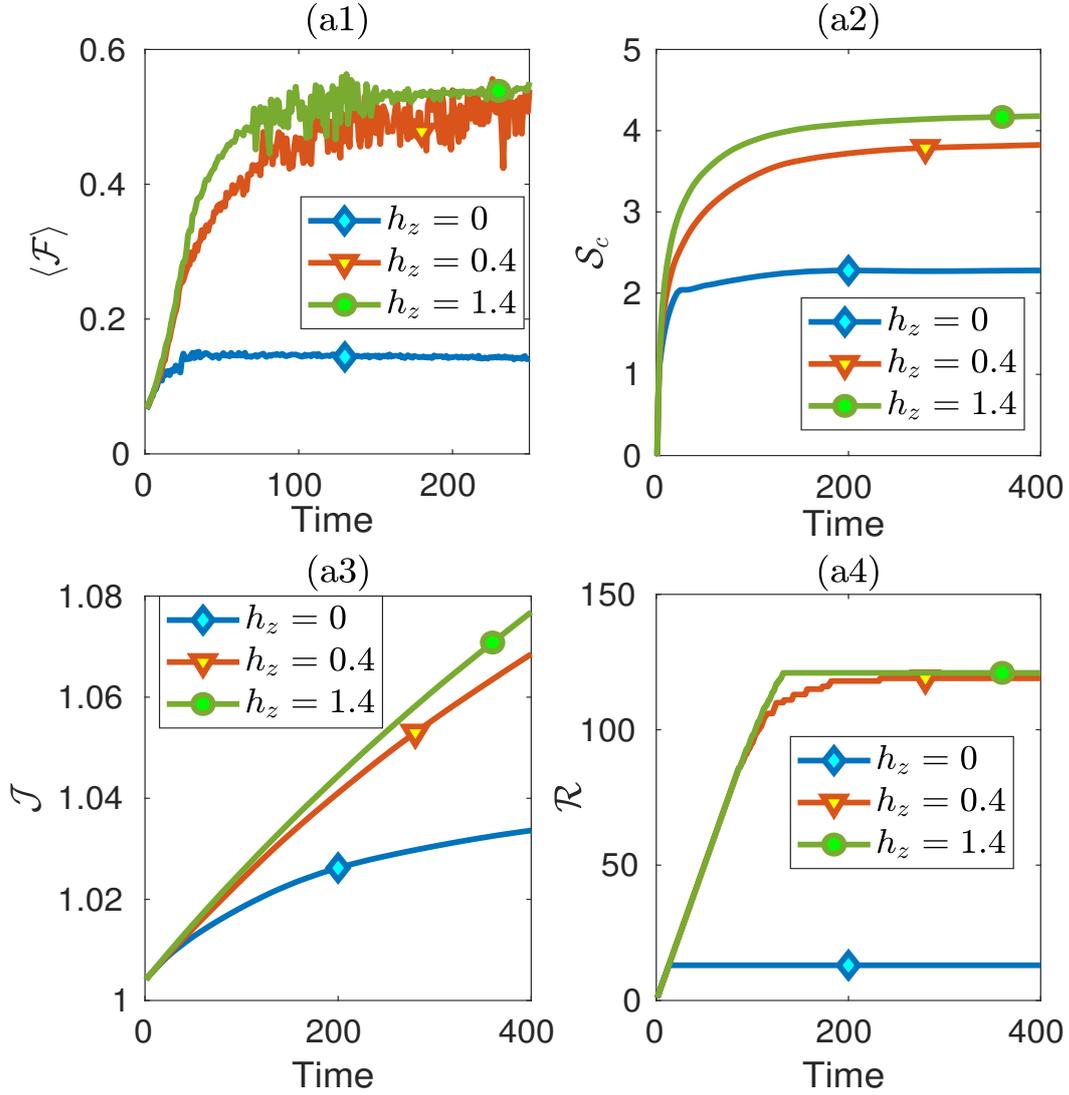

Figure 4.5: Quantifying operator spreading through various information-theoretic metrics as a function of time with an increase in the strength of the nonintegrability field. The time series of operators are generated by repeatedly applying the Floquet operator of the time-dependent tilted field kicked Ising model $U_{TKI}$ as visible in Eq. (4.7) for all plots. All numerical simulations are carried out for the Ising model of $L = 4$ spins with $J = 1$, $h_x = 1.4$. All plots plots are for the initial observable $S_x$. (a1) Average reconstruction fidelity $\langle \mathcal{F} \rangle$ as a function of time. (a2) The Shannon entropy $\mathcal{S}_C$. (a3) The Fisher information $\mathcal{J}$. (a4) Rank $\mathcal{R}$ of the covariance matrix. In all cases, the values of the quantifiers are higher for a higher nonintegrability parameter $h_z$.

### 4.3.2 Random matrix predictions for information-theoretic quantifiers

The tilted field Ising model has a discrete symmetry which makes the spin chain invariant under reflection (also known as "bit-reversal") about the centre of the spin chain [Karthik *et al.* (2007)]. Thus, the Floquet map for the kicked Ising model with a tilted magnetic field and the Hamiltonian for the time-independent tilted field Ising model are block diagonal in the eigenbasis of the reflection operator. The model is also invariant under time-reversal operation [Zhou and Luitz (2017); Pal and Lakshminarayan (2018)]. We generate some random matrices from circular orthogonal ensemble (COE) for the Floquet map and Gaussian orthogonal ensemble (GOE) for time-independent Hamiltonian, which are block diagonal in the basis of the reflection operator.

We consider a random local observable $O = u_r^\dagger s_1^y u_r$, where $u_r$ is a single-qubit random unitary chosen from Haar measure. The measurement record is generated by a time-dependent and time-independent tilted field Ising model to evaluate Shannon entropy $\mathcal{S}_c$, Fisher information $\mathcal{J}$, and rank $\mathcal{R}$. Also, we generate the time series of observables by the random unitary evolution picked from COE and random Hamiltonian time evolution picked from GOE having the desired block diagonal structure. Figure 4.6 shows the behavior of the information-theoretic quantifiers for both the models and the random matrix theory. We see excellent agreement between our predictions from random matrix theory and the calculation for the evolution by both time-dependent and time-independent models in the completely chaotic regime.

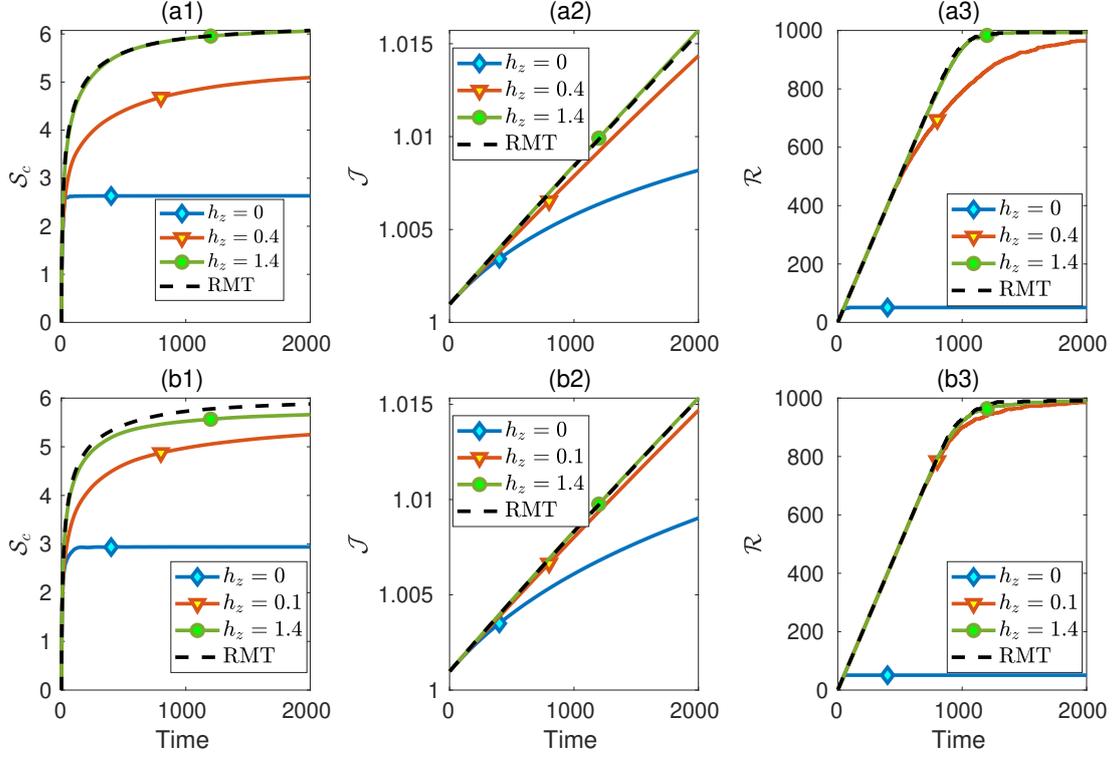

Figure 4.6: Quantifying operator spreading through various information-theoretic metrics as a function of time with an increase in the extent of chaos. The time series of operators are generated by repeatedly applying the Floquet operator of the time-dependent tilted field kicked Ising model $U_{TKI}$ as shown in Eq. (4.7) for plots (a1)-(a3). The unitary for time-independent tilted field Ising model Eq. (4.9) generates the time evolved operators for plots (b1)-(b3). All numerical simulations are carried out for the Ising model of $L = 5$ spins with $J = 1$, $h_x = 1.4$, and for an initial random local observable $u_r^\dagger s_1^y u_r$, where $u_r$ is a single-qubit Haar random unitary. (a1 and b1) The Shannon entropy $\mathcal{S}_c$. (a2 and b2) The Fisher information $\mathcal{J}$. (a3 and b3) Rank $\mathcal{R}$ of the covariance matrix. In all cases, the values of the quantifiers are higher for higher nonintegrability parameter $h_z$. In the strongly nonintegrable limit, $h_z = 1.4$, the results agree with values for measurement records obtained by a random matrix chosen from a suitable ensemble (dashed lines in all figures).

### 4.3.3 Operator spreading for different numbers of spins in the Ising model with a tilted magnetic field

We find the initial growth and the saturation value of average fidelity $\mathcal{F}$, Shannon entropy $\mathcal{S}_c$, Fisher information $\mathcal{J}$, and rank $\mathcal{R}$ of the covariance matrix are correlated with the degree of chaos in the dynamics. Thus, all these information-theoretic quantifiers are able to quantify operator spreading. The rank saturates at $\mathcal{R} = d^2 - d + 1$ for fully chaotic. The rank $\mathcal{R}$ for different numbers of spins is illustrated in Fig. 4.7 (a4) $L = 2$, $\mathcal{R} = 13$ (b4) $L = 3$, $\mathcal{R} = 57$, and (c4) $L = 4$, $\mathcal{R} = 241$. For all these figures, the initial observable is $O = s_1^y$, which does not respect the reflection symmetry about the centre of the tilted field Ising spin chain. It is interesting that even in the deep quantum regime, for $L = 2$ and $L = 3$, we can see the quantum signatures of chaos in the information-theoretic measures.

### 4.3.4 Results for Heisenberg XXZ spin chain with an integrability-breaking field

Here, we illustrate the operator spreading for the Heisenberg XXZ model for $L = 5$ number of spins with an integrability-breaking field. The single impurity $H_{si} = s_3^z$ is placed at the center of the spin chain, and the strength of the field $g$ is varied to drive the dynamics from integrable to chaotic. We choose the parameter set $g = \{0.0, 0.16, 0.94\}$, where the dynamics is fully chaotic and level statistics follow random matrix predictions for $g = 0.94$ [Rabinovici *et al.* (2022b)]. We have considered two initial observables $O = s_2^y + s_4^y$, and $O = s_2^y$. The observable $O = s_2^y + s_4^y$ respects the reflection symmetry about the center of the spin chain, whereas the operator $O = s_2^y$ does not respect the reflection symmetry. However, both the initial observables do not have support over the full spin chain. In Fig. 4.8, and Fig. 4.9 we have shown the average reconstruction fidelity $\langle \mathcal{F} \rangle$, the Shannon entropy $\mathcal{S}_c$, the Fisher information $\mathcal{J}$ and the rank of covariance matrix $\mathcal{R}$ for both the initial observables. We notice that the operator spreading is more when the dynamics becomes more chaotic irrespective of the choice of initial observable. We can see the effects of symmetries in the saturation value of all the quantities other than the Fisher information, which does not saturate. Fisher information, which is highly sensitive to vanishingly small eigenvalues associated with the covariance matrix, is not very sensitive to the chaoticity parameter, unlike the Shannon entropy. While we observe that the Fisher information computed for the fully chaotic and weakly nonintegrable dynamics can be close to each other, it certainly does preserve the correlation with the degree of chaos.

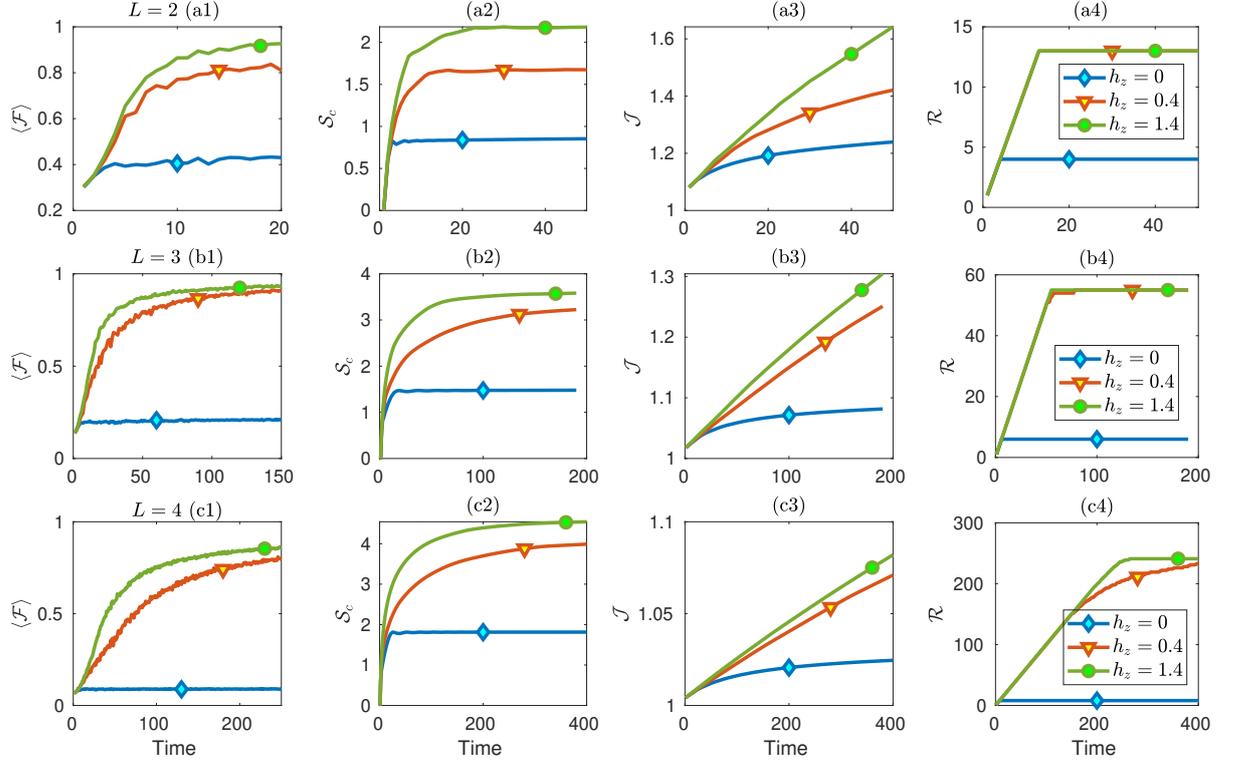

Figure 4.7: Quantifying operator spreading through various information-theoretic metrics as a function of time with an increase in the extent of chaos [The legends are same for all the plots; $h_z = 0$ (diamond), $h_z = 0.4$ (triangle) $h_z = 1.4$ (circle)]. The time series of operators are generated by repeatedly applying the Floquet operator of the time-dependent tilted field kicked Ising model $U_{TKI}$ as shown in Eq. (4.7) for all plots. The numerical simulations are carried out for the Ising model with $J = 1$, $h_x = 1.4$, and the initial observable $s_1^y$. The number of spins for the plots (a1 - a4) $L = 2$, (b1 - b4) $L = 3$, and (c1 - c4) $L = 4$. (a1, b1, and c1) Average reconstruction fidelity $\langle \mathcal{F} \rangle$ as a function of time. (a2, b2, and c2) The Shannon entropy $\mathcal{S}_C$. (a3, b3, and c3) The Fisher information $\mathcal{J}$. (a4, b4, and c4) Rank $\mathcal{R}$ of the covariance matrix. In all cases, the values of the quantifiers are more for a higher value of parameter $h_z$.

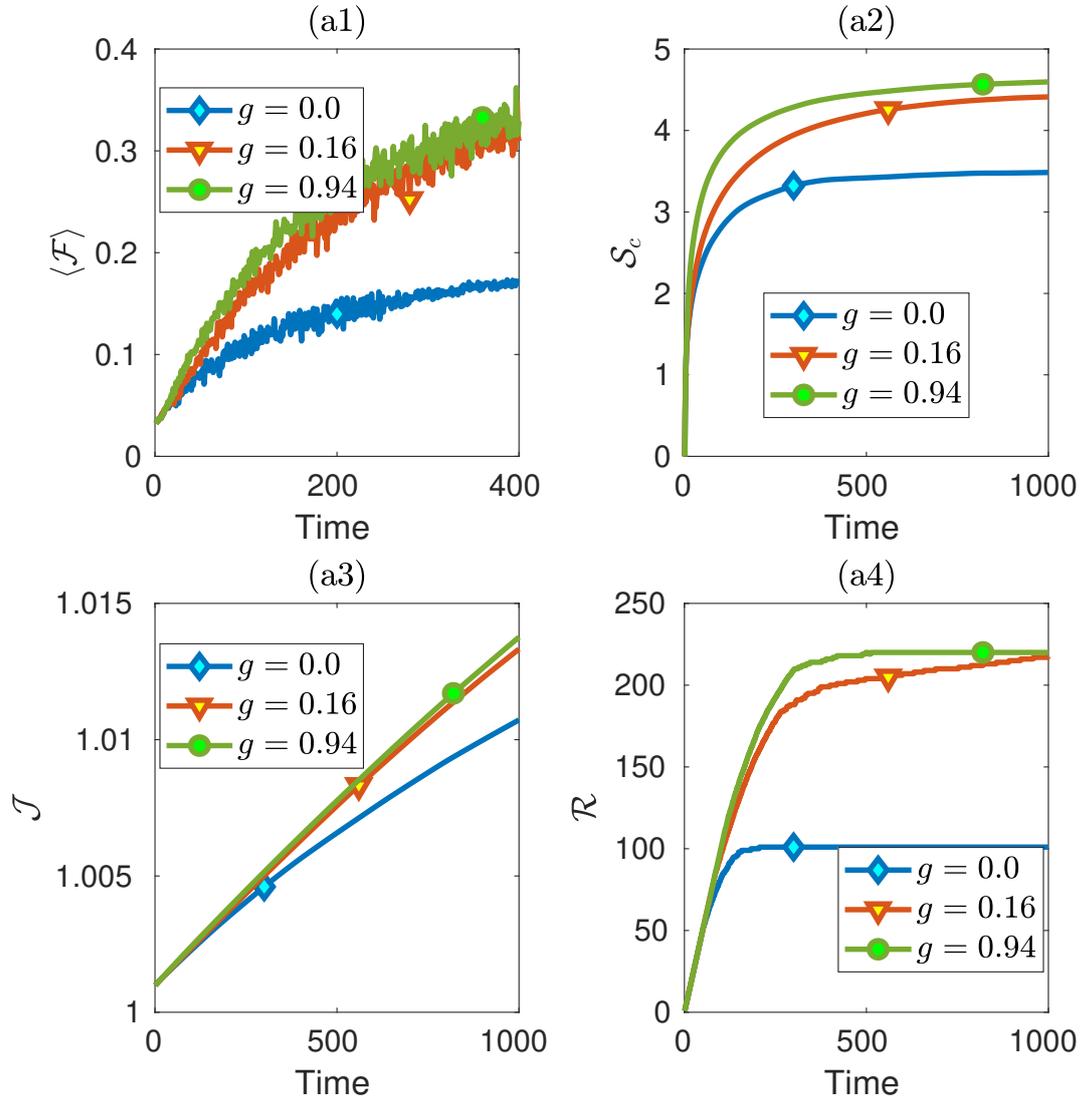

Figure 4.8: Operator spreading through various information-theoretic metrics as a function of time with an increase in the degree of chaos for Heisenberg XXZ spin chain with integrability breaking field $H_{si} = s_3^y$. All numerical simulations are carried out for the Hamiltonian $H_{HNI}$ of $L = 5$ spins with $J_{xy} = 1$, $J_{zz} = 1.1$. All plots are for the initial local observable $O = s_2^y + s_4^y$. (a1) Average reconstruction fidelity $\langle \mathcal{F} \rangle$ as a function of time. (a2) The Shannon entropy $\mathcal{S}_C$. (a3) The Fisher information $\mathcal{J}$. (a4) Rank $\mathcal{R}$ of the covariance matrix. In all cases, the values of the quantifiers are higher for a higher integrability breaking parameter $g$ value.

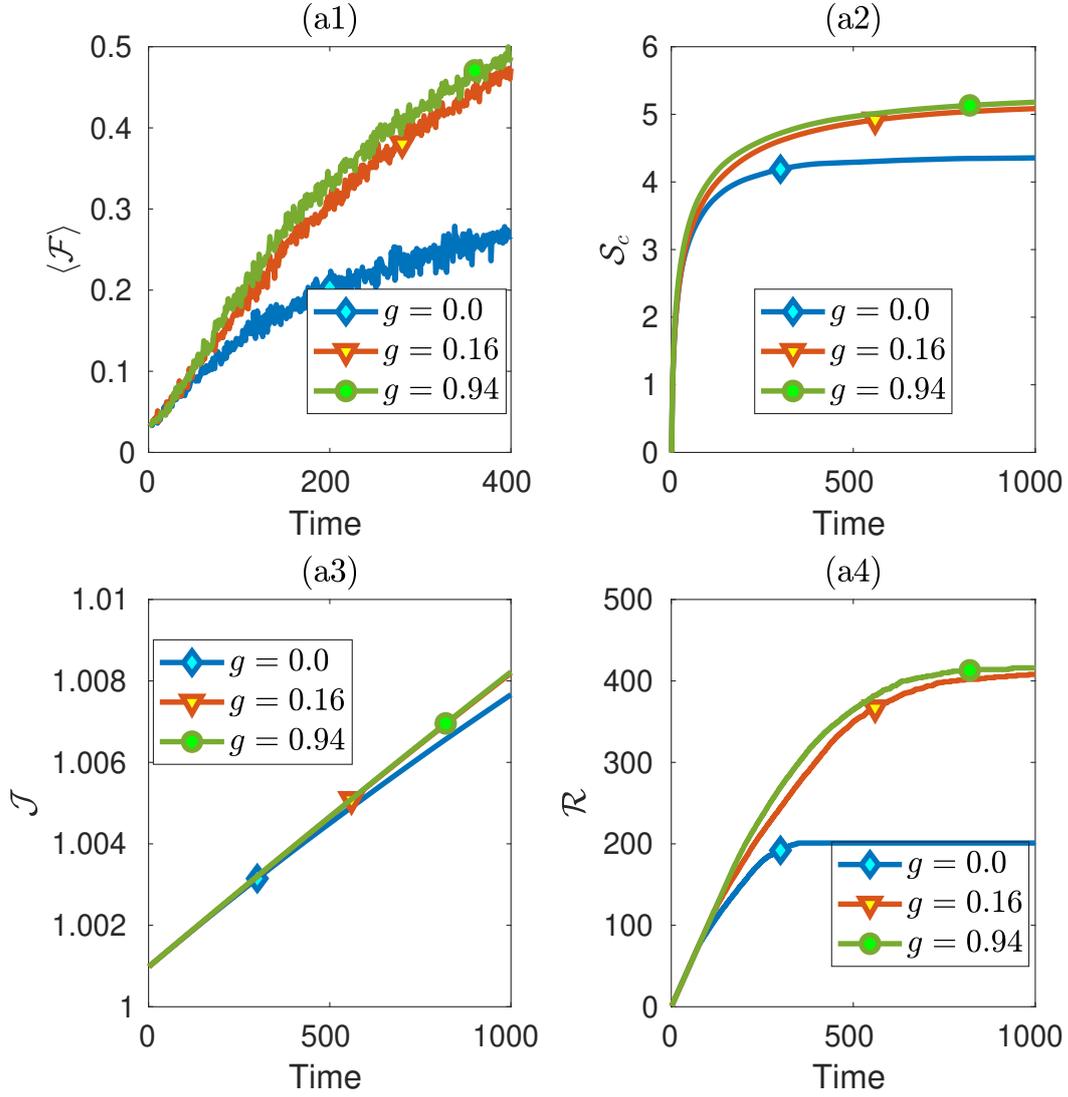

Figure 4.9: Operator spreading through various information-theoretic metrics as a function of time with an increase in the degree of chaos for Heisenberg XXZ spin chain with integrability breaking field $H_{si} = s_3^y$. All numerical simulations are carried out for the Hamiltonian $H_{HNI}$ of $L = 5$ spins with $J_{xy} = 1$, $J_{zz} = 1.1$. All plots (a1 - a4) are for the initial local observable $O = s_2^y + s_4^y$, and plots (b1 - b4) are for the initial observable $O = s_2^y$. (a1 and b1) Average reconstruction fidelity $\langle \mathcal{F} \rangle$ as a function of time. (a2 and b2) The Shannon entropy $\mathcal{S}_C$. (a3 and b3) The Fisher information $\mathcal{J}$. (a4 and b4) Rank $\mathcal{R}$ of the covariance matrix. In all cases, the values of the quantifiers are higher for a higher integrability breaking parameter $g$ value.

## 4.4 DISCUSSION

Characterizing chaos in the quantum world and its manifestations in quantum information processing is currently being vigorously pursued. In this chapter, we studied quantum chaos and its connections to operator spreading in many-body quantum systems. We linked operator spreading to the rate of information gain in quantum tomography - a protocol at the heart of quantum information processing. Our work in this chapter gives an operational interpretation of operator spreading and relates/contrasts to the Krylov complexity in the study of quantum chaos.

The operator, which is initially localized, will evolve under the chaotic dynamics and spread in the operator space as a more complex operator. Interestingly, various information-theoretic measures like Shannon entropy, Fisher information, and rank of the covariance matrix not only quantify the information gain but also support us in assessing the operator complexity. We show an unambiguous way of measuring operator complexity and operator scrambling as the dynamics becomes chaotic for the 1D Ising model with a tilted magnetic field and the 1D anisotropic Heisenberg XXZ spin chain with an integrability-breaking field. The rate of operator scrambling is positively correlated with the rate of information gain for random states, which increases the dynamics approaches the fully chaotic limit. Our information-theoretic quantifiers are suitable for both time-independent as well as time-dependent Hamiltonians.

The idea of operator spreading in the space of Hermitian observables can be compared to the exploration of classical trajectory in the classical phase space. Kolmogorov-Sinai (KS) entropy [Kolmogorov (1958, 1959); Sinai (1959); Pesin (1977)] is known to quantify the rate of exploration in the classical phase space, which is equal to the sum of positive Lyapunov exponents. Here, we relate the rate of the operator spreading to the degree of chaos in the dynamics through our information-theoretic metrics. Thus, the information gain in quantum tomography is connected to operator spreading in the operator space and the KS entropy. This is a tantalising direction which needs to be further pursued.

Simulating quantum chaos on a quantum computer [Lysne *et al.* (2020); Muñoz-Arias *et al.* (2020); Krithika *et al.* (2023); Maurya *et al.* (2022)] and exploring its information-theoretic signatures like operator scrambling is of significance both from a fundamental as well as from an applied point of view. These are concrete steps towards realizing a universal quantum simulator and eventually a quantum computer. After all, a quantum simulator is a many-body quantum system which would exhibit complex and possibly chaotic dynamics. The work presented in this chapter, connecting continuous measurement tomography and information scrambling, paves the way to realize and interpret such studies in the laboratory [Smith *et al.* (2006); Chaudhury *et al.* (2009)].

# CHAPTER 5

# ERROR SCRAMBLING IN NOISY TOMOGRAPHY

## 5.1 INTRODUCTION

The central goal of quantum chaos is to inform us about the properties of quantum systems whose classical counterpart is chaotic. Vigorous thrust in the understanding of quantum many-body dynamical systems through dynamically generated entanglement [Miller and Sarkar (1999); Bandyopadhyay and Lakshminarayan (2002); Wang *et al.* (2004); Trail *et al.* (2008); Furuya *et al.* (1998); Lakshminarayan (2001); Seshadri *et al.* (2018)], and quantum correlations [Madhok *et al.* (2015, 2018)], deeper studies in the ergodic hierarchy of quantum dynamical systems [Gomez and Castagnino (2014); Bertini *et al.* (2019); Aravinda *et al.* (2021); Vikram and Galitski (2022)] have been some recent milestones. Also, the out-of-time-ordered correlators (OTOCs) that attempt to capture operator growth and scrambling of quantum information have been very useful as a probe for chaos in quantum systems [Maldacena *et al.* (2016); Swingle *et al.* (2016); Hashimoto *et al.* (2017); Kukuljan *et al.* (2017); Swingle (2018); Wang *et al.* (2021); Sreeram *et al.* (2021); Varikuti and Madhok (2022)]. These, coupled with the traditional approach to studies of level statistics [Haake (1991)] and Loschmidt echo [Peres (1984, 1997); Goussev *et al.* (2012); Gorin *et al.* (2006)] and complemented by the ability to coherently control and manipulate many-body quantum systems in the laboratory [Gong and Brumer (2001, 2005); Brif *et al.* (2010); Smith *et al.* (2013); Mirkin and Wisniacki (2021)], have brought us to a fork in our path. On the one hand, this is a harbinger of the possibility of building quantum simulators, an important milestone in our quest for the holy grail - a many-body quantum computer. On the other hand, the same properties that make quantum systems generate complexity will make them sensitive to errors that naturally occur in implementing many-body Hamiltonians.

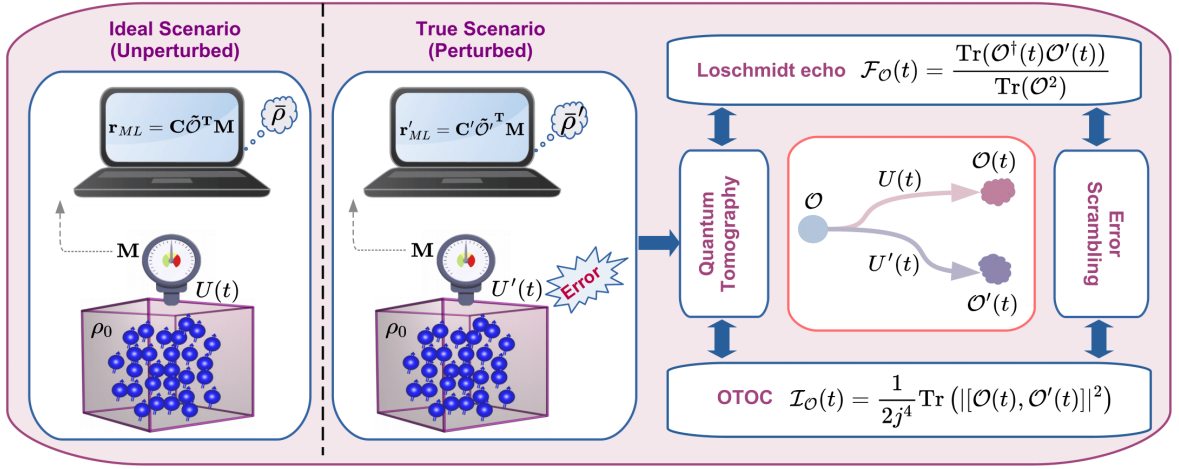

Figure 5.1: An illustration of continuous measurement tomography and its connection with various quantifiers of quantum chaos. In the ideal scenario where the experimentalist has complete knowledge of the true dynamics, the reconstructed state reaches the actual state with time. However, in reality, the experimentalist is ignorant about the true (perturbed) dynamics that leads to improper reconstruction of the quantum state. Thus, the reconstruction fidelity decays after some time and positively correlates with the operator Loschmidt echo. We quantify the scrambling of error as $\mathcal{I}_\mathcal{O}(t)$, and it helps to unify Loschmidt echo $\mathcal{F}_\mathcal{O}(t)$ and OTOC in a real quantum information processing task, quantum tomography.

While chaotic dynamics is a source of information quantified by the positive KS entropy [Pesin (1977)], it is sensitive to errors, as captured by Loschmidt echo. While the KS entropy enables a rapid information gain, Loschmidt echo will cause a rapid accumulation of errors, or *error scrambling* as we quantify. This interplay between KS entropy and Loschmidt echo is a generic feature of any many-body system, and we identify and quantify the crossover between these two competing effects. In many body systems, quantum or classical, we must expect the presence of both chaos and errors. In this chapter, we address this scenario; we go on to discover quantum signatures of chaos while shedding light on the larger question of many-body quantum simulations under unavoidable perturbations.

Quantum tomography gives us a window to study sensitivity to errors in quantum simulations of chaotic Hamiltonians [Lloyd (1996); Johnson *et al.* (2014)]. In Quantum tomography, one tries to estimate the actual state $\rho_0$ using the statistics of measurement records on an ensemble of identical systems. We consider continuous weak measurement tomography protocol [Silberfarb *et al.* (2005); Smith *et al.* (2006); Riofrío *et al.* (2011); Smith *et al.* (2013)], and the time series of operators can be generated by the Floquet map of a quantum dynamical system to investigate the role of chaos on the information gain in tomography [Madhok *et al.* (2014); Sreeram and Madhok (2021); Sahu *et al.* (2022a)]. The information gain in tomography quantifies the amount of new information added as one follows the trajectory of operators generated by the dynamics in the Heisenberg picture. However, OTOC is the quantum analog of divergence of two trajectories, which Lyapunov exponents capture in the classical picture and operator incompatibility in the quantum counterpart [Larkin and Ovchinnikov (1969); Maldacena *et al.* (2016); Swingle (2018)]. Therefore, a natural direction is to connect the information acquisition in state reconstruction to the Lyapunov exponents, thereby unifying the connections between information gain, scrambling, and chaos and associating it with an actual physical process. We connect the operator spreading to the rate of state reconstruction - a protocol at the heart of quantum information processing. Our work in the previous chapter gives an operational interpretation of operator spreading and relates/contrasts to the Krylov complexity in the study of quantum chaos. The operator, which is initially localized, will evolve under the chaotic dynamics and spread in the operator space as a more complex operator. Here we unify two quantifiers of quantum chaos, namely Loschmidt echo and OTOC, through scrambling of errors in continuous weak measurement tomography as illustrated in Fig. 5.1.

The rest of this chapter is assembled in the following way. In Sec. 5.2, we briefly describe the protocol for tomography with continuous-time weak measurements. In Sec. 5.3 we outline the protocol for tomography with error in the dynamics. Then we quantify the effect of perturbation in information gain using operator Loschmidt echo, relative entropy, and operator incompatibility for for scrambling of error in Sec. 5.4. We finally conclude this chapter in Sec. 5.6 with discussions on our findings and some outlook.

## 5.2 CONTINUOUS MEASUREMENT TOMOGRAPHY

An ensemble of $N_s$ identical systems $\rho_0^{\otimes N_s}$ undergo separable time evolution by a unitary $U(t)$. A weakly coupled probe will generate the measurement record by performing weak continuous measurement of an observable $O$. For sufficiently weak coupling, the randomness of the measurement outcomes is dominated by the quantum noise in the probe rather than the measurement uncertainty, i.e., the projection noise. In this case, the quantum backaction is negligible, and the state remains approximately separable. Thus, we get the stochastic measurement record

$$M(t) = \text{Tr}\,(O(t)\rho_0) + W(t), \tag{5.1}$$

where $O(t) = U^\dagger(t)OU(t)$ is the time evolved operator in Heisenberg picture, and $W(t)$ is a Gaussian white noise with spread $\sigma/N_s$.

Any density matrix of Hilbert-space dimension $d$ can be realized as a generalized Bloch vector $\mathbf{r}$ by expanding $\rho_0 = \mathbb{1}/d + \Sigma_{\alpha=1}^{d^2-1} r_\alpha E_\alpha$ in an orthonormal basis of traceless Hermitian operators $\{E_\alpha\}$. We consider the measurement record at discrete times as $M_n = M(t_n) = \text{Tr}\,(O_n\rho_0) + W_n$, that allows one to express the measurement history

$$\mathbf{M} = \tilde{O}\mathbf{r} + \mathbf{W}, \tag{5.2}$$

where $\tilde{O}_{n\alpha} = \text{Tr}\,(O_n E_\alpha)$. Thus, the problem of quantum tomography is reduced to linear stochastic state estimation of $\rho_0$ given $\{M_n\}$. In the limit of negligible backaction, the probability distribution associated with measurement history $\mathbf{M}$ for a given state vector $\mathbf{r}$ is [Silberfarb *et al.* (2005); Smith *et al.* (2006)]

$$\begin{aligned} p(\mathbf{M}|\mathbf{r}) &\propto \exp\left\{-\frac{N_s^2}{2\sigma^2}\sum_i [M_i - \sum_\alpha \tilde{O}_{i\alpha} r_\alpha]^2\right\} \\ &\propto \exp\left\{-\frac{N_s^2}{2\sigma^2}\sum_{\alpha,\beta}(\mathbf{r} - \mathbf{r_{ML}})_\alpha\, C^{-1}_{\alpha\beta}\,(\mathbf{r} - \mathbf{r_{ML}})_\beta\right\}. \end{aligned} \tag{5.3}$$

In the weak backaction limit, the fluctuations around the mean are Gaussian distributed, and hence the maximum likelihood estimate of the Bloch vector components is the

least-squared fit as

$$\mathbf{r}_{ML} = \mathbf{C}\tilde{O}^{\mathbf{T}}\mathbf{M}, \tag{5.4}$$

where $\mathbf{C} = (\tilde{O}^{\mathbf{T}}\tilde{O})^{-1}$ is the covariance matrix and the inverse is Moore-Penrose pseudo inverse [Ben-Israel and Greville (2003)] in general. The estimated Bloch vector $\mathbf{r}_{ML}$ may not always represent a positive density matrix because of the finite signal-to-noise ratio. Therefore we impose the constraint of positive semidefiniteness on the reconstructed density matrix and obtain the physical state closest to the maximum-likelihood estimate. To do this, we employ a convex optimization procedure [Vandenberghe and Boyd (1996)] where the final estimate of the Bloch vector $\bar{\mathbf{r}}$ is obtained by minimizing the argument

$$||\mathbf{r}_{ML} - \bar{\mathbf{r}}||^2 = (\mathbf{r}_{ML} - \bar{\mathbf{r}})^T \mathbf{C}^{-1} (\mathbf{r}_{ML} - \bar{\mathbf{r}}) \tag{5.5}$$

subject to the constraint

$$\mathbb{1}/d + \Sigma_{\alpha=1}^{d^2-1} \bar{r}_\alpha E_\alpha \geq 0.$$

## 5.3 TOMOGRAPHY WITH IMPERFECT KNOWLEDGE

The above description represents an ideal scenario where the experimentalist has complete knowledge of the *true* dynamics (which is symbolized as unprimed variables describing the observables, $O_n$, and covariance matrix, $\mathbf{C}$, thus generated over time) and they can properly reconstruct the state using Eq. (5.4). However, in reality, one never knows the true underlying dynamics, and there is always a departure from the ideal case due to inevitable errors and perturbations to the *true* dynamics. Thus, the experimentalist, oblivious to these, models their estimation using a covariance matrix, $\mathbf{C}' = (\tilde{O}'^{\mathbf{T}}\tilde{O}')^{-1}$. Here the primed variables represent the experimentalist's knowledge of the dynamics in the laboratory, and as a result, they end up reconstructing an incorrect state as $\bar{\rho}'$ from

$$\mathbf{r}'_{ML} = \mathbf{C}'\tilde{O}'^{\mathbf{T}}\mathbf{M}. \tag{5.6}$$

In the above equation, the measurement record is obtained from the measurement device (probe), and the experimentalist is ignorant about the *true* dynamics (which is accompanied by perturbations relative to the idealized dynamics as assumed by the experimentalist), given by the unitary $U(t)$, that has generated this record. However, the covariance matrix is uniquely determined from the experimentalist's version of the dynamics given by the unitary $U'(t)$ and the initial observable $O$. Thus, the ignorance about the error in the dynamics directs the operator trajectory away from the actual one, leading to an improper reconstruction of the state $\rho_0$.

Our goal is to study the effect of the perturbation on the information gain in tomography in the presence of chaos. To accomplish this, we implement the quantum kicked top [Haake *et al.* (1987); Haake (1991); Chaudhury *et al.* (2009)] described by the Floquet map $F_{QKT} = e^{-i\lambda J_z^2/2J} e^{-i\alpha J_x}$ as the unitary for a period $\tau$ for simplicity, and the unitary at $n^{\text{th}}$ time step is $U(n\tau) = U_\tau^n$. The measurement record obtained by repeatedly applying such a Floquet operator is not informationally complete, and it leaves out a subspace of dimension $\geq d-2$, out of $d^2-1$ dimensional operator space. For our current work of this chapter, we fix the linear precision angle $\alpha = 1.4$ and choose the kicking strength $\lambda$ as the chaoticity parameter. The classical dynamics change from highly regular to fully chaotic as we vary $\lambda$ from 0 to 7. The dynamics that represents the *true* evolution is perturbed relative to the idealized dynamics given by $F_{QKT}$, and we choose a small variation in the kicking strength, $\lambda + \delta\lambda$, and the perturbed unitary becomes $U_\tau = e^{-i(\lambda+\delta\lambda)J_z^2/2J} e^{-i\alpha J_x}$. For our analysis, we consider the dynamics of quantum kicked top for a spin $j = 10$, and perturbation strength $\delta\lambda = 0.01$.

The connection between chaos and information gain depends on the localization properties of the state, i.e. their inverse participation ratio, the degree of chaos, as well as how well the state is aligned with time-evolved measurement observables [Sahu *et al.* (2022a)]. The inverse participation ratio (IPR) is a standard terminology in the quantum chaos literature. IPR quantifies the localization of a state $|\psi\rangle$ in a given basis $\{|i\rangle\}_{i=1}^d$ as $IPR = \sum_{i=1}^d |\langle i|\psi\rangle|^4$. The state $|\psi\rangle$ is localized when it is one of the basis $\{|i\rangle\}$ vectors, and the maximum value of IPR is 1. The minimum value of IPR is $1/d$ when the state $|\psi\rangle$ is a uniform superposition of all the basis vectors, which makes the state fully delocalized in that basis. This concept is not specific to the kicked top but is employed to study quantum chaos regardless of the underlying dynamics. More specifically, for the kicked top, we can determine the participation ratio with respect to the basis of the Floquet operator or the initial observable we are choosing for generating the measurement record. Previously we have shown that the rate of information gain in tomography depends on the dynamics, the choice of initial observable and the localization of the states [Sahu *et al.* (2022*a*)]. Localization of the state can be studied in two ways. One is the localization of the state in a given basis which is quantified by the participation ratio, and the other one is the localization of the spin coherent state in the phase space as quantified by Husimi entropy.

We aim to study the effect of the degree of chaos on the performance of noisy tomography purely. Therefore, we consider random initial states measured via random initial observables (generated by rotating $J_x$ through random unitary) picked from the appropriate Haar measure. We apply our reconstruction protocol on an ensemble of 100 random pure states sampled from the Haar measure on $SU(d)$, where $d = 2j + 1 = 21$. We choose one random initial observable and generate the measurement record by repeatedly applying the Floquet operator of the quantum kicked top. The fidelity of the reconstructed state $\bar{\rho}'$ obtained from Eq. (5.6) is determined relative to the actual state $|\psi_0\rangle$, $\mathcal{F} = \langle\psi_0|\bar{\rho}'|\psi_0\rangle$ as a function of time. We notice that the reconstruction fidelity increases in the beginning despite the errors, and after a certain period of time, it starts decaying. The rise in fidelity during the initial time period is because any information, even if partially inaccurate, about a completely unknown random state offsets the presence of errors in its estimation. However, as time progresses, the effect of errors becomes significant. Beyond a certain time, we observe a decline in fidelity as the dynamics continues to accumulate errors that dominate the archive of information present in the measurement record. Most interestingly, the rate of this fidelity decay is inversely correlated with the degree of chaos in the dynamics. This is the analog of an interplay between the rapid information gain due to Lyapunov divergence, a "quantum" analog of the classical KS entropy, and Loschmidt echo leading to errors that cause fidelity decay.

## 5.4 QUANTIFYING THE ROLE OF CHAOS IN RECONSTRUCTION IN THE PRESENCE OF ERROR

We now quantify the role of chaos in tomography when the error in the dynamics influences our ability to reconstruct the random quantum states. It is evident from Fig. 5.2a that the rate of drop in fidelity decreases with an increase in the strength of chaos for small perturbations in the dynamics. To understand the foregoing discussion, we define certain quantities to quantify effect of error in the information gain.

### 5.4.1 Operator Loschmidt echo

We define the operator Loschmidt echo $\mathcal{F}_O$ as the Hilbert-Schmidt inner product of the operators $O_n$, and $O'_n$ generated from repeated application of the Floquet map for *true* (perturbed) dynamics $U_\tau$ and ideal (unperturbed) dynamics $U'_\tau$ of the kicked top on the initial observable $O$

$$\mathcal{F}_O(t_n) = \frac{\text{Tr}\,(O_n^\dagger O'_n)}{\text{Tr}\,(O^2)}. \tag{5.7}$$

The operator Loschmidt echo that captures the overlap of the operators $O_n$ and $O'_n$, decays with time. We can see from Fig. 5.2b that the operator Loschmidt echo decays much slower when the dynamics is chaotic than when it is regular. This behavior of the operator Loschmidt echo correlates positively with the rate of drop in reconstruction fidelity as demonstrated in Fig. 5.2a and Fig. 5.2b. The greater the distance between the operators at a given time, the greater the difference between the expectation values with respect to the state and the archive of the measurement record obtained through the time series. Our results give an *operational interpretation* of the operator Loschmidt echo by connecting it to a concrete physical task of continuous measurement quantum tomography. This also points to a beneficial way to probe these quantities in experiments using current techniques.

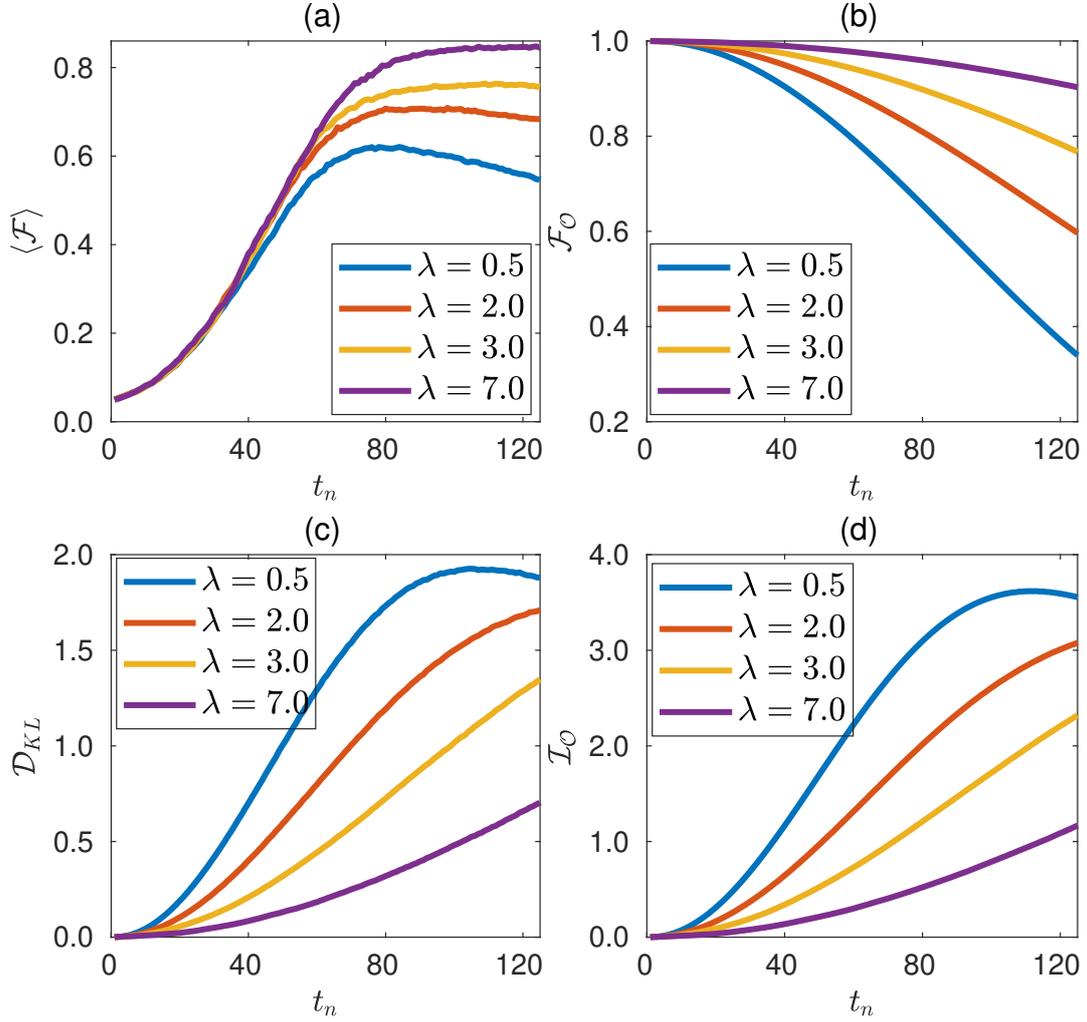

Figure 5.2: Effect of perturbation on tomography quantified by different metrics as a function of time with an increase in the level of chaos. The kicked top Floquet map of *true* (perturbed) dynamics $U_\tau = e^{-i(\lambda+\delta\lambda)J_z^2/2J}e^{-i\alpha J_x}$ with $\delta\lambda = 0.01$, and ideal (unperturbed) dynamics $U'_\tau = e^{-i\lambda J_z^2/2J}e^{-i\alpha J_x}$ generate the time series of operators for a spin $j = 10$ and fixed $\alpha = 1.4$. (a) Average reconstruction fidelity $\langle \mathcal{F} \rangle$ of the state $\bar{\rho}'$ derived from Eq. (5.6) relative to the actual state $|\psi_0\rangle$, where the average is taken over 100 Haar random states. (b) The operator Loschmidt echo $\mathcal{F}_O$ between two operators. (c) The quantum relative entropy $\mathcal{D}_{KL}$ of regularized operator evolved under unperturbed dynamics to the operator evolved under perturbed dynamics. (d) The operator incompatibility $\mathcal{I}_O$ quantifies the scrambling of errors.

### 5.4.2 Quantum relative entropy

Quantum relative entropy is a measure of distance between two quantum states. Here we use this metric to measure the distance between two operators $O_n$ and $O'_n$. To treat both observables as density operators, we regularize them in the following way. First, we diagonalize the time-evolved operator $O_n$ as

$$O_n = V_n D_n V_n^\dagger, \tag{5.8}$$

where $V_n$ is the unitary matrix that diagonalizes $O_n$. In the second step, we take the absolute value of the eigenvalues of $D_n$ and divide it by its trace to get $\tilde{D}_n = |D_n|/\text{Tr}(|D_n|)$. We, then, construct a positive operator with unit trace that behaves as a density matrix while keeping the eigenvectors of the observable $O_n$ intact as

$$\rho_{O_n} = V_n \tilde{D}_n V_n^\dagger. \tag{5.9}$$

Now we can determine the quantum relative entropy

$$\mathcal{D}_{KL}(\rho_{O_n} || \rho_{O'_n}) = \text{Tr}(\rho_{O_n}(\ln \rho_{O_n} - \ln \rho_{O'_n})), \tag{5.10}$$

where $\rho_{O_n}$ and $\rho_{O'_n}$ are positive operators of unit trace obtained from the regularization of operators $O_n$ and $O'_n$ respectively. We can see clearly from Fig. 5.2c that the distance between the two operators increases rapidly when the level of chaos is less in the dynamics. This indicates the operator becomes less prone to error in the Hamiltonian with the rise in the level of chaos. Ultimately, this makes quantum state tomography more immune to error in the presence of chaos, as we see in Fig. 5.2a.

### 5.4.3 Operator incompatibility and Error scrambling

To further elucidate the decline rate of reconstruction fidelity, we connect the operator incompatibility to the information gain. The operators at time $t$ be $O(t) = U^\dagger(t) O U(t)$ and $O'(t) = U'^\dagger(t) O U'(t)$, where $U(t)$ is the unitary for the unperturbed dynamics and $U'(t)$ is the unitary for perturbed dynamics. Thus, the operator incompatibility of two operators $O_n$ and $O'_n$ is

$$\mathcal{I}_O(t) = \frac{1}{2j^4}\text{Tr}\left(|[O(t), O'(t)]|^2\right). \tag{5.11}$$

Now one can simplify this expression using the Hermiticity property of the physical observables $O(t) = O^\dagger(t)$.

$$\begin{aligned}
\text{Tr}\left(|[O(t), O'(t)]|^2\right) &= \text{Tr}\left([O(t), O'(t)]^\dagger [O(t), O'(t)]\right) \\
&= \text{Tr}\left((O(t)O'(t) - O'(t)O(t))^\dagger (O(t)O'(t) - O'(t)O(t))\right) \\
&= \text{Tr}\left((O'(t)O(t) - O(t)O'(t))(O(t)O'(t) - O'(t)O(t))\right) \\
&= \text{Tr}\bigl(O'(t)O(t)O(t)O'(t) - O'(t)O(t)O'(t)O(t) \\
&\quad - O(t)O'(t)O(t)O'(t) + O(t)O'(t)O'(t)O(t)\bigr)
\end{aligned} \tag{5.12}$$

We can further simplify the above expression by exploiting the cyclic property of trace operation and the property of a unitary operator $U^\dagger U = UU^\dagger = \mathbb{1}$. Let us consider the first term in the above expression

$$\begin{aligned}
\text{Tr}\left(O'(t)O(t)O(t)O'(t)\right) &= \text{Tr}\left(U'^\dagger(t)OU'(t)U^\dagger(t)OU(t)U^\dagger(t)OU(t)U'^\dagger(t)OU'(t)\right) \\
&= \text{Tr}\left(U^\dagger(t)U(t)U'^\dagger(t)OU'(t)U^\dagger(t)OOU(t)U'^\dagger(t)OU'(t)\right) \\
&= \text{Tr}\left(U(t)U'^\dagger(t)OU'(t)U^\dagger(t)OOU(t)U'^\dagger(t)OU'(t)U^\dagger(t)\right) \\
&= \text{Tr}\left(\mathcal{U}^\dagger(t)O\mathcal{U}(t)OO\mathcal{U}^\dagger(t)O\mathcal{U}(t)\right),
\end{aligned} \tag{5.13}$$

where we define the error unitary $\mathcal{U}(t) = U'(t)U^\dagger(t)$. Similarly, we can modify the remaining terms of the RHS in Eq. (5.12) and rewrite Eq. (5.11) as

$$\mathcal{I}_O(t) = \frac{1}{2j^4}\text{Tr}\left(|[O, \mathcal{U}^\dagger(t)O\mathcal{U}(t)]|^2\right), \tag{5.14}$$

where the error unitary $\mathcal{U}(t) = U'(t)U^\dagger(t)$. This is the expression for error scrambling, which helps us to connect the operator incompatibility (out-of-time-order correlator) and Loschmidt echo. If the error enters through the Hamiltonian, then the error unitary for time-independent Hamiltonian becomes $\mathcal{U}(t) = e^{iH't}e^{-iHt}$ and for time-dependent kicked Hamiltonian (e.g. kicked top) $\mathcal{U}_n = U_\tau'^n U_\tau^{\dagger n}$, where $U_\tau$ is the Floquet map. Thus, the operator incompatibility, which can be realized as Eq. (5.14), is very general, and we do not assume any specific form of the error or any particular form of Hamiltonian (whether time-dependent or time-independent), as we can see below. The growth of out-of-time-order correlator (OTOC) has been studied extensively as a quantifier for information scrambling under chaotic dynamics [Maldacena *et al.* (2016); Swingle *et al.* (2016); Hashimoto *et al.* (2017); Kukuljan *et al.* (2017); Swingle (2018); Wang *et al.* (2021); Sreeram *et al.* (2021); Varikuti and Madhok (2022)]. Similarly, growth of $\mathcal{I}_O$ implies *scrambling of errors* with time. It is apparent from Fig. 5.2d that the rate of error scrambling decreases with an increase in the value of the chaoticity parameter $\lambda$. This signifies that the measurement record is less affected by the error in the dynamics when one approaches a greater extent of chaos. In Eq. (5.6), the measurement record **M** is obtained from the *true* (perturbed) dynamics, but the covariance matrix **C**′, and $\tilde{O}'$ are determined from the experimentalist's version of the dynamics (ideal or unperturbed). Thus, a higher rate of error scrambling for regular dynamics leads to a faster decay of reconstruction fidelity as the measurement record is more vulnerable. How errors scramble across a chaotic system, as given by Eq. (5.14), is itself an interesting quantifier of quantum chaos.

## 5.5 UNDERSTANDING INFORMATION GAIN IN TOMOGRAPHY WITH ERRORS

Here, we describe information gain and the general behavior of reconstruction fidelity as shown in the main text. Later in this section, we will demonstrate the effect of the magnitude of perturbation on the fidelity obtained in tomography. We observe that independent of the degree of chaos in the dynamics, the fidelity initially rises despite errors and then starts to decline after attaining a peak. As described in the main text, we consider the density matrix of Hilbert-space dimension $d$ that can be realized as a generalized Bloch vector **r** by expanding $\rho_0 = \mathbb{1}/d + \Sigma_{\alpha=1}^{d^2-1} r_\alpha E_\alpha$ in an orthonormal basis of traceless Hermitian operators $\{E_\alpha\}$.

The probability of reconstructing a state $\rho_0$ is [Sahu *et al.* (2022*a*)]

$$p(\rho_\mathbf{0}|\mathbf{M}, \mathcal{L}, \mathcal{M}) = A\, p(\mathbf{M}|\rho_\mathbf{0}, \mathcal{L}, \mathcal{M})\, p(\rho_0|\mathcal{L}, \mathcal{M})\, p(\mathcal{L}, \mathcal{M}), \quad (5.15)$$

where $A$ is a normalization constant. The first term $p(\mathbf{M}|\rho_\mathbf{0}, \mathcal{L}, \mathcal{M})$, is the probability of acquiring a measurement record **M**, given an initial state $\rho_0$, the dynamics $\mathcal{L}$ (choice of unitaries), and the measurement process $\mathcal{M}$ (choice of operators $O$ for generating measurement record). This term contains the errors due to shot noise and helps one to quantify the signal-to-noise ratio in various directions in the operator space independent of the state to be estimated. Thus, in the limit of negligible backaction $p(\mathbf{M}|\rho_\mathbf{0}, \mathcal{L}, \mathcal{M})$ is identical to the probability distribution corresponding to the measurement history **M** for a given Bloch vector **r** [Silberfarb *et al.* (2005); Smith *et al.* (2006); Merkel *et al.* (2010); Madhok *et al.* (2014); Sreeram and Madhok (2021)],

$$\begin{aligned} p(\mathbf{M}|\mathbf{r}) &\propto \exp\left\{-\frac{N_s^2}{2\sigma^2}\sum_i [M_i - \sum_\alpha \tilde{O}_{i\alpha} r_\alpha]^2\right\} \\ &\propto \exp\left\{-\frac{N_s^2}{2\sigma^2}\sum_{\alpha,\beta}(\mathbf{r}-\mathbf{r_{ML}})_\alpha\, C^{-1}_{\alpha\beta}\, (\mathbf{r}-\mathbf{r_{ML}})_\beta\right\}. \end{aligned} \quad (5.16)$$

Therefore, this term estimates the information gained, given a density matrix, in different directions in the operator space. Now we have the second term $p(\rho_0|\mathcal{L}, \mathcal{M})$, which is the posterior probability distribution relating the knowledge of the dynamics and the measurement operators. On that account, in the limit of vanishing shot noise and with complete knowledge of the system dynamics for given measurement observables $\{E_\alpha\}$, this conditional probability is continuously updated and ultimately becomes a product of Dirac-delta functions. Once we obtain an informationally complete measurement record, each Dirac-delta function identifies a particular Bloch vector component. The term $p(\mathcal{L}, \mathcal{M})$ in Eq. (5.15) can be absorbed in the constant as it gives the prior information about the choice of dynamics and measurement operators. Thus, Eq. (5.15) separates the probability of quantum state estimation into a product of two terms (up to a constant) [Sahu *et al.* (2022a)].

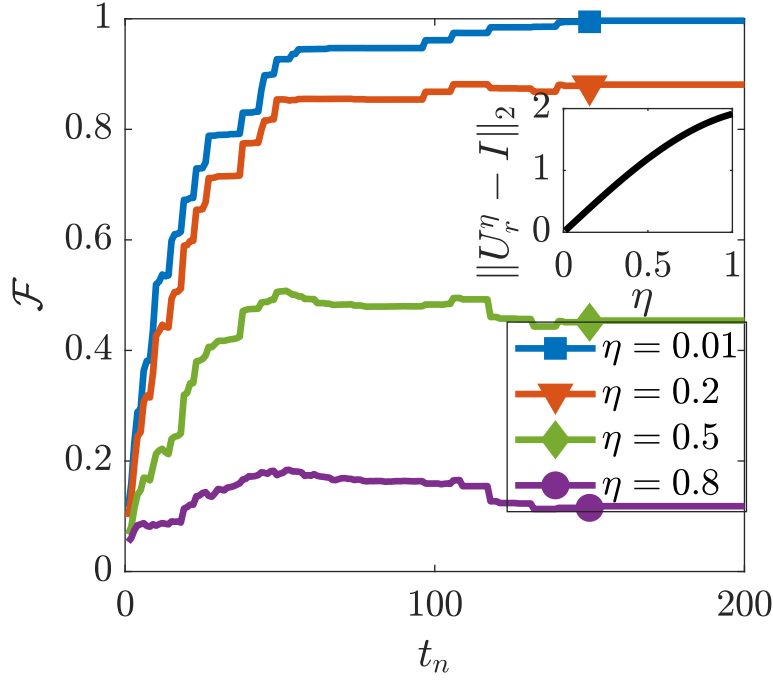

Figure 5.3: Reconstruction fidelity as a function of time in the limit of vanishing shot noise for an increase in the perturbation of $\{E_\alpha\}$. The measurement operators are the perturbed ordered set $\{E'_1, E'_2, ..., E'_k\}$ with ordered Bloch vector components of the initial state $\rho_0$ (i.e. that corresponds to the Bloch vector components, $r_\alpha = \mathrm{Tr}\,[\rho_0 E_\alpha]$ in a particular order of their magnitudes). The perturbed operators $\{E'_\alpha\}$ are generated by applying a fractional power $\eta$ of a random unitary $U_r$. The inset figure shows the Euclidean norm of the difference between $U_r^\eta$ and Identity $I$, which increases with an increase in the value $\eta$.

$$\begin{aligned}
p(\rho_\mathbf{0}|\mathbf{M},\mathcal{L},\mathcal{M}) &\propto \exp\left\{-\frac{N_s^2}{2\sigma^2}\sum_i[M_i - \sum_\alpha O_{i\alpha}r_\alpha]^2\right\} p(\rho_0|\mathcal{L},\mathcal{M}) \\
&\propto \exp\left\{-\frac{N_s^2}{2\sigma^2}\sum_{\alpha,\beta}(\mathbf{r}-\mathbf{r_{ML}})_\alpha\, C^{-1}_{\alpha\beta}\,(\mathbf{r}-\mathbf{r_{ML}})_\beta\right\} p(\rho_0|\mathcal{L},\mathcal{M})
\end{aligned} \tag{5.17}$$

In the limit of zero shot-noise, the errors due to the first term are zero, and we may purely focus on the conditional probability distribution, $p(\rho_0|\mathcal{L}, \mathcal{M})$. In terms of the observables in continuous measurement tomography, one can express $p(\rho_0|\mathcal{L}, \mathcal{M}) = p(\mathbf{r}|O_1, O_2, ..., O_n)$, giving the conditional probability of the density matrix parameters $\mathbf{r}$ till the time step $n$. For example, consider the measurement operator at the first $k$ time steps are the ordered set $\{E_1, E_2, ..., E_k\}$, giving precise information about Bloch vector components $\{r_1, r_2, ..., r_k\}$. The conditional probability distribution at time $k$ is,

$$p(\mathbf{r}|E_1, E_2, ..., E_k) = \delta(r_1 - \text{Tr}[E_1\rho_0]) \, \delta(r_2 - \text{Tr}[E_2\rho_0]) \, ... \, \delta(r_k - \text{Tr}[E_k\rho_0])$$
$$\delta\left(\sum_{\alpha \neq 1,2,...k}^{d^2-1} r_\alpha^2 = 1 - 1/d - r_1^2 - r_2^2 ... - r_k^2\right).$$
(5.18)

Each noiseless measurement above gives us complete information in one of the orthogonal directions. For example, after the first measurement,

$$p(\mathbf{r}|E_1) = \delta(r_1 - \text{Tr}[E_1\rho_0]) \, \delta\left(\sum_{\alpha \neq 1}^{d^2-1} r_\alpha^2 = 1 - 1/d - r_1^2\right). \quad (5.19)$$

Therefore, once $r_1$ is determined, the rest of the $d^2 - 2$ Bloch vector components are constrained to reside on a surface given by the equation $\sum_{\alpha \neq 1}^{d^2-1} r_\alpha^2 = 1 - 1/d - r_1^2$. The state estimation procedure shall select a state based on incomplete information consistent with $r_1$ as determined precisely by the first measurement and the remaining Bloch vector components from a point on this surface. Therefore, qualitatively, the average fidelity of the estimated state is proportional to the area of this surface. After $k$ time steps, the error is proportional to the area of the surface consistent with the equation $\sum_{\alpha \neq 1}^{d^2-1} r_\alpha^2 = 1 - 1/d - r_i^2 - r_j^2 - ... - r_k^2$. This area, quantifying the average error, decreases with each subsequent measurement.

To see it in another way, consider the fidelity between the actual and reconstructed state.

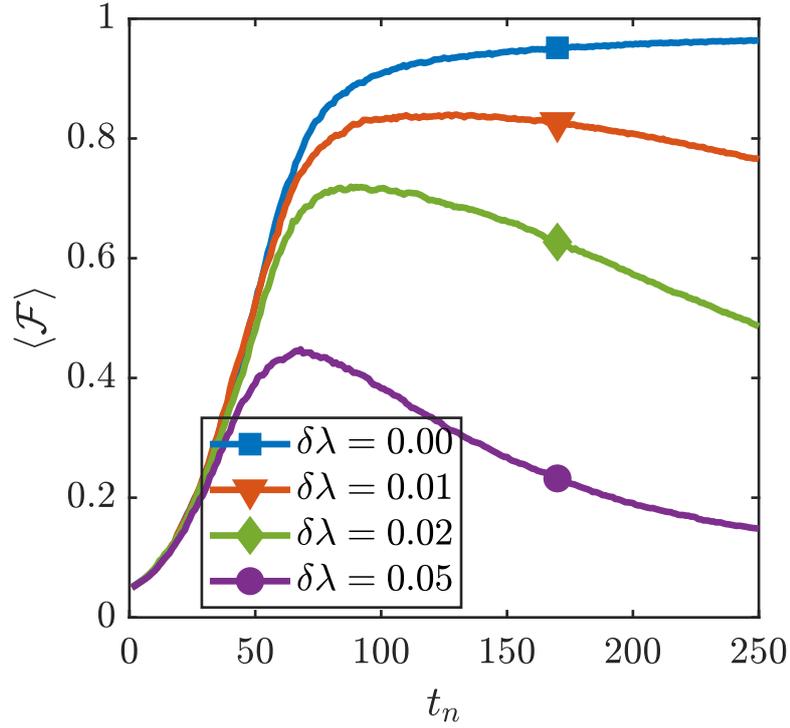

Figure 5.4: Reconstruction fidelity as a function of time for an increase in perturbation strength. The measurement record is generated for spin $j = 10$. Here we consider rotation angle $\alpha = 1.4$ and kicking strength $\lambda = 7.0$ for the quantum kicked top.

The fidelity $\mathcal{F} = \langle \psi_0 | \bar{\rho} | \psi_0 \rangle$,

$$\mathcal{F} = 1/d + \Sigma_{\alpha=1}^{d^2-1} \bar{r}_\alpha r_\alpha \qquad (5.20)$$

Here $r_\alpha$ and $\bar{r}_\alpha$ are the Bloch vectors for $\rho_0$ and $\bar{\rho}$ respectively. As one makes measurements, $E_1, E_2, ..., E_k$ and gets information about the corresponding Bloch vector components (with absolute certainty in the case of zero noise for example), one can express the fidelity as

$$\mathcal{F} = 1/d + \Sigma_{i=1}^{k} r_i^2 + \Sigma_{\alpha \neq 1,2,...k}^{d^2-1} \bar{r}_\alpha r_\alpha \qquad (5.21)$$

The term $\Sigma_{i=1}^{k} r_i^2$ puts a lower bound on the fidelity obtained after $k$ measurements. Here $\bar{r}_\alpha$ for $\alpha \neq 1, 2, ..., k$ represent the state estimator's guess for the unmeasured Bloch vector components consistent with the constraint $\left(\sum_{\alpha \neq 1,2,...,k}^{d^2-1} r_\alpha^2 = 1 - 1/d - r_1^2 - r_2^2... - r_k^2\right)$. It is this guess that picks a point from the surface with an area consistent with the above constraint.

Consider the same scenario but now with perturbations to the system dynamics. The estimate of the density matrix gets modified as

$$p(\mathbf{r}|E'_1, E'_2, ..., E'_k) = \delta(r'_1 - \text{Tr}\,[E'_1\rho_0])\,\delta(r'_2 - \text{Tr}\,[E'_2\rho_0])\,...\,\delta(r'_k - \text{Tr}\,[E'_k\rho_0])$$
$$\delta\left(\sum_{\alpha \neq 1,2,...k}^{d^2-1} r'^2_\alpha = 1 - 1/d - r'^2_1 - r'^2_2... - r'^2_k\right). \tag{5.22}$$

Here, $E'_1, E'_2, ..., E'_k$ are the perturbed operators leading to a slightly inaccurate estimate of the Bloch vector components $r'_1, r'_2, ..., r'_k$ respectively. The operators $\{E'_\alpha\}$ are obtained by rotating $\{E_\alpha\}$ by a unitary $U_r^\eta$, where $U_r$ is a random unitary and $\eta$ is a fractional power which makes $U_r^\eta$ close to identity. The Euclidean norm of the operator $U_r^\eta - I$ is less when $\eta$ is small. Thus, $\eta$ serves as the strength of perturbation in this analysis of Bloch vector components.

Despite the perturbation, the uncertainty of the Bloch vector components $r_\alpha$ for $\alpha \neq 1, 2, ..., k$ reduces to the area of the surface consistent with the equation

$$\sum_{\alpha \neq 1,2,...k}^{d^2-1} r'^2_\alpha = 1 - 1/d - r'^2_1 - r'^2_2... - r'^2_k.$$

The fidelity between the original and the estimated state now reads as

$$\mathcal{F} = 1/d + \Sigma_{i=1}^{k} r'_i r_i + \Sigma_{\alpha \neq 1,2,...k}^{d^2-1} \bar{r}_\alpha r_\alpha, \tag{5.23}$$

that we can see in Fig. 5.3. We know that the overlap between two Bloch vectors is maximum only when they are exactly aligned in the same direction, and the overlap decreases when they move far from each other. Comparing the second terms of Eq.

(5.21) and Eq. (5.23) it is now clear why with an increase in perturbation, the initial rise in fidelity is less. Therefore, the drop in fidelity is more, and the fidelity saturates at a lower value if the perturbation is more, as illustrated in Fig. 5.4. For relatively weaker perturbations, the fidelity will continue to increase when there is an information gain despite such errors to the measurement operators. The partially inaccurate information about the $j$th Bloch vector owing to perturbations to the dynamics still offsets the estimator's guess of the Bloch components of the unmeasured $j$th direction in the operator space determined by $E_j$.

In this chapter, we address how errors have consequences for the entire dynamics and affect our simulation and quantum information processing protocols. Even in the limit of vanishing errors, which can be treated as perturbations and will certainly not randomise the state, there are consequences for the fidelities obtained. We have shown that the error propagates much faster when the dynamics is regular. The operators evolved in the Heisenberg picture follow a trajectory of operators in the operator Hilbert space. Perturbation in the dynamics will direct the trajectory of time evolved operator away from the actual trajectory, and the errors in the measurement record increase with time. Thus, the error scrambles in the measurement record because of perturbation. Here we have considered the kicked top and calculated the error scrambling for the whole dynamics of operators and not just for a part of the process. We show that the measurement record differs because of the perturbation, so the reconstruction fidelity drops.

Here we notice the correlation between scrambling of errors as captured by the incompatibility between the operator and its time evolution through the *error unitary* in Eq. (5.14) and operator Loschmidt echo, as viewed from the lens of quantum tomography under chaotic dynamics. This links two fundamental quantifiers of quantum chaos, complements findings in [Yan *et al.* (2020)] and provides a different but more intuitive connection. It is indeed remarkable that we have a rather elegant expression capturing operator incompatibility that also connects to an operational interpretation of Loschmidt echo for operators.

## 5.6 DISCUSSION

We find dynamical signatures of chaos that quantify the scrambling of errors across a many-body quantum system that has consequences on the performance of quantum information and simulation protocols. We also give an operational interpretation of the operator Loschmidt echo by connecting it to the growth of distance between operators evolved in continuous measurement quantum tomography. Our results linking Loschmidt echo, error scrambling, and OTOCs will be helpful to the condensed matter community as well and in addressing broader issues involving non-integrable quantum systems [Pandey *et al.* (2020)].

Our work paves the way for further studies in the performance of quantum simulations under inadvertent noise. In the era of noisy, intermediate-scale quantum (NISQ) devices [Preskill (2018)], the accuracy of an analog quantum simulator will decay after just a few time steps. The reliability of such analog quantum simulators is highly questionable even for state-of-the-art architecture when it is likely to exhibit quantum chaos [Hauke *et al.* (2012); Lysne *et al.* (2020)]. On the contrary, the digital quantum simulation is often associated with the inherent Trotter errors [Heyl *et al.* (2019)] because of the discretization of the time evolution of a quantum many-body system as a sequence of quantum gates. Thus, a better understanding of errors in simulating many-body quantum systems and information processing protocols that exploit such rich dynamics is paramount. These signatures of chaos can be further explored using state-of-the-art experimental techniques involving cold atoms interacting with lasers and magnetic fields [Chaudhury *et al.* (2009)]. In future work, we hope to further build upon our results to develop quantum analogs of the "classical shadowing lemma" that guarantee a *true* classical trajectory in the neighbourhood of any arbitrary simulated trajectory of a chaotic system in the presence of truncation errors due to finite precision [Anosov (1967);

Grebogi *et al.* (1990); Sauer and Yorke (1991); Sauer *et al.* (1997); Vaníček (2004)].

# CHAPTER 6

# CONCLUSIONS AND OUTLOOK

## 6.1 SUMMARY

This thesis connects information gain, operator spreading, and sensitivity to perturbations with chaos in quantum dynamical systems. We studied quantum signatures of chaos in both single-body as well as many-body quantum systems. Continuous weak measurement tomography is a common thread and a powerful paradigm that has given us a window to explore various aspects of non-integrable/chaotic quantum dynamics that we summarise below.

In this thesis, we present a complete story relating chaos and information gain in quantum tomography. Continuous weak measurement tomography is a very powerful paradigm to study chaos and explore its effect on information gain. As the degree of chaos in the dynamics goes up, the rate of information gain for spin coherent states goes down, showing entirely opposite behavior to that of random states. This is because, in addition to the dynamics, coherent states are highly localized and have useful prior information that also needs to be taken into account for the purpose of state reconstruction. We incorporated the prior knowledge of the state and explained the decline in fidelity rate as the dynamics becomes increasingly chaotic. Angular momentum operators get delocalized in the phase space as we evolve them with chaotic dynamics. We saw that the degree of delocalization of operators increases with chaos. This results in a decreased fidelity in reconstructing highly localised spin coherent states. Furthermore, we showed that the ordering of operators whose expectations values are sequentially measured in tomography also plays a role in the reconstruction rate. Meanwhile, random states are highly delocalized, spreading over the entire phase space with a uniform prior and the dynamics which delocalizes the operators rapidly in the phase space causes maximal gain in information.

Information scrambling in many-body quantum systems is intimately tied to non-integrability and chaos, and the question of witnessing and quantifying this via delocalization of operators is of fundamental importance. However, this has been a contentious issue and existing ways to quantify operator spreading do not behave consistently with the degree of chaos in the system. In this thesis, we have given an unambiguous criterion for the above by connecting scrambling to information gain in quantum tomography. Employing our criterion based on tomography, we clearly saw the transition from integrability to complete chaos and demonstrated its applicability by considering examples of many-body systems like spin chains that may exhibit chaos. Our work gives an operational interpretation to operator spreading that can be realized in the laboratory via weak continuous measurements. Interestingly, various information-theoretic measures like the Shannon entropy, Fisher information, and the rank of the covariance matrix not only quantify the information gain but also quantify operator complexity.

Lastly, we unraveled dynamical signatures of chaos that quantify the scrambling of errors across a many-body quantum system that has consequences on the performance of quantum information and simulation protocols. We also gave an operational interpretation of the operator Loschmidt echo by connecting it to the growth of distance between operators evolved in continuous measurement quantum tomography. Our results linking Loschmidt echo, error scrambling, and OTOCs will be helpful to the condensed matter community as well and in addressing broader issues involving non-integrable quantum systems [Pandey *et al.* (2020)]. These signatures of chaos can be further explored using state-of-the-art experimental techniques involving cold atoms interacting with lasers and magnetic fields [Chaudhury *et al.* (2009)].

**6.2 OUTLOOK**

At a fundamental level, chaos implies unpredictability and sensitivity to initial conditions in classical systems. One of the primary goals of studying chaos in the quantum domain is to inform us about the signatures of chaos in quantum dynamical systems whose classical counterpart is chaotic. Various signatures of chaos in quantum many-body systems, such as out-of-time-order correlators (OTOCs), information gain in tomography, and hypersensitivity to perturbation in the Hamiltonian, have been explored in this thesis. Information theory has not only enabled us to characterize as well as quantify these signatures but also provided new insights into the notion of complexity and its connection to chaos. Moreover, from the point of view of application, one can ask the consequences of the presence of chaos on the reliability of quantum information processing and quantum simulations. This leads to interesting new directions that we describe below.

Quantum tomography and quantum control are two sides of the same coin. Generating an informationally complete record requires sufficient non-integrability in the dynamics. This is the very same resource that drives a fiducial state to a target state. Therefore, an interesting exploration is the quantum control of arbitrary states in the Hilbert space and whether or not chaos is a resource or nuisance towards achieving this. For example, control of well-localized states, like coherent states, is possible using regular quantum dynamics. More specifically, one can accomplish quantum control by taking Gaussian states to Gaussian states with pure rotations. One would, in general, need non-integrable quantum maps to take initial coherent states to target states that are non-Gaussian or random in nature. Since chaotic quantum maps are hyper-sensitive to errors, one needs to quantify how much chaos is optimal in order to achieve robust control.

A major thrust in quantum information science is to simulate many-body systems on a quantum device and on the related issues involving quantifying the complexity of these simulations, benchmarking these simulations in the presence of errors, and exploring connections to quantum chaos in these systems. Simulating quantum chaos on a quantum computer [Lysne *et al.* (2020); Muñoz-Arias *et al.* (2020); Krithika *et al.* (2023); Maurya *et al.* (2022)] and exploring its information-theoretic signatures like operator scrambling are of great interest. Our work, connecting continuous measurement tomography and information scrambling, paves the way to realize and interpret such experiments in the laboratory [Smith *et al.* (2006); Chaudhury *et al.* (2009)] as well as to understand and find new quantum signatures of chaos like hypersensitivity to perturbations in quantized chaotic systems and its implications on efficient simulations of such systems on a quantum computer in the presence of errors. Furthermore, designing new algorithms to simulate quantum chaotic systems and efficient calculation of semi-classical formulas on a quantum computer are related goals in the near future.

Unitarity makes the quantum systems insensitive to initial conditions, yet they do show sensitivity to perturbation in the Hamiltonian. This leads to the question of the robustness of quantum tomography and, more excitingly, quantum simulations in general. How does our ability to implement quantum tomography get affected by the perturbations in the dynamics? In what circumstances does our ability to execute quantum simulations get affected by the chaos in the dynamics? Despite truncation errors due to finite precision, the classical shadowing lemma assures us of a true classical trajectory in the neighborhood of any arbitrary simulated trajectory of a hyperbolic system. Our findings open the possibility of developing a quantum analog of the shadowing lemma, where almost all quantum systems are subject to imperfections. One can further build upon our results to develop quantum analogs of the "classical shadowing lemma" that guarantee a *true* classical trajectory in the neighbourhood of any arbitrary simulated trajectory of a chaotic system in the presence of truncation errors due to finite precision [Anosov (1967); Grebogi *et al.* (1990); Sauer and Yorke (1991); Sauer *et al.* (1997); Vaníček (2004)].

Our work on chaos has exciting connections to thermalization in closed quantum systems. We have evidence from the literature that the rate of thermalization in quantum systems is closely related to non-integrability, chaos, and symmetries respected by the system. Could we find a "thermalization witness" based on the measurement record and properties of the eigenstates of the Hamiltonian? Many-body chaos and its connections to thermalization in both closed and open quantum systems is another avenue that can be explored.

Another promising direction is the emergence of chaos in high-dimensional phase space. What consequences does dimensionality of the phase space and scaling of Lyapunov exponents of many body systems in high dimensional phase space have under environmental decoherence when treated quantum mechanically? What quantitative connections can we make about the Ehrenfest regime, Lyapunov exponents, and thermalization in the context of both open and closed many-body quantum systems? The interplay between entanglement, equilibration, and thermalization in the context of many-body systems is a subject of further exploration. For interacting many-body systems, the quantum states are high dimensional, and interference effects do not have a real space analog. We can use the techniques developed to construct semi-classical path integrals using the stationary phase approximation to explore these issues.

Lastly and perhaps most importantly, this thesis has an intimate connection to what can be achieved in the laboratory through the current state-of-the-art experimental quantum technology. Atomic-molecular-optical systems are particularly well suited to the task given the ever-increasing quantum-control toolbox, drawing from traditional coherent spectroscopy and newer developments, including non-classical light, cavity QED, laser cooling and trapping, and the production of degenerate quantum gases. All of these techniques are being integrated to provide new handles for manipulating complex systems in a clean and well-characterized geometry. Complex systems, such as the large spin ensemble, provide a unique opportunity to explore the quantum/classical interface. A standard paradigm for quantum chaos is the kicked top, recently observed in experiments in the context of internal spin control. This also opens a window to explore experimentally the Ehrenfest break time and hypersensitivity to perturbations in quantum chaotic systems.

Studying the role of decoherence and measurement in the emergence of the classical world is a promising prospect, and exploring the possibility of such experiments will be an exciting avenue.

# APPENDIX A

# MATLAB CODES FOR SIMULATION

### A.1 CONTINUOUS WEAK MEASUREMENT TOMOGRAPHY (CMT)

```
% code basically calculates the average fidelities of p random states where the dynamics
is Kicked top

F = 6; %Angular momentum
d = 2 * F + 1; %Hilbert space dimension
p = 20; %number of random states
k = 20; %total number of time steps in the evolution
z = [0.5, 2.5, 3, 7]; %chaoticity parameter
[Fx, Fy, Fz] = AngMomentum(F); %calling angular momentum function
avgfidelity = zeros(k, 1, length(z)); %initializing average fidelity
fidelityr = zeros(p, k, 1, length(z)); %initializing fidelity
hb = hermitian_basis_S(d); %hermitian basis without the identity in superoperator notation

for l = 1 : length(z)
[Fl1] = floquetop(z(l), Fx, Fz); %Floquet operator for the kicked top
OPc = eloutionRandom(Fl1, k, Fy); % time evolution of operator Fy for k time steps

for t = 1 : p
rho0 = pure_rand(d); %generates a random density matrix of dimension d
sorho0 = reshape(rho0.', d^2, 1);
for N = 1 : k
OP = OPc(:, 1 : N);
OP2 = OPc2(:, 1 : N);
```

```matlab
M = zeros(N, 1); % initialization of measurement records as a column mztrix of all zeros
for j = 1 : N
M(j) = sorho0' * OP(:, j);
end

M = real(M); %measurement record
M = M + 0.1 * randn(N, 1);
NewOP = real(hb' * OP);
NewR = NewOP * NewOP';
E = zeros(d, d, d^2 − 1);
for j = 1 : d^2 − 1
E(:, :, j) = reshape(hb(:, j)', d, d);
end

rhoest = pinv(NewR) * NewOP * (M); %least squares estimator
%Convex program: you'll need to install CVX in your MATLAB for this to work

cvx_begin
cvx_precision high
variable x(d^2 − 1);
s = zeros(d, d);
for j = 1 : d^2 − 1
s = s + x(j) * squeeze(E(:, :, j));
end
minimize(quad_form(x − rhoest, NewR));
subject to
eye(d)/d + s == hermitian_semidefinite(d);
cvx_end
%End of convex program
```

```
%put back the Identity component in the density matrix and return to standard matrix
representation from superoperator notation
s = zeros(d, d);
for j = 1 : d^2 − 1
s = s + x(j) ∗ squeeze(E(:, :, j))
end
rho = eye(d)/d + s;
fidelity = (trace(sqrtm(sqrtm(rho) ∗ rho0 ∗ sqrtm(rho)))).^2;

fidelityr(t, N, 1, l) = fidelity;
end

end

for N = 1 : k
avgfidelity(N, 1, l) = real(mean(fidelityr(:, N, 1, l)));
end
end
```

## A.2  FUNCTIONS REQUIRED FOR CMT

### A.2.1  Angular momentum operators

```
function[Fx, Fy, Fz] = AngMomentum(F)
%calculates the operator for the angular momentum along the x and y directions, in
mvalues = −F : 1 : F;
dim = 2 ∗ F + 1;

Fplus = zeros(dim);
Fminus = zeros(dim);

for j = 1 : (dim − 1)
```

```matlab
m = mvalues(j);
Fplus(j, j + 1) = sqrt((F − m) * (F + m + 1));
end

for j = 2 : (dim)
m = mvalues(j);
Fminus(j, j − 1) = sqrt((F + m) * (F − m + 1));
end

Fx = 0.5 * (Fplus + Fminus);
% Fplus and Fminus are the $J_+$ and $J_-$ matrices of the angular momentum.
Fy = (0.5/1i) * (Fplus − Fminus);
Fz = −diag(mvalues);
```

### A.2.2 Generalized Gellmann matrices

```matlab
function[out] = hermitian_basis_S(d)
out = zeros(d, d, d^2 − 1);
for ii = 1 : d − 1
out(:, :, ii) = diag([ones(1, ii), −ii, zeros(1, d − 1 − ii)])./sqrt(ii + ii.^2);
end
s = d * (d − 1)./2;
forii = 2 : d
for jj = 1 : ii − 1
n = (ii − 2) * (ii − 1)./2 + jj + d − 1;
out(ii, jj, n) = 1./sqrt(2);
out(jj, ii, n) = 1./sqrt(2);
out(ii, jj, n + s) = 1i./sqrt(2);
out(jj, ii, n + s) = −1i./sqrt(2);
end
```

end

out = reshape(out, [d^2, d^2 − 1]);

return

### A.2.3 Floquet map for kicked top

function[Fl1] = floquetop(lambda1, alpha1, Jx, Jz)

jota = (0.5 ∗ (length(Jz) − 1));

Fl1 = expm(−1i ∗ lambda1 ∗ Jz^2/(2 ∗ jota)) ∗ expm(−1i ∗ alpha1 ∗ Jx);

% lambda1 and alpha1 are two parameters of the kicked top

### A.2.4 Time evolved operators

function[OP] = eloutionRandom(U, N, O0)

F = floor(0.5 ∗ (length(U) − 1));

OP = zeros((2 ∗ F + 1)^2, N);

U1 = eye(2 ∗ F + 1);

for j = 1 : N

O = U1′ ∗ O0 ∗ U1;

OP(:, j) = reshape(O.′, (2 ∗ F + 1)^2, 1);

% it gives the operator in a column matrix of 2*F+1 elements

U1 = U ∗ U1;

end

### A.2.5 Haar random state generation

function rho = pure_rand(d)

%this funciton generates a pure random density matrix chosen from the haar measure.

%input is the dimension of the matrix.

rho1 = randn(d, 1) + 1i ∗ randn(d, 1);

rho1 = rho1/sqrt(rho1′ ∗ rho1);

rho = rho1 ∗ rho1′;

# CURRICULUM VITAE

**NAME**                Abinash Sahu

**DATE OF BIRTH**       01 July 1996

**EDUCATION QUALIFICATIONS**

| 2016 | **Bachelor of Science** | |
|---|---|---|
| | Institution | Govt Autonomous College, Angul, Odisha |
| | Specialization | Physics |

| 2018 | **Master of Science** | |
|---|---|---|
| | Institution | Pondicherry University, Puducherry |
| | Specialization | Physics |

| 2024 | **Doctor of Philosophy** | |
|---|---|---|
| | Institution | Indian Institute of Technology Madras, Chennai |
| | Specialization | Physics |
| | Registration Date | 09 July 2018 |

# DOCTORAL COMMITTEE

**Chairperson**       Prof. Arul Lakshminarayan
                      Professor and Head
                      Department of Physics, IIT Madras

**Guide(s)**          Dr. Vaibhav Madhok
                      Associate Professor
                      Department of Physics, IIT Madras

**Member(s)**         Dr. Pradeep Kiran Sarvepalli
                      Associate Professor
                      Department of Electrical Engineering, IIT Madras

                      Dr. Sunethra Ramanan
                      Associate Professor
                      Department of Physics, IIT Madras

                      Dr. Sivarama Krishnan
                      Associate Professor
                      Department of Physics, IIT Madras